\documentclass[opre]{informs4}
\RequirePackage{tgtermes}
\RequirePackage{newtxtext}
\RequirePackage{newtxmath}
\RequirePackage{bm}
\RequirePackage{endnotes}

\OneAndAHalfSpacedXI

\usepackage{tabularx,booktabs}
\newcolumntype{Y}{>{\centering\arraybackslash}X}
\usepackage{diagbox}
\usepackage{multirow}

\usepackage{titlesec}
\usepackage{algorithm}
\usepackage{algpseudocode}
\usepackage{tikz}
\usetikzlibrary{positioning, arrows.meta, decorations.markings, calc, intersections,decorations.pathmorphing,patterns}
\usepackage[skip=0pt]{caption}
\usepackage{subcaption}
\usepackage{graphicx}
\usepackage{amsmath}
\usepackage{amsfonts}
\usepackage{amssymb}
\usepackage{amsbsy}
\usepackage{color,soul}
\def\R{{\mathbb{R}}}   

\def\N{{\mathbb{N}}}
\def\E{{\mathbb{E}}}
\def\P{{\mathbb P}}

\def\1{{\mathbf{1}}}

\DeclareMathOperator*{\diag}{diag}

\DeclareMathOperator*{\diver}{div}
\DeclareMathOperator*{\dist}{dist}

\DeclareMathOperator*{\esssup}{ess\,sup}
\newcommand*{\QEDB}{\null\nobreak\hfill\ensuremath{\square}}%

\usepackage{natbib}
 \bibpunct[, ]{(}{)}{,}{a}{}{,}%
 \def\bibfont{\small}%
\usepackage{cleveref}

\EquationsNumberedThrough    
\crefname{assumption}{assumption}{assumptions}
\crefformat{equation}{(#2#1#3)}
\Crefname{equation}{}{}

\TheoremsNumberedThrough     
\ECRepeatTheorems  %

\MANUSCRIPTNO{MOOR-0001-2024.00}

\begin{document}


 \RUNAUTHOR{Ren, Fu, and L'Ecuyer}

\RUNTITLE{Stochastic Derivative Estimation through Leibniz Integration}

\TITLE{Stochastic Derivative Estimation for Discontinuous Sample Performances: A Leibniz Integration Perspective}

\ARTICLEAUTHORS{\vspace*{-1em}%
\AUTHOR{Xingyu Ren}
\AFF{Department of Electrical and Computer Engineering \& Institute for System Research,
University of Maryland, College Park, \EMAIL{renxy@umd.edu}}

\AUTHOR{Michael C. Fu}
\AFF{Robert H. Smith School of Business \& Institute for System Research,
University of Maryland, College Park, \EMAIL{mfu@umd.edu}}

\AUTHOR{Pierre L'Ecuyer}
\AFF{Département d'Informatique et de Recherche Opérationnelle,
Université de Montréal, \EMAIL{lecuyer@iro.umontreal.ca}}
\vspace*{-1em}} 

\ABSTRACT{
Based on the multidimensional Leibniz integral rule, we propose a novel unbiased stochastic derivative estimation framework for sample performance functions that are discontinuous in the parameter of interest. In this framework, for general discontinuities, a change of variables followed by the Leibniz integral rule generally yields a multi-run estimator, whereas for indicator-type discontinuities, a suitable transformation of the sample performance leads directly to a single-run estimator. The new framework includes existing likelihood ratio (LR)-based approaches as special cases and applies more broadly. In particular, it encompasses generalized LR (GLR) methods and the new formulation permits us to obtain much milder and simpler sufficient regularity conditions for these estimators to be unbiased, compared with the conditions given in previous papers. We also provide guidance on how to construct efficient single-run estimators in settings where existing methods may require the simulation of multiple sample paths or may be inapplicable. We validate the effectiveness and robustness of the proposed framework through an inventory-price control problem.
\vspace*{-0.5em}
}%





\KEYWORDS{Stochastic derivative estimation, Leibniz integral rule, Perturbation analysis, Likelihood ratio\vspace*{-1em}} 

\maketitle
\vspace*{-1em}
\section{Introduction}\label{sec:intro}
Evaluating the derivative of a stochastic system's output with respect to (w.r.t.) its parameters is crucial for sensitivity analysis and optimization. We consider a model with an output sample performance $\psi(X,\theta)$, where $\psi:\R^n\times\Theta\mapsto\R$ is real-valued, $\Theta$ is an open interval in $\R$, $\theta \in \Theta$ is a parameter of interest, and $X$ is an input random vector with a cumulative distribution function (CDF) $F(\cdot,\theta)$ and support $\Omega\subseteq\R^n$, assumed to be independent of $\theta$. This assumption is without loss of generality: when the support is $\theta$-dependent, the dependence can be represented through an indicator function and absorbed into the sample performance $\psi$, so that the model is equivalently formulated over a $\theta$-independent support. Let $dF(\cdot,\theta)$ denote the induced probability measure. Our goal is to estimate the derivative w.r.t. $\theta$ of
\vspace*{-0.5em}
\begin{align*} 
    \E(\psi(X,\theta))=\int_{\Omega} \psi(x,\theta) dF(x,\theta).
\end{align*}
For example, in a queueing system, $X$ may include interarrival and service times, $\psi(X,\theta)$ could be the average queue length, and $\theta$ could be the service rate. For vectors of parameters, we simply estimate the derivative w.r.t. each parameter separately, which gives a gradient estimator. Commonly used methods for derivative estimation include finite-difference (FD) approximations, perturbation analysis (PA), likelihood ratio (LR), and weak derivatives (WD), aka measure-valued differentiation (MVD) \citep{glasserman1990gradient,oLEC91a,fu2012conditional,glynn1987likelilood,fu2008you,pflug2012optimization,heidergott2006measure,fu2006gradient,fu2006sensitivity,fu2015stochastic}. FD methods are straightforward to implement, since they do not require analytical derivatives of the sample performance or the input distribution; however, they introduce bias and require simulating multiple sample paths \citep{glasserman2003monte}. The most fundamental PA method is infinitesimal PA (IPA) \citep{suri1988perturbation,heidelberger1988convergence,glasserman1990gradient,glasserman1991structural,jiang2015estimating}, which directly computes the pathwise stochastic derivative $\frac{d}{d\theta}\psi(X,\theta)$ for a structural parameter $\theta$. When it applies, IPA is unbiased and usually efficient (single-run and low variance), but it cannot handle distributional parameters or discontinuous sample performance functions \citep{cao1985convergence,glasserman1991structural,broadie1996estimating}. The LR method \citep{glynn1987likelilood,reiman1989sensitivity}, aka the score function (SF) method \citep{rubinstein1986score}, estimates derivatives by differentiating the underlying probability measure w.r.t. $\theta$ instead of the sample performance. Similar to IPA, under appropriate conditions, the LR method is unbiased and single-run, but it cannot handle structural parameters and usually produces higher variance than IPA \citep{l1990unified,cui2020variance}. A unified theoretical framework integrating both IPA and LR methods is introduced in \citet{l1990unified}. The domains where IPA and LR methods are applicable overlap, allowing them to be combined or selected interchangeably depending on the problem formulation.

In this paper, we focus on a class of LR-based methods for handling discontinuous sample performance functions via a change of variables, aka ``push-out'' \citep{rubinstein1992sensitivity,pflug2002inventory}. Through push-out, the parameter $\theta$ is shifted from within the discontinuous sample performance $\psi$ to the underlying probability measure. 
From a broader perspective \citep{l1990unified}, the push-out technique avoids differentiating a discontinuous sample performance by redefining the underlying probability space. To illustrate, consider a univariate input $X$ with density $f(x,\theta)$. Suppose there exists a real-valued function $g(x,\theta)$, invertible w.r.t. $x$ for each $\theta$, and differentiable in both arguments. Let $g^{-1}(y,\theta)$ denote the inverse of $g(x,\theta)$ w.r.t. $x$. Additionally, assume the sample performance takes the form $\psi(x,\theta)= \varphi (g(x,\theta))$,
where $\varphi:\R\mapsto\R$ may be discontinuous. Let $\partial_y$ denote differentiation w.r.t. $y$, and similarly for other variables throughout the paper. With the change of variables $Y = g(X, \theta)$, the transformed density is $\tilde f(y, \theta) = f(g^{-1}(y, \theta), \theta)\left| \partial_y g^{-1}(y, \theta) \right|$ supported on $\tilde \Omega = g(\Omega, \theta)$. If $\tilde \Omega$ is independent of $\theta$, the expected performance becomes:
\vspace*{-0.5em}
\begin{align*}
    \E(\psi(X,\theta))=\int_{\tilde\Omega}  \varphi(y) \tilde f(y,\theta) dy = \E(\varphi(Y)),
\end{align*}
where the dependence on $\theta$ is fully transferred to the transformed density $\tilde{f}$, and the LR method applies. The support-independent unified LR-IPA (SLRIPA) method \citep{wang2012new} is closely related to the push-out LR method but does not require fully removing $\theta$ from the sample performance. Instead, it only requires that the transformed sample performance function is differentiable w.r.t. $\theta$, and the IPA-LR method applies \citep{l1990unified}. Another closely related method is the generalized LR (GLR), introduced by \citet{peng2018new,peng2020generalized,heidergott2023gradient}. Unlike push-out LR and SLRIPA, GLR does not require an explicit change of variables and is more generally applicable, assuming only local invertibility of $g$. 
By applying integration by parts, GLR avoids differentiating discontinuous functions and expresses the derivative of the expected performance as the sum of a volume integral and a surface integral. 
\citet{peng2018new} study cases where the surface integral vanishes and show that GLR then coincides with push-out LR. \citet{peng2020generalized} consider scenarios where the input random vector has i.i.d. standard uniform components, for which the surface integral can be estimated by a single sample path. GLR presents two important challenges. First, the surface integral can significantly complicate the estimator and, when it is nonvanishing and the input is not i.i.d., may require the simulation of multiple sample paths, making estimation computationally expensive. Second, the analyses in \citet{peng2018new,peng2020generalized} rely on assumptions that are restrictive and often difficult to verify in practice. One of the main goals of this paper is to identify ways to avoid the surface integral and relax these assumptions.

In this paper, we develop derivative estimation methods for discontinuous sample performance functions based on the \textit{Leibniz integral rule}, which enables differentiation under the integral sign when both the integrand and the \textit{integration domain} depend on the parameter, providing greater flexibility for designing new estimators and extending existing methods to a broader range of problems. To illustrate, the classical one-dimensional Leibniz integral rule states that for differentiable functions $a,b:\Theta\mapsto\R$, and $G:\Omega\times\Theta\mapsto\R$,
\begin{align*}
    \frac{d}{d\theta}\int^{b(\theta)}_{a(\theta)}G(x,\theta)dx=b'(\theta)G(b(\theta),\theta)-a'(\theta)G(a(\theta),\theta) + \int^{b(\theta)}_{a(\theta)}\partial_\theta G(x,\theta) dx,
\end{align*}
where the boundary terms arise from differentiating the limits of integration. We use multidimensional extensions of this formula, where the boundary terms generalize to surface integrals. We also consider an alternative formulation, called the \textit{Leibniz divergence rule}, which, under more restrictive conditions, applies the divergence theorem to convert the surface integral into a single volume integral. 
The distinction between these two forms is especially important in stochastic derivative estimation: evaluating surface integrals may require sampling from multiple conditional distributions, which becomes costly for high-dimensional, correlated inputs, whereas volume integrals allow for simpler, single-run estimation. Technical details are provided in \Cref{sec1:leibniz}. We summarize the main contributions of our work as follows:
\begin{itemize}
    \item \textbf{We develop a stochastic derivative estimation framework that is substantially more general than existing approaches.} Our framework unifies and extends existing methods such as IPA, LR, push-out LR, SLRIPA, and GLR, and thus applies to a broader class of discontinuous sample performance functions and problem settings. In particular, for general discontinuous sample performance functions that admit a change of variables, we combine the push-out LR and SLRIPA methods with the \textit{Leibniz integral rule} to handle parameter-dependent domains. The resulting push-out Leibniz method offers greater flexibility in the choice of change of variables, thereby expanding the range of problems that can be handled.
    \item \textbf{Our methods can avoid the surface integral and thereby enable efficient single-run estimation in many settings where GLR becomes computationally burdensome.} Although the push-out Leibniz estimator shares the same general mathematical form as GLR methods, the Leibniz framework provides constructive geometric insight into when the surface integral vanishes, for example, when the transformed input has parameter-independent support. Moreover, for sample performance functions whose discontinuities arise from indicator functions, we show that, under suitable conditions, the \textit{Leibniz divergence rule} yields a novel single-run unbiased derivative estimator. This is especially valuable when the input is correlated or otherwise non-i.i.d., since evaluating the GLR surface integral in such settings may require simulating multiple sample paths and can be computationally costly or even prohibitive.
    \item \textbf{Our theoretical justification relies on substantially milder and more easily verifiable assumptions than existing GLR analyses.} In particular, our unbiasedness proofs are established under weaker regularity conditions than those imposed in previous GLR papers, thereby showing that these methods apply in settings beyond those previously justified.
\end{itemize}

The rest of this paper is organized as follows. In \Cref{sec1:leibniz}, we introduce the Leibniz integral rule and the Leibniz divergence rule, and illustrate their applications and advantages through examples that motivate the estimators developed in the subsequent sections. \Cref{sec2:indicator} focuses on sample performance functions with discontinuities caused by indicator functions. In \Cref{sec2:inv}, we examine general discontinuous sample performance functions that admit a change of variables. In \Cref{sec3:relate}, we discuss implementation issues and extensions, including surface integral estimation for hyperrectangle-supported inputs, relaxation and simplification of regularity conditions, local changes of variables, and unbounded sample spaces. 
In \Cref{sec:compare}, we compare the Leibniz and GLR estimators, demonstrating that the Leibniz framework requires milder regularity conditions than GLR and enables the elimination of the surface integral term. In \Cref{sec:simulation}, we demonstrate the effectiveness and robustness of the proposed estimators through an inventory-price control problem, where existing methods either suffer from high variance or become inapplicable because of correlated inputs or violated regularity conditions. \Cref{sec:conc} concludes.

Due to space constraints, some technical details and simulation results are relegated to an online electronic companion providing the following supplemental materials. \ref{appd:geo-leibniz} presents a geometric explanation of the Leibniz integral rules. The derivations of the estimators in \Cref{sec1:leibniz,sec2:indicator,sec2:inv} are given in \ref{appd:estimator-derive}. 
\ref{appd:ipalr-leibniz} and \ref{appd:div-discont} provide detailed discussions of the material in \Cref{sec:ipalr-leibniz}, explaining how the Leibniz framework is connected to IPA-LR. Additional simulation experiments related to \Cref{exp2:log,exp4:leibniz}, involving correlated input random variables with dependence structures modeled by copulas, are provided in \ref{appd:exp-copula} and \ref{appd:copula}. \ref{appd:details-simu} provides the derivation of the estimators and additional simulation results for the inventory-price model in \Cref{sec:simulation}. Although we focus on LR-based methods, PA extensions like Smoothed PA (SPA) \citep{fu2012conditional} and Conditional Monte Carlo/Change-of-Variables (CMC-CoV) \citep{feng2016conditional} can also handle discontinuities. These PA extensions may have lower variance, but their implementation is highly problem-specific. For further details and comparisons, see \ref{appd:prac-implem} and \ref{appd:cmccov-leibniz}.

\section{Leibniz Rules in $\R^n$ with Motivating Examples}\label{sec1:leibniz}



In this section, we introduce the Leibniz integral rule and the Leibniz divergence rule, and illustrate their application to stochastic derivative estimation using two simple examples.
\vspace*{-0.5em}
\begin{lemma}\label{thm:leibniz}
    \textcolor{red}{Let $U\subset \R^n$ be a Lipschitz domain and let $\overline{U}$ denote its closure. Suppose $\phi:U\times\Theta\mapsto\R^n$ is continuously differentiable in both arguments and, for each $\theta$, $\phi(\cdot,\theta)$ extends to a diffeomorphism on an open neighborhood of $\overline U$, with inverse denoted by $\phi^{-1}(\cdot,\theta)$. Assume that $\partial_\theta\phi(u,\theta)$ is continuously differentiable w.r.t. $u$ for each $\theta\in\Theta$. Let $J_\phi(u,\theta)$ denote the Jacobian matrix of $\phi$, given by
    \vspace*{-0.5em}
    \begin{align*}
        J_\phi(u,\theta)=\begin{bmatrix}
        \partial_{u_1} \phi_1(u,\theta) &\partial_{ u_2 } \phi_1(u,\theta)&\cdots &\partial_{ u_n} \phi_1(u,\theta)\\
        \partial_{ u_1 } \phi_2(u,\theta)&\partial_{ u_2} \phi_2(u,\theta) &\cdots &\partial_{ u_n} \phi_2(u,\theta)\\
        \cdots &\cdots &\ddots &\cdots\\
        \partial_{ u_1} \phi_n(u,\theta) &\partial_{ u_2} \phi_n(u,\theta) &\cdots &\partial_{ u_n} \phi_n(u,\theta)
        \end{bmatrix}.
    \end{align*}
    Assume that $\det J_\phi(u,\theta)\neq 0$ for all $(u,\theta)\in U\times\Theta$. Denote $D_\theta=\phi(U,\theta)$.} Let $\Omega\subseteq\R^n$ be an open set containing $\overline D_\theta$ for every $\theta\in\Theta$, and let $G:\Omega\times\Theta\mapsto \R$ be a scalar-valued function continuously differentiable in both arguments. 
    Then the following equation, called the \textbf{Leibniz integral rule}, holds:
    \vspace*{-0.5em}
    \begin{align}\label{eq:leibniz-sec1}
        \frac{d}{d\theta}\int_{D_\theta} G(x,\theta)dx&=\int_{\partial D_\theta}G(x,\theta) \left(\vec v(x,\theta) \cdot\vec n(x,\theta) \right) d\sigma \ +\int_{D_\theta}\partial_\theta G(x,\theta)dx,
    \end{align}
    where $\vec v(x,\theta) = \partial_\theta \phi(u,\theta)|_{u=\phi^{-1}(x,\theta)}$, $\vec n(x,\theta)$ is the outward unit normal vector to $\partial D_\theta$, and $\sigma$ denotes the surface measure on $\partial D_\theta$, defined as the $(n-1)$-dimensional Hausdorff measure induced by the Lebesgue measure on $\R^n$. Applying the divergence theorem transforms the first integral on the right-hand side (RHS), which is taken over $\partial D_\theta$ and is therefore referred to as a surface integral, into a volume integral, yielding the following equivalent form, called the \textbf{Leibniz divergence rule}:
    \vspace*{-0.5em}
    \begin{align}\label{eq:leibniz-div-sec1}
        \frac{d}{d\theta}\int_{D_\theta} G(x,\theta)dx=\int_{D_\theta} \left( \diver(G(x,\theta)  \vec v (x,\theta)) + \partial_\theta G(x,\theta) \right) dx,
    \end{align}
    where $\diver$ is the divergence operator, defined as $\diver(\vec v)=\sum_{i=1}^n \partial_{x_i}\vec v_i$ for a vector field $\vec v:\R^n\mapsto\R^n$ with components $\vec v_i$. 
    
\end{lemma}
\vspace*{-0.5em}
Both \Cref{eq:leibniz-sec1,eq:leibniz-div-sec1} are special cases of the general Leibniz integral rule given in \citet[Section 7 and 8]{flanders1973differentiation}, with a broader formulation given in \citet[Theorem 2.11, Chapter XII]{amann2005analysis}. Since we use both forms throughout the paper and wish to distinguish them clearly, we assign separate names to each. On the RHS of \Cref{eq:leibniz-sec1}, the volume integral results from differentiating the integrand, while the surface integral arises from differentiating w.r.t. the boundary of the integration domain (see \ref{appd:geo-leibniz} for a geometric interpretation). The velocity vector $\vec v(x,\theta)$ describes how each point $x\in D_\theta$ moves w.r.t. $\theta$
and can be computed explicitly as follows. Since $\phi(\phi^{-1}(x,\theta),\theta)=x$, by (implicit) differentiation, $0=\frac{d}{d\theta}\phi(\phi^{-1}(x,\theta),\theta)=\partial_\theta \phi(\phi^{-1}(x,\theta),\theta) + J_\phi(\phi^{-1}(x,\theta),\theta)\partial_\theta \phi^{-1}(x,\theta)$, 
i.e., $\vec v(x) = - J_\phi(\phi^{-1}(x,\theta),\theta)\partial_\theta \phi^{-1}(x,\theta)$.

Distinctions between \Cref{eq:leibniz-div-sec1,eq:leibniz-sec1} are particularly important for stochastic derivative estimation:
\begin{enumerate} 
    \item Applying the Leibniz integral rule \Cref{eq:leibniz-sec1} may introduce a surface integral, whereas the Leibniz divergence rule \Cref{eq:leibniz-div-sec1} does not. Volume integrals can typically be expressed as expectations and estimated from a single sample path. In contrast, as we will show in \Cref{sec3-1:GLR}, when the support of the input is a hyperrectangle, the surface integral can be written as a weighted sum of conditional expectations, with the number of terms scaling linearly with the input dimension. Estimating these terms may require simulating multiple sample paths, particularly when the input components are dependent. Furthermore, if the support lacks a simple structure (e.g., is not a hyperrectangle), deriving the sampling distributions and computing the normal and velocity vectors becomes more challenging. Therefore, it is desirable to avoid the surface integral when the support is complex or when the input random vector has dependent components.
    \item Applying the Leibniz divergence rule \Cref{eq:leibniz-div-sec1} requires differentiability of $G(x,\theta)$ w.r.t. both $x$ and $\theta$ over $\Omega\times\Theta$. In contrast, under milder conditions, the Leibniz integral rule \Cref{eq:leibniz-sec1} only requires differentiability w.r.t. $\theta$, which makes \Cref{eq:leibniz-sec1} applicable to broader classes of problems.
\end{enumerate}

Based on the above distinctions, we apply \Cref{eq:leibniz-div-sec1,eq:leibniz-sec1}, respectively, to the following two scenarios involving discontinuous sample performance functions:
\begin{enumerate}
    \item In the first scenario, when discontinuities arise from \textit{indicator} functions, we absorb the indicator functions directly into the integration domain, leaving a differentiable integrand. If the resulting domain admits a suitable parametrization, the \textbf{Leibniz divergence rule} \Cref{eq:leibniz-div-sec1} applies and yields a single-run estimator.
    \item In the second scenario, we consider a \textit{general discontinuous} sample performance that permits a change of variables (“push-out”), transferring the parameter $\theta$ from the sample performance into the sample space and the underlying probability measure. Unlike the first scenario, the transformed sample performance remains discontinuous in the (transformed) input random vector, making the Leibniz divergence rule \Cref{eq:leibniz-div-sec1} inapplicable. Applying the \textbf{Leibniz integral rule} \Cref{eq:leibniz-sec1} in this case yields both a volume integral and a surface integral, with the latter potentially requiring simulation from multiple sample paths.
\end{enumerate}

The assumption in the first scenario that discontinuities arise from indicator functions encompasses many practical applications, including inventory management, control charts, option pricing, and distribution sensitivity \citep{peng2018new,peng2020generalized,l2022monte}. The following toy example, a slight generalization of the density estimation problem in \citet[Example 3]{puchhammer2022likelihood}, gives an illustration.
\vspace*{-0.5em}
\begin{example}\label{exp1:max}
    Let $X=(X_1,X_2)$ be a random vector supported on $[0,1]^2$, with a differentiable density $f:[0,1]^2\mapsto \R_+$ (parameter-independent). Consider the sample performance $\psi(X,\theta)=\1\{\max\{X_1,X_2\}\leq \theta\}=\1\{X_1\leq \theta\}\1\{X_2\leq \theta\},~\theta\in(0,1).$ Note that $\frac{d}{d\theta}\E(\psi(X,\theta))$ yields the density of the random variable $\max\{X_1,X_2\}$ at $\theta$. We absorb the indicator function into the integration domain as follows:
    \begin{align*}
        \E (\psi(X,\theta)) = \int_{(0,1)^2}\1\{x_1\leq\theta\}\1\{x_2\leq\theta\} f(x)dx = \int_{D_\theta}  f(x) dx,
    \end{align*}
    where $D_\theta=[0,\theta]^2$. This reformulation yields a differentiable integrand over a $\theta$-dependent domain. Define $\phi(u,\theta) = \theta u$ for $u \in \R^2$, so that $D_\theta = \phi(U,\theta)$ with $U = [0,1]^2$. Applying the Leibniz divergence rule \Cref{eq:leibniz-div-sec1} yields (see \ref{appd:estimator-derive} for additional details):
    \begin{align*}
        \frac{d}{d\theta}\E (\psi(X,\theta)) = \int_{D_\theta}  \left(\sum_{i=1}^2\frac{x_i\partial_{x_i}f(x)}{\theta f(x)}+\frac{2}{\theta} \right) f(x) dx = \E\left(\psi(X,\theta)\left(\sum_{i=1}^2\frac{X_i\partial_{x_i}f(X)}{\theta f(X)}+\frac{2}{\theta}\right)\right),
    \end{align*}
    which leads to a single-run unbiased derivative estimator given inside the last expectation. \QEDB
    \vspace*{-0.5em}
\end{example}
In this example, since na\"ive changes of variables such as $Y = X - \theta$ or $Y = X/\theta$ lead to $\theta$-dependent support for $Y$, methods like push-out LR or SLRIPA are inapplicable unless a more suitable (and often nontrivial) transformation is identified. This example can be addressed by the GLR method, which fits the distribution sensitivity estimation setting in \citet[Section 4.1]{peng2020generalized};
however, in addition to a volume integral (which can be expressed as a standard expectation), the GLR method requires evaluating a surface integral, which, as we will show in \Cref{sec3-1:GLR}, can be rewritten as a weighted sum of conditional expectations:
\vspace*{-0.2em}
\begin{align}\label{eq:glr-exp1}
    \frac{d}{d\theta}\E (\psi(X,\theta)) 
    =\E\left(\psi(X,\theta)\frac{\sum_{i=1}^2 \partial_{x_i} f(X)}{f(X)}\right)+\underbrace{\sum_{i=1}^2 f_{X_i}(0) \E \left(\psi(X) | X_i=0 \right)}_{\text{``surface integral''}},
    \vspace*{-0.5em}
\end{align}
where $f_{X_i}$ is the marginal density of $X_i$. Estimating the derivative from \Cref{eq:glr-exp1} requires additional samples from the conditional densities $f_{X|X_1=0}$ and $f_{X|X_2=0}$, where $f_{X|X_i=0}$ denotes the density of $X$ conditional on $X_i = 0$.
The GLR method can avoid this extra sampling in two special cases: (1) when the density $f$ vanishes on $\partial \Omega$, the surface integral disappears \citep{peng2018new}; or (2) when $X$ has independent components, the surface integral can be estimated using the same single sample path as the volume integral \citep{peng2020generalized}. For instance, \citet[Example 3]{puchhammer2022likelihood} considers a special case of \Cref{exp1:max} with i.i.d. uniform inputs. Their resulting estimator coincides with the GLR estimator in \citet{peng2020generalized}, and the independence allows the surface integral to be estimated without additional sampling. However, as in \Cref{exp1:max}, various practical problems may not satisfy these conditions. Although representing $X$ as a function of independent uniforms via inverse CDFs, as in \citet[Example 1]{peng2020generalized}, is theoretically possible, the inverse of (conditional) CDFs and their derivatives required for constructing GLR estimators could be unavailable in closed form or analytically intractable, necessitating additional numerical methods for approximation (e.g., for Gamma or Chi-squared distributions).

To illustrate the second scenario where the Leibniz integral rule \Cref{eq:leibniz-sec1} is combined with the push-out LR method, consider the following example. 
\vspace*{-0.5em}
\begin{example}\label{exp2:log}
    Let $X=(X_1,X_2)$ be a random vector supported on $\R^2_+$, with a differentiable density $f:\R^2_+\mapsto \R_+$ (parameter-independent). Consider the sample performance $\psi(X,\theta)=\1\{\sum_{i=1}^{2}\log(X_i+\theta)<q\}$, where the parameter of interest is $\theta\in(1,\infty)$ and $q>0$ is a constant. This form appears in the expression for the regenerative cycle length of the inventory-price model discussed in \Cref{sec:simulation}---with $\theta$ as the price, $\log(X_i + \theta)$ as the per-period demand, and $q$ as the replenishment gap. Let $g=(g_1,g_2)$ with $g_i(x,\theta)=\log(x_i+\theta)$, $i=1,2$, and define $\varphi(y)=\1\{y_1+y_2<q\}$ with $y=(y_1,y_2)$, so that $\psi(X,\theta)=\varphi(g(X,\theta))$. Consider the change of variables $Y=g(X,\theta)$ and define the density of $Y$ as $\tilde f(y, \theta) = f(g^{-1}(y, \theta))| \det J_{g^{-1}}(y, \theta) |$. The expected performance becomes $\E(\varphi(Y))=\int_{g(\R_+^2,\theta)}\varphi(y) \tilde f(y,\theta) dy$. Since $\varphi$ is discontinuous, only the Leibniz integral rule \Cref{eq:leibniz-sec1} applies. Applying \Cref{eq:leibniz-sec1} and reversing the change of variables, we obtain an expression for $\frac{d}{d\theta}\E(\psi(X,\theta))$ that coincides in form with \Cref{eq:glr-exp1}, yielding the following unbiased estimator (see \Cref{sec2:inv} and \ref{appd:estimator-derive} for details):
    \begin{align}\label{eq:exp2-leibniz}
        -\psi(X,\theta)\frac{\sum_{j=1}^2\partial_{x_j}f(X)}{f(X)}\bigg|_{X\sim f}-\sum_{i=1}^{2}\psi(X,\theta)f_i(0,\theta)\bigg|_{X\sim f_{X|X_i=0}},
    \end{align}
    which 
    necessitates additional sampling from $f_{X|X_1=0}$ and $f_{X|X_2=0}$. \QEDB
\end{example}
Similar to \Cref{exp1:max}, the indicator function can be absorbed into the integration domain; however, parameterizing the resulting domain is not straightforward. In \Cref{sec2:indicator}, we present a method for doing so and apply the Leibniz divergence rule \Cref{eq:leibniz-div-sec1} to derive a single-run estimator. Although applying the GLR method with the same $g$ and $\varphi$ yields the same estimator as \Cref{eq:exp2-leibniz}, the assumptions made in \citet{peng2018new,peng2020generalized} to prove the correctness of the GLR estimators are often difficult to verify or are violated in practical applications \citep{puchhammer2022likelihood}. Nonetheless, GLR estimators often perform well empirically even when these assumptions fail, suggesting that such assumptions may be overly restrictive. In contrast, the assumptions used in this paper are milder and easier to verify (see \Cref{sec3:relate}). Our push-out Leibniz framework also provides a clear geometric interpretation: the surface integral arises from differentiating the parameter-dependent domain, a mechanism not readily apparent in the GLR method. This insight naturally motivates selecting transformations that yield parameter-independent domains, thereby eliminating the surface integral and reducing simulation costs. We discuss these advantages in more detail in \Cref{sec:compare}.

\section{Discontinuous Sample Performances due to Indicator Functions}\label{sec2:indicator}
In this section, we focus on a class of sample performances involving indicator functions that can be absorbed into the integration domain (sample space), yielding a domain that is parametrizable and thus allows direct application of the Leibniz divergence rule \Cref{eq:leibniz-div-sec1}. Specifically, we consider a sample performance of the form $\psi(X,\theta)=\varphi(X,\theta)\1\{g(X,\theta)\in U\}, \theta\in\Theta$ under the following two assumptions.
\vspace*{-0.5em}
\begin{assumption}\label{assump:thm1-a1}
    Let $\Theta \subset \R$ be a bounded open interval and $U \subseteq \R^n$ a Borel set. Let $X$ be an $n$-dimensional random vector supported on $\Omega \subseteq \R^n$, with a density $f: \Omega \times \Theta \to \R$ that is continuously differentiable in both arguments. The function $\varphi: \Omega \times \Theta \to \R$ is also continuously differentiable in both arguments.
    \vspace*{-0.5em}
\end{assumption}
\begin{assumption}\label{assump:thm1-a2}
    There exists a bounded \textcolor{red}{Lipschitz domain} $V\subset \R^n$ and a function $h:V\times\Theta\mapsto \R^n$ such that, for each $\theta\in\Theta$, $h(\cdot,\theta)$ is twice continuously differentiable \textcolor{red}{and extends to a diffeomorphism on an open neighborhood of $\overline V$}. Moreover, $h(v,\theta)$ is continuously differentiable w.r.t. $\theta$, \textcolor{red}{and $\partial_\theta h(v,\theta)$ is continuously differentiable w.r.t. $v$}. Finally, $\{x\in\Omega: g(x,\theta)\in U\} = h(V,\theta)$, i.e., $g(X,\theta)\in U \Leftrightarrow X\in h(V,\theta)$.
    \vspace*{-0.5em}
\end{assumption}
\Cref{assump:thm1-a2} is a geometric regularity condition on the discontinuity region. It is often reasonable in applications where the region is characterized by a few smoothly varying geometric parameters, such as hyperrectangles, ellipsoids, or threshold-type stopping and control regions; see \Cref{exp5:inventory,exp6:glr-cov-smart} and the inventory-price model in \Cref{sec:simulation}. We will show in \Cref{prop:relax-bound-V} that under suitable conditions, the boundedness requirement on $V$ can be relaxed. The indicator function $\1\{g(X, \theta) \in U\}$  defines the region of the sample space that contributes to the expected performance.
By absorbing it into the sample space $\Omega$, we obtain the restricted domain $\{x \in \Omega : g(x, \theta) \in U\}$. Under \Cref{assump:thm1-a2}, this domain is bounded and can be parametrized by a differentiable mapping $h$ over a set $V$, allowing the expected performance to be written as
\vspace*{-0.5em}
\begin{align}\label{eq:trans-performance}
    \E(\psi(X,\theta))=\int_{h(V,\theta)} \varphi(x,\theta) f(x,\theta) dx,
\end{align} 
to which the Leibniz divergence rule \Cref{eq:leibniz-div-sec1} applies directly. 
In practice, constructing $h$ and $V$ is problem-dependent. We illustrate this with two examples frequently encountered in applications.
\vspace*{-0.5em}
\begin{example}\label{exp4:glr=leibniz}
    Assume $g(x,\theta)$ is continuously differentiable in $\theta$, and for each $\theta$, it is invertible and twice continuously differentiable in $x$, with $U\subseteq g(\Omega,\theta)$. Let $g^{-1}(\cdot,\theta)$ denote the inverse of $g(\cdot,\theta)$. Then, the expected performance can be expressed as $\E(\psi(X,\theta))=\int_{g^{-1}(U,\theta)} \varphi(x,\theta) f(x,\theta) dx$, as in \Cref{eq:trans-performance}.
    We can then take $h:=g^{-1}$ and $V:=U$. A typical example arises when the sample performance includes an indicator function of the form $\1\{0 \leq p(x)\leq w(\theta)\}$, where $p:\R^n\mapsto\R^n$ is invertible, $w:\Theta\mapsto\R^n_+$, and $[0,w(\theta)]\subseteq p(\Omega)$. This can be recast as $\1\{g(X,\theta)\in U\}$ by defining $U=[0,1]^n$ and $g_i(x,\theta)=p_i(x)/w_i(\theta)$ for each $i$, where $g_i$, $p_i$, and $w_i$ denote the $i^{\text{th}}$ component of the functions $g$, $p$, and $w$, respectively. Clearly, $g(x,\theta)$ is invertible in $x$, and $U\subseteq g(\Omega,\theta)$, since $[0,1]\subseteq p_i(\Omega)/w_i(\theta)=g_i(\Omega,\theta)$ for each $i$. This formulation commonly appears when $\theta$ represents target points for density estimation \citep{puchhammer2022likelihood}, control charts limits, or barrier levels in option pricing \citep{peng2018new}. \QEDB
    \vspace*{-0.5em}
\end{example}

\begin{example}\label{exp5:inventory}
    Let $X=(X_1,\cdots,X_n)$ be supported on a product of open intervals $\Omega = \Omega_1 \times \cdots \times \Omega_n$, where $\Omega_i = (a_i, b_i)$. For each $i$, let $z_i(x_i, \theta)$ be a real-valued function continuously differentiable in $\theta$, twice continuously differentiable and invertible in $x_i$, and, without loss of generality, strictly increasing in $x_i$. Also, let $\varphi_i(x_1, \dots, x_i, \theta)$ be a real-valued function continuously differentiable in all arguments. Consider a sample performance of the form $\psi(X,\theta)=\sum_{m=1}^n \psi_m(X,\theta)$, where
    \vspace*{-0.5em}
    \begin{align*}
        \psi_m(X,\theta)=\varphi_m(X_1,\cdots,X_m,\theta)\prod_{j=1}^{m}\1\Bigg\{\sum_{i=1}^{j}z_i(X_i,\theta)< q_j\Bigg\},
    \end{align*}
    and $q_1,\cdots,q_n$ are real numbers. This formulation extends \Cref{exp2:log} and often arises in sequential decision-making problems, such as option pricing \citep{wang2012new} and inventory management \citep{huh2008s}. For example, in the $(s,S)$ inventory-price system discussed in \Cref{sec:simulation}, the sample performance $\psi$ may represent the total cost incurred over a regenerative cycle, where each $\psi_m$ denotes the cost at period $m$, $\varphi_m$ is the corresponding cost function, $\{z_i(X_i, \theta)\}$ represents a sequence of demands driven by the random variables $\{X_i\}$, $\theta$ is the price parameter, and for each $j$, $q_j=S-s$ is the replenishment gap. 
    Each $\psi_m$ depends only on the first $m$ components of $X$ and can be treated separately. For instance, in $\psi_n$, the product of indicator functions can be written as $\1\{g(X,\theta)\in U\}$, where $U=(-\infty,q_1)\times\cdots\times(-\infty,q_n)$, and the $j^{\text{th}}$ component of $g$ is given by $g_j(x,\theta)=\sum_{i=1}^{j} z_i(x_i,\theta)$. Assume $\sum_{i=1}^{j} z_i(a_i,\theta) < q_j$ for each $j$; otherwise, $\1\{g(X,\theta)\in U\}$ is identically zero over $\Omega$. To characterize the region defined by $\1\{g(X,\theta)\in U\}$, the feasible range of each $z_i(x_i,\theta)$ can be recursively determined as $z_i(x_i,\theta)\in\left( z_i(a_i,\theta) , q_i- \sum_{k=1}^{i-1} z_k(x_k,\theta) - \sum_{k=i+1}^{n} z_k(a_k,\theta)\right),~i=1,\cdots,n.$ Solving for $x_i$, we obtain $x_i\in\left( a_i , z_i^{-1}\left( q_i- \sum_{k=1}^{i-1} z_k(x_k,\theta) - \sum_{k=i+1}^{n} z_k(a_k,\theta),\theta\right)\right),~i=1,\cdots,n$, where $z^{-1}(\cdot,\theta)$ denotes the inverse of $z(\cdot,\theta)$.
    To construct a differentiable mapping $h$ and domain $V$ satisfying $\{x \in \Omega: g(x,\theta) \in U\}=h(V,\theta)$, we match the feasible range of each $x_i$ to $h_i(V,\theta)$. Specifically, define $V=(0,1)^n$ and construct the mapping $h$ recursively by setting, for each $i=1,\cdots,n$,
    \vspace*{-0.5em}
    \begin{align*}
        h_i(v,\theta)=\left(z_i^{-1}\left(q_i- \sum_{k=1}^{i-1} z_k(h_k(v,\theta),\theta) - \sum_{k=i+1}^{n} z_k(a_k,\theta),\theta\right) - a_i\right)v_i + a_i.
    \end{align*}
    Here, the interval $(0,1)$ for each $v_i$ is chosen so that $h_i(v,\theta)$ spans the full feasible range of $x_i$. Specifically: when $v_i=0$, we have $h_i(v,\theta)=a_i$, matching the lower bound; when $v_i=1$, we have $h_i(v,\theta)= z_i^{-1}\left(q - \sum_{j=1}^{i-1} z_j(h_j(v,\theta), \theta) - \sum_{j=i+1}^{n} z_j(a_j, \theta), \theta\right)$, matching the upper bound. Moreover, since each $h_i(v, \theta)$ depends only on $v_1, \cdots, v_i$, the mapping $x = h(v, \theta)$ can be inverted recursively: for any given $x\in\Omega$, one can solve for $v$ by sequentially inverting each equation $h_i(v, \theta) = x_i$, using previously computed values $v_1, \ldots, v_{i-1}$. The strict monotonicity of $z_i$ in $x_i$ guarantees that each inversion yields a unique solution, ensuring that $h$ invertible. See \Cref{exp4:leibniz} for an application of this construction. \QEDB
    \vspace*{-0.5em}
\end{example}


Applying the Leibniz divergence rule \Cref{eq:leibniz-div-sec1} to the expected performance \Cref{eq:trans-performance} yields the following result.
\vspace*{-0.5em}
\begin{theorem}\label{thm:glr}
    Consider $\psi(X,\theta)=\varphi(X,\theta)\1\{g(X,\theta)\in U\}$, and let $l(x,\theta) = \partial_\theta \log f(x,\theta)$ denote the LR term. Under \Cref{assump:thm1-a1,assump:thm1-a2}, we have:
    \vspace*{-0.5em}
    \begin{align*}
        \frac{d}{d\theta}\E(\psi(X,\theta))
        =\int_{h(V,\theta)} \left( (\partial_\theta\varphi(x,\theta)  + \varphi(x,\theta)l(x,\theta))f(x,\theta) + \diver\left( \varphi(x,\theta) f(x,\theta) \partial_\theta h(v,\theta)|_{v=h^{-1}(x,\theta)} \right) \right) dx.
    \end{align*}
\end{theorem}
From \Cref{thm:glr}, we obtain the following \textbf{Leibniz divergence estimator}:
\vspace*{-0.5em}
\begin{align}\label{estimator:leibniz-div}
   \1\{g(X,\theta)\in U\}\left(\partial_\theta\varphi(X,\theta)  + \varphi(X,\theta)l(X,\theta)
    +\diver\left( \varphi(X,\theta) f(X,\theta) \partial_\theta h(v,\theta)|_{v=h^{-1}(X,\theta)}\right)/f(X,\theta)\right),
\end{align}
where $X$ has density $f(\cdot, \theta)$. The term $\partial_\theta\varphi(X,\theta) + \varphi(X,\theta)l(X,\theta)$ corresponds to the standard IPA-LR estimator \citep{l1990unified}, while the divergence term---arising from ``differentiation'' of the domain $h(V, \theta)$---captures the effect of discontinuities introduced by the indicator function. 
\vspace*{-0.5em}
\begin{remark}
    \Cref{assump:thm1-a2} can be relaxed by allowing multiple mappings $h_1,\cdots,h_k$ and corresponding sets $V_1,\cdots,V_k$, such that the images $\{h_i(V_i,\theta)\}$ are mutually disjoint and $g(\Omega,\theta)\cap U=\cup_{i=1}^k h_i(V_i,\theta)$. We can then apply \Cref{thm:glr} to each $h_i(V_i,\theta)$ individually and sum their contributions to form the final estimator.
\end{remark}
Note that applying the Leibniz divergence estimator \Cref{estimator:leibniz-div} requires the output dimension of the function $g$ to match that of the input $X$. When $g$ has lower dimension than $X$, we can augment it using auxiliary functions. Specifically, consider an indicator function $\1\{g_1(X,\theta)\in U_1\}$, where $g_1:\R^n\times\Theta\mapsto \R^m$, $U_1\subseteq\R^m$, and $m<n$. Suppose there exists an auxiliary function $g_2:\R^{n}\times\Theta\mapsto \R^{n-m}$ with $U_2\subseteq\R^{n-m}$, such that for each $\theta\in\Theta$, the inclusion $\{x\in\Omega:g_1(x,\theta)\in U_1\}\subseteq\{x\in\Omega:g_2(x,\theta)\in U_2\} $ holds. Then, defining the $n$-dimensional augmented function $g=(g_1,g_2)$ and domain $U=U_1\times U_2\subseteq\R^n$, we have $\1\{g_1(X,\theta)\in U_1\}=\1\{g(X,\theta)\in U\}$. This allows us to work with the reformulated indicator function $\1\{g(X,\theta) \in U\}$, to which the estimator \Cref{estimator:leibniz-div} can be directly applied. The following example illustrates this approach.
\vspace*{-0.5em}
\begin{example}\label{exp4:leibniz}
    Consider the sample performance from \Cref{exp2:log}. Let $X=(X_1,X_2)$ and $\psi(X,\theta)=\1\{\sum_{i=1}^{2}\log(X_i+\theta)<q\},~\theta\in(1,\infty)$, which can be written as $\1\{g_1(X,\theta)\in U_1\}$, where $g_1(x,\theta)=\sum_{i=1}^{2}\log(x_i+\theta)$ and $U_1=(-\infty,q)$. To bring this into the form required by \Cref{estimator:leibniz-div}, we introduce an auxiliary function $g_2(x,\theta)=\log(x_1+\theta)$ and define $U_2=U_1$. Since $g_2(x,\theta)\leq g_1(x,\theta)$, it follows that $g_1(x,\theta)\in U_1\implies g_2(x,\theta)\in U_2$. Defining the augmented function $g=(g_1,g_2)$ and domain $U=U_1\times U_2$, we obtain $\1\{g_1(X,\theta)\in U_1\}=\1\{g(X,\theta)\in U\}$. Therefore, we can equivalently work with the augmented indicator function $\1\{g(X,\theta)\in U\}$, which is covered by \Cref{exp5:inventory}. Specifically, we define $V = (0,1)^2$,
    $h_1(v,\theta)=\left(e^q/\theta-\theta\right)v_1,\text{ and }h_2(v,\theta)=\big(((e^q/\theta-\theta)v_1+\theta)^{-1}e^q-\theta\big)v_2.$
    The Leibniz divergence estimator \Cref{estimator:leibniz-div} becomes
    \vspace*{-0.5em}
    \begin{align}\label{exp4:leibniz-div-estimator}
        \1\{g(X,\theta)\in U\}\sum_{i=1}^{2}\left((\partial_{x_i}\log f(X))\partial_\theta h_i(v,\theta)|_{v=h^{-1}(X,\theta)} + \partial_{x_i}\partial_\theta h_i(v,\theta)|_{v=h^{-1}(X,\theta)}\right).
    \end{align}
    Compared to the \Cref{exp2:log} estimator (formally introduced in \Cref{sec2:inv} as the Leibniz integral estimator), the Leibniz divergence estimator avoids estimating a surface integral and is thus single-run. In this example, the surface $\partial h(V, \theta)$ lacks a simple shape, making the associated surface integral challenging to estimate and thus the application of the Leibniz integral rule \Cref{eq:leibniz-sec1} less appealing. \QEDB
    \vspace*{-0.5em}
\end{example}

The boundedness requirement for $V$ in \Cref{assump:thm1-a2} can be relaxed under the following integrability conditions. Let $\gamma(x,\theta)=(\partial_\theta\varphi(x,\theta)  + \varphi(x,\theta)l(x,\theta))f(x,\theta) + \diver\left( \varphi(x,\theta) f(x,\theta) \textcolor{red}{\partial_\theta} h(v,\theta)|_{v=h^{-1}(x,\theta)} \right)$.
\begin{assumption}\label{assump:thm1-a3} (1) For each $\theta$, $\E(|\psi(X,\theta)|)<\infty$, and (2)\hspace*{-1pt} $\int_{V} \sup_{\theta\in\Theta}\left| \gamma(h(v,\theta),\theta) \cdot \det (J_h(v,\theta)) \right| dv<\infty$.
\end{assumption}
\begin{proposition}\label{prop:relax-bound-V}\vspace*{-0.25em}
    If \Cref{assump:thm1-a3} holds, then \Cref{thm:glr} remains valid without the boundedness assumption on $V$ in \Cref{assump:thm1-a2}. 
\end{proposition}
\begin{proof}{Proof.}\vspace*{-0.25em}
    \textcolor{red}{Since $V$ is a Lipschitz domain, it admits} an increasing sequence of \textcolor{red}{bounded Lipschitz subdomains $\{V_n\}_{n\geq 1},$ with $V_{n}\subseteq V_{n+1}\subseteq V$ and $V=\cup_{n\geq 1} V_n$ \citep[Proposition 8.2.1]{daners2008domain}}. Define $\psi_n(x,\theta):=\varphi(x,\theta)\1\{x\in h(V_n,\theta)\}$. Then, for each $\theta\in\Theta$ and $x\in \Omega$, $\lim_{n\rightarrow\infty}\psi_n(x,\theta)=\psi(x,\theta)$, and by the dominated convergence theorem (DCT), $\lim_{n\rightarrow\infty}\E(\psi_n(X,\theta))=\E(\psi(X,\theta))$. 
    By \Cref{thm:glr}, $\frac{d}{d\theta}\E(\psi_n(X,\theta))=\int_{h(V_n,\theta)}  \gamma(x,\theta) dx$ for each $n$. To pass to the limit, note that $\left| \frac{d}{d\theta}\E(\psi_n(X,\theta)) -\int_{h(V,\theta)}  \gamma(x,\theta) dx \right| \leq \int_{h(V\setminus V_n,\theta)}  |\gamma(x,\theta)| dx$. Using the change of variables $x=h(v,\theta)$ and applying the DCT again, we obtain $\sup_{\theta\in\Theta}\int_{h(V\setminus V_n,\theta)}  |\gamma(x,\theta)| dx
    \leq\int_{V\setminus V_n}\sup_{\theta\in\Theta}\Big|\gamma(h(v,\theta),\theta) \cdot \det (J_h(v,\theta))\Big|dv\rightarrow 0$ as $n\rightarrow\infty$.
    By \citet[Theorem 4, Section 16.3.5]{zorich2004mathematical}, we conclude that $\frac{d}{d\theta}\E(\psi(X,\theta))=\int_{h(V,\theta)}  \gamma(x,\theta) dx$. \QEDB
\end{proof}


\section{Combining the Leibniz Integral Rule with the Push-out LR Method}\label{sec2:inv}
While the Leibniz divergence estimator \Cref{estimator:leibniz-div} applies to many practical scenarios, it requires that the intersection of the sample space and the region defined by the indicator function be parameterized by a differentiable mapping. Identifying such a parameterization is problem-specific. In this section, we introduce the \textit{push-out Leibniz approach} for general discontinuous sample performances. By a change of variables, we shift the parameter $\theta$ from within the sample performance into the density and (possibly) the integration domain. Unlike in \Cref{sec2:indicator},
the integrand remains discontinuous w.r.t. the input vector $X$ even after the change of variables, making the Leibniz divergence rule \Cref{eq:leibniz-div-sec1} inapplicable. Instead, we apply the Leibniz integral rule \Cref{eq:leibniz-sec1} to avoid differentiation w.r.t. $X$. While \Cref{thm:leibniz} still assumes the differentiability w.r.t. $X$, we propose verifiable conditions under which \Cref{eq:leibniz-sec1} remains valid without it.

To define the setting, consider a sample performance of the form $\psi(X,\theta) = \varphi(g(X,\theta))$, where $\varphi$ could be discontinuous, but $g$ is relatively well-behaved. For each $\theta$, we apply a change of variables $Y = g(X,\theta)$ such that the transformed performance $\varphi(Y)$ does not depend explicitly on $\theta$. We make the following assumption.
\begin{assumption}\label{assump:thm2-a1}
    Suppose $\Theta\subseteq\R$ is a bounded open interval. Let $X$ be an $n$-dimensional random vector supported on a bounded \textcolor{red}{Lipschitz domain} $\Omega\subset\R^n$, with density $f:\Omega\times\Theta\mapsto \R_+$, continuously differentiable in both arguments. The function $\varphi:\tilde{\Omega}\mapsto \R$ is $L^2$-integrable over $\tilde\Omega:=\cup_{\theta\in\Theta} g(\Omega,\theta)$. The function $g:\Omega\times\Theta\mapsto \R^n$ is such that, for each $\theta\in\Theta$, $g(\cdot,\theta)$ is twice continuously differentiable \textcolor{red}{and extends to a diffeomorphism on an open neighborhood of $\overline{\Omega}$, with inverse denoted by $g^{-1}(\cdot,\theta)$}. Moreover, $g(x,\theta)$ is continuously differentiable w.r.t. $\theta$, \textcolor{red}{and $\partial_\theta g(x,\theta)$ is continuously differentiable w.r.t. $x$. Finally, $g$ admits a continuous extension to $\overline{\Omega}\times\overline{\Theta}$.}
    \vspace*{-0.5em}
\end{assumption}
\Cref{assump:thm2-a1} parallels \Cref{assump:thm1-a1} but allows $\varphi$ to be a general (possibly) discontinuous function. Rather than defining $\varphi$ over all of $\R^n$, we restrict it to the domain $\tilde\Omega=\cup_{\theta\in\Theta} g(\Omega,\theta)$. \textcolor{red}{Since $g$ admits a continuous extension to the compact set $\overline{\Omega}\times\overline{\Theta}$, the domain $\tilde\Omega$ is bounded.} Bounded discontinuous functions that are not $L^2$-integrable over $\R^n$ are still $L^2$-integrable over $\tilde \Omega$---such as indicator functions---making them admissible under \Cref{assump:thm2-a1}. In \Cref{sec3:relate}, we show that assumptions about the (global) invertibility of $g$ and the boundedness of $\Omega$ can be relaxed under suitable conditions. To apply the Leibniz integral rule \Cref{eq:leibniz-sec1} to $\E(\varphi(g(X,\theta)))$, we present the key steps below, with the final result stated in \Cref{thm:glr1}.

\textbf{Step 1: Change of variables.} By the change of variables $Y = g(X, \theta)$, we can express the expected performance as $\E(\varphi(Y))=\int_{g(\Omega,\theta)} \varphi(y) \tilde f (y,\theta) dy$, where $\tilde f (y,\theta)=f(g^{-1}(y,\theta),\theta) \left|\det(J_{g^{-1}}(y,\theta))\right|$. 

\textbf{Step 2: Smooth approximation.} Since $\varphi$ could be discontinuous, the Leibniz integral rule \Cref{eq:leibniz-sec1} is not directly applicable. To proceed, we substitute the smooth approximation of $\varphi$. The existence of a smooth approximation is guaranteed by the following lemma \textcolor{red}{\citep[Theorem 8.14]{folland1999real}}.
\begin{lemma}\label{prop:dense}\vspace*{-0.5em}
    \textcolor{red}{Let $\varphi\in L^p(\R^m)$, $1\leq p<\infty$, and let
    $q:\R^m\to\R_+$ be a nonnegative, compactly supported smooth function
    satisfying $\int_{\R^m} q(z)dz=1$. Define $q_n(z)=n^mq(nz),~\varphi_n=\varphi*q_n.$ Then $\varphi_n$ is smooth, and $\varphi_n\to\varphi\text{ in }L^p(\R^m)$ as $n\to\infty$. Moreover, if $\varphi$ is bounded and continuous on an open set
    $U\subseteq\R^m$, then $\varphi_n\to\varphi$ uniformly on every
    compact subset of $U$.}\vspace*{-0.5em}
\end{lemma}
\textcolor{red}{A standard choice is $\phi(z) = C \exp\left( -1/\left(1 - \|z\|_2^2\right) \right) \1\{\|z\|_2 < 1\}$, $z\in\R^n$, where $\|\cdot\|_2$ denotes the Euclidean norm and $C$ is a normalization constant. Since $\varphi\in L^2(\tilde\Omega)$, it admits an $L^2(\R^n)$ extension, for example, by setting it equal to zero outside $\tilde\Omega$.  We fix such an extension and, with a slight abuse of notation, continue to denote it by $\varphi$. We apply \Cref{prop:dense} to obtain a sequence of smooth functions $\{\varphi_n\}_{n\in\N}$ approximating $\varphi$ in $L^2(\R^n)$.}


\textbf{Step 3: Apply the Leibniz divergence rule.} Substituting $\varphi_n$ for $\varphi$ in $\E(\varphi(Y))$, we apply the Leibniz divergence rule \Cref{eq:leibniz-div-sec1} and obtain:
\vspace*{-0.5em}
\begin{align}\label{eq:leibniz-apr-y}
        \frac{d}{d\theta}\E(\varphi_n(Y))=
        \int_{g(\Omega,\theta)} \varphi_n(y) \frac{d}{d\theta} \tilde f (y,\theta) dy
        +\int_{g(\Omega,\theta)} \diver(\varphi_n(y) \tilde f (y,\theta)  \vec v(y)) dy,
\end{align}
where $\vec v(y)=\partial_\theta g(x,\theta)|_{x=g^{-1}(y,\theta)}$. We apply the Leibniz divergence rule rather than the integral rule, as it involves only volume integrals, making the reversal of the change of variables in the next step more straightforward. The final result will take the form of the Leibniz integral rule and will be recovered later via the divergence theorem.

\textbf{Step 4: Reverse the change of variables.} Reversing the change of variables is preferable for several reasons. First, in Step 5, we convert the second volume integral in \Cref{eq:leibniz-apr-y} into a surface integral via the divergence theorem. Estimating the surface integral over $\partial g(\Omega,\theta)$ can be challenging, but if $\Omega$ has a simple shape such as a hyperrectangle, then reversing the change of variables before applying the divergence theorem yields a surface integral over $\partial \Omega$ that is more tractable. Second, although $g$ is assumed to be invertible, it may not admit a closed-form inverse. Third, in the more general setting of \Cref{sec2-1:locoinv}, $g$ is only locally invertible, and a global change of variables $Y = g(X,\theta)$ may not be valid. For these reasons, we reverse the change of variables using the proposition below, previously established in \citet{ren2024generalizing}.
\begin{proposition}\label{prop:cov}\vspace*{-0.5em}
    Define scalar-valued functions $d(x,\theta)= \diver(-f(x,\theta)J_g^{-1}(x,\theta)\partial_\theta g(x,\theta))/f(x,\theta)$ and $l(x,\theta)=\partial_\theta \log f(x,\theta)$, and a vector-valued function $s(x,\theta)= J_g^{-1}(x,\theta)\partial_\theta g(x,\theta)$. For $y=g(x,\theta)$, we have:\vspace*{-0.5em}
    \begin{align*}
        \frac{d}{d\theta} \tilde f (y,\theta)
        &=\! \left|\det(J_{g^{-1}}(y,\theta))\right|(d(x,\theta)+l(x,\theta)) f(x,\theta),\\
        \diver(\varphi_n(y) \tilde f (y,\theta) \vec v(y)) \nonumber\
        &=\! \left|\det(J_{g^{-1}}(y,\theta))\right| \diver(\varphi_n(g(x,\theta)) f(x,\theta)s(x,\theta)).
    \end{align*}
\end{proposition}
\vspace*{-0.5em}
By \Cref{prop:cov}, we reverse the change of variables on the right-hand side of \Cref{eq:leibniz-apr-y}:\vspace*{-0.25em}
\begin{align}\label{eq:glr-div-approx}
    \frac{d}{d\theta}\E(\varphi_n(g(X,\theta)))
    =\int_\Omega  \varphi_n(g(x,\theta))(d(x,\theta)+l(x,\theta)) f(x,\theta) + \int_\Omega\diver (\varphi_n(g(x,\theta))f(x,\theta) s(x,\theta))  dx.
\end{align}



\vspace*{-0.25em}
\textbf{Step 5: Apply the divergence theorem.} Applying the divergence theorem to the second integral in \Cref{eq:glr-div-approx} yields the following result, which can be interpreted as the result of applying the \textbf{Leibniz integral rule \Cref{eq:leibniz-sec1}} to $\E(\varphi_n(Y))$ followed by reversing the change of variables:
\vspace*{-0.25em}
\begin{align}\label{eq:glr-approx}
    \begin{split}
        \frac{d}{d\theta}\E(\varphi_n(g(X,\theta)))
        =\int_\Omega \varphi_n(g(x,\theta)) (d(x,\theta)+l(x,\theta)) f(x,\theta)dx
        + \int_{\partial\Omega} \varphi_n(g(x,\theta))s(x,\theta)^T \vec n(x)f(x,\theta) d\sigma,
    \end{split}
\end{align}
where $\vec n(x)$ is the outward unit normal vector to the surface $\partial\Omega$.

\textbf{Step 6: Take $n\rightarrow\infty$ in \Cref{eq:glr-approx}.} Under \Cref{assump:thm2-a2,assump:thm2-a4} below, both integrals in \Cref{eq:glr-approx} converge uniformly over $\theta \in \Theta$ to their counterparts with $\varphi_n$ replaced by $\varphi$ as $n\rightarrow\infty$, leading to the final result \Cref{eq:glr} in \Cref{thm:glr1}.

\begin{assumption}\label{assump:thm2-a2}\vspace*{-0.5em}
    $\lim_{n\rightarrow\infty}\sup_{\theta\in\Theta}\int_{\partial\Omega}\left|(\varphi(g(x,\theta))-\varphi_n(g(x,\theta)))s(x,\theta)^T \vec n(x)f(x,\theta)\right| d\sigma=0$.
\end{assumption}
\Cref{assump:thm2-a2} is general but hard to verify; a more practical alternative is provided in \Cref{sec:easy-verifiable}.

\begin{assumption}\label{assump:thm2-a4}
    Let $\tilde f(y,\theta)=f(g^{-1}(y,\theta),\theta) \left|\det(J_{g^{-1}}(y,\theta))\right|$. Suppose $\left\| \tilde f(y,\theta)\right\|_{L^2}<\infty$ for each $\theta\in\Theta$, and $\sup_{\theta\in\Theta}\left\|\frac{d}{d\theta} \tilde f(y,\theta)\right\|_{L^2}<\infty$, where $\|\cdot\|_{L^2}$ denotes the $L^2$ norm.
\end{assumption}
\begin{theorem}\label{thm:glr1}
    Consider $\psi(X,\theta)=\varphi(g(X,\theta))$. Let $l(x,\theta)=\partial_\theta \log f(x,\theta)$, $s(x,\theta)= J_g^{-1}(x,\theta)\partial_\theta g(x,\theta) $, and $d(x,\theta)= \diver(-f(x,\theta)s(x,\theta))/f(x,\theta)$. Under \Cref{assump:thm2-a1,assump:thm2-a2,assump:thm2-a4},
    \begin{align}\label{eq:glr}
        \begin{split}
            \frac{d}{d\theta}\E(\varphi(g(X,\theta)))
            = \int_\Omega \varphi(g(x,\theta)) (d(x,\theta)+l(x,\theta)) f(x,\theta)dx
            +\int_{\partial\Omega}\varphi(g(x,\theta))s(x,\theta)^T \vec n(x)f(x,\theta) d\sigma.
        \end{split}
    \end{align}
\end{theorem}
\begin{proof}{Proof.}
    By the Cauchy–Schwarz inequality, the following convergence result holds for each $\theta\in\Theta$:
    \begin{align}\label{eq:thm:glr1-1}
        \lim_{n\rightarrow\infty}\left|\E(\varphi(g(X,\theta)))-\E(\varphi_n(g(X,\theta)))\right|\leq \lim_{n\rightarrow\infty}\left\|\varphi_n(y)-\varphi(y)\right\|_{L^2}\left\| \tilde f(y,\theta)\right\|_{L^2}=0,
    \end{align}
    where the last equality follows from the assumptions that $\varphi_n\rightarrow\varphi$ in $L^2$ and $\| \tilde f(y,\theta)\|_{L^2}<\infty$. Similarly, by the assumption $\sup_{\theta\in\Theta}\left\|\frac{d}{d\theta} \tilde f(y,\theta)\right\|_{L^2}<\infty$, the following convergence result holds:
    \begin{align}
        &~~~~\lim_{n\rightarrow\infty}\sup_{\theta\in\Theta}\left|\int_\Omega \varphi(g(x,\theta)) (d(x,\theta)+l(x,\theta)) f(x,\theta)dx-\int_\Omega \varphi_n(g(x,\theta)) (d(x,\theta)+l(x,\theta)) f(x,\theta)dx\right| \nonumber \\
        &\leq\lim_{n\rightarrow\infty}\sup_{\theta\in\Theta}\int_{g(\Omega,\theta)} \left|\varphi_n(y)-\varphi(y)\right| \left|\frac{d}{d\theta} \tilde f (y,\theta)\right| dy \leq  \lim_{n\rightarrow\infty}\|\varphi_n(y)-\varphi(y)\|_{L^2} \sup_{\theta\in\Theta} \left\|\frac{d}{d\theta} \tilde f (y,\theta)\right\|_{L^2}=0, \label{eq:thm:glr1-2}
    \end{align}
    where we make the change of variables $y = g(x,\theta)$ based on \Cref{prop:cov}. Combining \Cref{eq:thm:glr1-1,eq:thm:glr1-2} with \Cref{assump:thm2-a2} and invoking \citet[Theorem 4, Section 16.3.5]{zorich2004mathematical}, \Cref{thm:glr1} follows. \QEDB
\end{proof}
An unbiased estimator for the volume integral in \Cref{eq:glr} is given by
\vspace*{-0.5em}
\begin{align}\label{eq:glr1}
    \varphi(g(X,\theta)) (d(X,\theta)+l(X,\theta)), 
\end{align}
which can be interpreted as a (generalized) LR term, as it results from differentiating the density $\tilde f$ of the transformed variable $Y=g(X,\theta)$. In \Cref{sec3-1:GLR}, we focus on the case where the sample space $\Omega$  is a hyperrectangle, which allows the surface integral in \Cref{eq:glr} to be expressed as a sum of conditional expectations. An estimator for the surface integral is provided in \Cref{eq:surface} of \Cref{sec3-1:GLR}. Combining \Cref{eq:glr1} and \Cref{eq:surface} yields what we refer to as the \textbf{Leibniz integral estimator}.



While sharing the same general mathematical form as GLR methods, the Leibniz integral estimator provides a distinct geometric perspective: it explicitly links the surface integral to the $\theta$-dependent domain boundary. This perspective allows us to obtain much milder regularity conditions for GLR than those given in previous papers, and also strategies to eliminate the surface integral that yield efficient single-run estimators. We defer a detailed comparison and discussion to \Cref{sec:compare}. In the next section, we discuss estimating the surface integral and extend the Leibniz integral estimator to accommodate unbounded sample spaces and local changes of variables. We conclude this section by highlighting two straightforward extensions of the Leibniz integral estimator:
\begin{itemize}
    \item The Leibniz integral estimator naturally extends to sample performances of the form $\varphi(g(x,\theta),\theta)$, as long as the outer function $\varphi(y,\theta)$ is differentiable in $\theta$. This extension simply adds an extra IPA term $\partial_\theta \varphi(y,\theta)|_{y=g(X,\theta)}$ to the existing estimator. With this extension, the Leibniz integral estimator encompasses the SLRIPA estimator \citep{wang2012new} as a special case.
    \item The Leibniz integral estimator can also be extended to cases where the dimension of $X$ exceeds that of $g$, i.e., $g:\R^n\times\Theta\mapsto \R^m$ with $m<n$. By conditioning on $n-m$ variables, we can apply the Leibniz integral estimator to the remaining $m$, provided that $g$ is invertible w.r.t. those $m$ variables (or, more generally, that the corresponding $m\times m$ submatrix of $J_g(x,\theta)$ is invertible, as we will discuss in \Cref{sec2-1:locoinv}).
\end{itemize}

\section{Extensions of the Leibniz Integral Estimator}\label{sec3:relate}
In this section, we discuss extensions and implementation aspects of the Leibniz integral estimator. \Cref{sec3-1:GLR} investigates the case where $\Omega$ is a hyperrectangle, so the boundary $\partial\Omega$ has a simple shape, making surface integral estimation more tractable. \Cref{sec2-1:locoinv} extends \Cref{thm:glr1} to settings where $g$ is locally invertible. \Cref{sec:easy-verifiable} proposes an alternative to \Cref{assump:thm2-a2} that is easier to verify. \Cref{sec2-2:ubdd} addresses unbounded sample spaces.
\subsection{Hyperrectangle Support $\Omega$}\label{sec3-1:GLR}
Consider $\Omega=[a_1,b_1]\times\cdots\times[a_n,b_n]$, a hyperrectangle in $\R^n$, with boundary given by a union of $2n$ faces $\partial \Omega = \cup_{i=1}^n (\Omega_{a_i}\cup\Omega_{b_i})$, where $\Omega_{x_i}: = [a_1,b_1]\times\cdots\times\{x_i\}\times\cdots\times[a_n,b_n],~x_i=a_i \text{ or }b_i$ .
For each $i$, the normal vectors to faces $\Omega_{a_i}$ and $\Omega_{b_i}$ are $-e_i$ and $e_i$, respectively, where $e_i$ is the unit vector in the $i^{\text{th}}$ direction. Let $x_{-i}$ denote the vector $x$ with its $i^{\text{th}}$ coordinate removed, $\prod_{j \ne i} [a_j, b_j]$ the projection of $\Omega$ along the $i^\text{th}$ direction, and $f_{X_i}:[a_i,b_i]\times \Theta\mapsto \R_+$ the marginal density of $X_i$. The surface integral over each of these faces reduces to a standard $(n-1)$-dimensional multivariate (volume) integral, which can be expressed as a conditional expectation as follows:
\vspace*{-0.5em}
\begin{align}
    &~~~~\int_{\partial\Omega}\varphi(g(x,\theta))s(x,\theta)^T \vec n(x)f(x,\theta) d\sigma=\sum_{i=1}^n\int_{\prod_{j \ne i} [a_j, b_j]} \varphi(g(x,\theta))
    s(x,\theta)^Te_i f(x,\theta) dx_{-i} \Big| ^{b_i}_{x_i=a_i} \nonumber\\
    &=\sum_{i=1}^n \left(\E(\varphi(g(X,\theta))s(X,\theta)^Te_i|X_i=b_i)f_{X_i}(b_i,\theta)
    -\E(\varphi(g(X,\theta))s(X,\theta)^Te_i|X_i=a_i)f_{X_i}(a_i,\theta)\right). \label{eq:surface-orig}
\end{align}
An unbiased estimator for the surface integral is given by
\vspace*{-0.5em}
\begin{align}\label{eq:surface}
    \begin{split}
        \sum_{i=1}^n \left(\varphi(g(X,\theta))f_{X_i}(b_i,\theta)s(X,\theta)^Te_i  \big|_{X\sim f_{X|X_i=b_i}}
        -\varphi(g(X,\theta))f_{X_i}(a_i,\theta)s(X,\theta)^Te_i  \big|_{X\sim f_{X|X_i=a_i}}\right),
    \end{split}
\end{align}
where $f_{X|X_i}$ is the density of $X$ conditional on $X_i$. Sampling \Cref{eq:surface} requires generating a sample path from each conditional density. If $X=(X_1,\cdots,X_n)$ has independent components, as in \citet{peng2020generalized}, the conditional density $f_{X|X_i}$ reduces to a product of marginal densities, allowing \Cref{eq:surface} to be estimated by a single sample path, concurrently with sampling \Cref{eq:glr1}.
\citet{peng2018new} considers another special case where the density $f$ vanishes at the boundary of the support. For this case, the marginal densities $f_{X_i}(a_i,\theta)$ and $f_{X_i}(b_i,\theta)$ are zero, and the surface integral vanishes. 

\subsection{Local Change of Variables}\label{sec2-1:locoinv}
In this section, we relax the global invertibility condition on $g$ in \Cref{assump:thm2-a1} and \textcolor{red}{instead require its Jacobian $J_g$ to be invertible almost everywhere (a.e.) on $\Omega$. By the inverse function theorem, $g$ is then locally invertible at every point where $J_g$ is invertible \citep[Section 8.6]{zorich2004mathematical1}.} As a result, \Cref{assump:thm2-a4}, which relies on global invertibility, no longer applies. \textcolor{red}{We replace it with \Cref{assump:prop3-a0,assump:prop3-a1}; the latter,} like \Cref{assump:thm2-a2}, is general but difficult to verify. A more practical alternative is given in \Cref{sec:easy-verifiable}.
\vspace*{-0.5em}
\textcolor{red}{
\begin{assumption}\label{assump:prop3-a0}
    The families $\{f(\cdot,\theta)\}_{\theta \in \Theta}$ and $\{\partial_\theta f(\cdot,\theta)\}_{\theta \in \Theta}$ are uniformly integrable on $\Omega$.
    \vspace*{-0.5em}
\end{assumption}
Since $\Omega$ is bounded, a sufficient condition for either family to be
uniformly integrable is that it be uniformly bounded in $L^p(\Omega)$ for
some $p>1$. Alternatively, the existence of an integrable pointwise envelope
is also sufficient (e.g.,  $\|\sup_{\theta \in \Theta} f(\cdot, \theta)\|_{L^1} < \infty$).}
\vspace*{-0.5em}
\begin{assumption}\label{assump:prop3-a1}
    $\lim_{n\rightarrow\infty}\sup_{\theta\in\Theta}\E\left(\left|(\varphi(g(x,\theta))-\varphi_n(g(x,\theta))) (d(x,\theta)+l(x,\theta))\right|\right)=0$.
    \vspace*{-0.5em}
\end{assumption}
\begin{assumption}\label{assump:prop3-a2}
    \textcolor{red}{The condition in \Cref{assump:thm2-a1} that, for each $\theta\in\Theta$, $g(\cdot,\theta)$ extends to a diffeomorphism on an open neighborhood of $\overline{\Omega}$ is relaxed as follows: $g(\cdot,\theta)$ admits a continuously differentiable extension to an open neighborhood of $\overline{\Omega}$, and its Jacobian $J_g(\cdot,\theta)$ is invertible a.e. on $\Omega$. Moreover, we additionally require that $\partial_\theta g$ admit a continuous extension to $\overline{\Omega}\times\overline{\Theta}$. All other conditions are maintained.}
    \vspace*{-0.5em}
\end{assumption}
\textcolor{red}{Define the singular set $N_g(\theta):=\{x\in\Omega : \det(J_g(x,\theta))=0 \}$. Under \Cref{assump:prop3-a2}, $N_g(\theta)$ has Lebesgue measure zero for every $\theta\in\Theta$. We impose the following assumption to ensure that the local identities obtained away from $N_g(\theta)$ extend across the singular set and that the divergence theorem can be applied globally \citep{evans2015measure}.
\vspace*{-0.5em}
\begin{assumption}\label{assump:prop3-a3}
    For every $\theta\in\Theta$, $s(\cdot,\theta)f(\cdot,\theta)$, initially defined on $\Omega\setminus N_g(\theta)$, admits an extension in $W^{1,1}(\Omega;\R^n)$, where $W^{1,1}(\Omega;\R^n)$ is the Sobolev space of vector-valued functions with integrable weak derivatives. With a slight abuse of notation, we continue to denote this extension by $s(\cdot,\theta)f(\cdot,\theta)$.
    \vspace*{-0.5em}
\end{assumption}
}
\begin{theorem}\label{prop:local-inv}
    \textcolor{red}{Under \Cref{assump:thm2-a2,assump:prop3-a0,assump:prop3-a1,assump:prop3-a2,assump:prop3-a3}}, \Cref{thm:glr1} remains valid.
\end{theorem}
\vspace*{-0.5em}
\begin{proof}{Proof.}
\textcolor{red}{
Fix $\theta_0\in\Theta$ and consider any $x_0\in\Omega\setminus N_g(\theta_0)$. Then $\det J_g(x_0,\theta_0)\neq0$. By the inverse function theorem and the regularity conditions maintained from \Cref{assump:thm2-a1}, there exist open neighborhoods $U(x_0,\theta_0)\subseteq\Omega$ of $x_0$ and $V(x_0,\theta_0)\subseteq\Theta$ of $\theta_0$ such that $g(\cdot,\theta)$ is invertible on $U(x_0,\theta_0)$ for every $\theta\in V(x_0,\theta_0)$. Hence, on $U(x_0,\theta_0)\times V(x_0,\theta_0)$, Steps 1--4 of \Cref{sec2:inv} can be applied using the corresponding local inverse of $g$. In particular, the local counterpart of \Cref{eq:glr-div-approx} remains valid on this neighborhood. We now verify the underlying pointwise differential identity in \Cref{eq:glr-div-approx}:
\begin{align}\label{eq:pt-dif}
\partial_\theta
\left(
\varphi_n(g(x,\theta))f(x,\theta)
\right)
=
\varphi_n(g(x,\theta))
(d(x,\theta)+l(x,\theta))f(x,\theta)
+
\diver
\left(
\varphi_n(g(x,\theta))
s(x,\theta)f(x,\theta)
\right),
\end{align}
Let $\nabla_x$ denote the gradient operator w.r.t. $x$. By the chain rule,
\begin{align*}
    \partial_\theta\varphi_n(g(x,\theta)) = \nabla_y\varphi_n(y)^T|_{y=g(x,\theta)} \partial_\theta g(x,\theta)
    =\nabla_y\varphi_n(y)^T|_{y=g(x,\theta)} J_g(x,\theta)s(x,\theta)
    =\nabla_x\varphi_n(g(x,\theta))^T s(x,\theta).
\end{align*}
For any differentiable scalar-valued function $h(x)$ and vector-valued function $\vec v(x)$, the product rule for divergence, $\diver(h(x)\vec v(x))=\nabla_x h(x)^T \vec v(x) + h(x)\diver( \vec v(x))$ holds \citep{frankel2011geometry}.
For notational simplicity, we suppress the function arguments below. Therefore,
\begin{align*}
    \partial_\theta\left(\varphi_n(g)f\right)&= \partial_\theta\varphi_n(g)f + \varphi_n(g)\partial_\theta f
    = -\varphi_n(g)\diver(sf) + \varphi_n(g)\partial_\theta f + \nabla_x\varphi_n(g)^Tsf + \varphi_n(g)\diver(sf)\\
    &= \varphi_n(g)(d+l)f + \diver(\varphi_n(g)sf),
\end{align*}
which verifies \Cref{eq:pt-dif}. Since $x_0$ is arbitrary, every $x\in\Omega\setminus N_g(\theta_0)$ belongs to a neighborhood on which \Cref{eq:pt-dif} holds. Consequently, \Cref{eq:pt-dif} holds for every $x\in\Omega\setminus N_g(\theta_0)$ and hence a.e. on $\Omega$. Since $\theta_0$ is arbitrary, \Cref{eq:pt-dif} holds for a.e. $x\in\Omega$ for every $\theta\in\Theta$.
}

\textcolor{red}{
It remains to extend the divergence term in \Cref{eq:pt-dif}, initially defined away from $N_g(\theta)$, to all of $\Omega$. By \Cref{assump:prop3-a3}, $s(\cdot,\theta)f(\cdot,\theta)\in W^{1,1}(\Omega;\R^n)$. Since $\varphi_n$ is smooth and $g$ is continuously differentiable on an open neighborhood of $\overline{\Omega}$, $\varphi_n(g(\cdot,\theta))s(\cdot,\theta)f(\cdot,\theta) \in W^{1,1}(\Omega;\R^n)$. Hence, the divergence term $\diver(\varphi_n(g)sf)$ in \Cref{eq:pt-dif} is well defined on $\Omega$ in the weak sense, and the divergence theorem applies \citep{evans2015measure}. To recover \Cref{eq:glr-div-approx} by integrating \Cref{eq:pt-dif} over $\Omega$, it remains to justify differentiation under the integral sign:
\begin{align}\label{eq:inter-d-i}
    \partial_\theta\int_\Omega\varphi_n(g(x,\theta))f(x,\theta)\,dx
    =\int_\Omega\partial_\theta\left(\varphi_n(g(x,\theta))f(x,\theta)\right)dx.
\end{align}
Fix $\theta_0\in\Theta$ and define $H_n(x,\theta)=\varphi_n(g(x,\theta))f(x,\theta).$ Then, $\partial_\theta H_n(x,\theta)
=\nabla\varphi_n(g(x,\theta))^T\partial_\theta g(x,\theta)f(x,\theta)
+\varphi_n(g(x,\theta))\partial_\theta f(x,\theta),$ and hence $|\partial_\theta H_n(x,\theta)| \leq C\|\nabla\varphi_n\|_\infty f(x,\theta)+
\|\varphi_n\|_\infty|\partial_\theta f(x,\theta)|$, where $C=\sup_{(x,\theta)\in\overline \Omega\times\overline\Theta}\|\partial_\theta g(x,\theta)\|_\infty$. By \Cref{assump:prop3-a0}, together with the boundedness of $\partial_\theta g$ on $\overline\Omega\times\overline\Theta$ and the smoothness of $\varphi_n$, the family $\{\partial_\theta H_n(\cdot,\theta):\theta\in \overline\Theta\}$ is uniformly integrable over $\Omega$. For sufficiently small $h$, define $Q_h(x):=(H_n(x,\theta_0+h)-H_n(x,\theta_0))/h.$ By the fundamental theorem of calculus, $Q_h(x)=\int_0^1\partial_\theta H_n(x,\theta_0+th)dt.$ The family $\{Q_h\}$ is also uniformly integrable. Indeed, for any measurable $A\subseteq\Omega$,
\begin{align*}
\int_A|Q_h(x)|dx\leq\int_0^1\int_A|\partial_\theta H_n(x,\theta_0+th)|dxdt.
\end{align*}
Thus, when $\mu(A)$ is sufficiently small, the RHS can be made arbitrarily small uniformly over sufficiently small $h$. Moreover, $Q_h(x)\rightarrow\partial_\theta H_n(x,\theta_0)$ for a.e. $x$ as $h\rightarrow0$. Therefore, by the Vitali convergence theorem \citep[Theorem 3.3.11]{benedetto2009integration}, $Q_h\rightarrow\partial_\theta H_n(\cdot,\theta_0)$ in $L^1(\Omega)$, which justifies \Cref{eq:inter-d-i}.
}

\textcolor{red}{
Integrating \Cref{eq:pt-dif} over $\Omega$ therefore recovers \Cref{eq:glr-div-approx}, with the divergence interpreted in the weak sense. Applying the divergence theorem then yields \Cref{eq:glr-approx}. The remainder of the proof follows from the same limiting argument as in Step 6 of \Cref{sec2:inv}, with \Cref{assump:prop3-a1} replacing \Cref{assump:thm2-a4}.\QEDB}
\end{proof}

\vspace*{-2em}
\textcolor{red}{\begin{remark}
A similar argument can be used to relax the condition in \Cref{assump:thm1-a2} that the Jacobian matrix $J_h(\cdot,\theta)$ be invertible on $V$ to requiring invertibility a.e. on $V$ for all $\theta\in\Theta$, while maintaining the validity of \Cref{thm:glr}.
\end{remark}}
\vspace*{-1em}



\subsection{Easier-to-Verify Alternative Conditions} \label{sec:easy-verifiable}
\Cref{assump:thm2-a2,assump:prop3-a1} are difficult to verify, as they involve convergence related to an infinite sequence of smooth approximations $\{\varphi_n\}$ that may lack closed-form expressions. In this section, we introduce more practical and easily verifiable alternatives, which focus on the structure of discontinuities in $\varphi$.
\begin{assumption}\label{assump:thm2-a3}
    \textcolor{red}{The extension $\varphi\in L^2(\R^n)$ is bounded, and its discontinuity set $D_\varphi$ is closed.} 
\end{assumption}
For each $\theta\in\Theta$, define \textcolor{red}{$D_{\varphi\circ g}(\theta):=\{x\in\overline\Omega : g(x,\theta)\in D_\varphi\}$}, i.e., the set of discontinuities of the sample performance $\varphi(g(\cdot,\theta))$, for a fixed $\theta$. \textcolor{red}{Let $L^1(\partial\Omega, d\sigma)$ denote the space of integrable functions on the boundary $\partial\Omega$ w.r.t. the surface measure $d\sigma$.}
\begin{assumption}\label{assump:thm2-a5}
    The boundary $\partial\Omega$ has finite surface measure, i.e., $\sigma(\partial \Omega)<\infty$. For each \textcolor{red}{$\theta\in\overline \Theta$}, the intersection $D_{\varphi\circ g}(\theta)\cap \partial\Omega$ has surface measure zero, i.e., $\sigma(D_{\varphi\circ g}(\theta)\cap \partial\Omega)=0$. \textcolor{red}{Moreover, letting $w(x,\theta)=s(x,\theta)^T \vec n(x)f(x,\theta)$, the mapping $\theta\to w(\cdot,\theta)$ is continuous from $\overline \Theta$ to $L^1(\partial\Omega, d\sigma)$.}
\end{assumption}
\textcolor{red}{In general, the $L^1$-continuity condition is mild. A simple sufficient condition is that $s(x,\theta)f(x,\theta)$ is continuous in $\theta$ for $\sigma$-a.e. $x\in\partial\Omega$, and that $\sup_{x\in\partial\Omega,\theta\in\overline\Theta}\|s(x,\theta)\| f(x,\theta)<\infty$.} For example, in \Cref{exp2:log}, if the sample space $\Omega$ is a bounded rectangle, \Cref{assump:thm2-a3,assump:thm2-a5} are readily verified: $\varphi$ is an indicator function, hence bounded; \textcolor{red}{$D_{\varphi}$ is the union of two straight lines within the $\R^2$ plane and is therefore closed; the discontinuity set is $D_{\varphi\circ g}(\theta)=\{x\in\overline\Omega : \log(x_1+\theta)+\log(x_2+\theta)=q  \}$, which defines a smooth curve in the $\R^2$ plane that intersects the boundary $\partial\Omega$ at only finitely many points. Finally, $w(x,\theta)=[1~1]^T n(x)f(x)$ is independent of $\theta$, and hence its $L^1$-continuity follows immediately.} 
\begin{proposition}\label{prop:relax-smooth-appr}
    If \Cref{assump:thm2-a1} holds, then \Cref{assump:thm2-a3,assump:thm2-a5} imply \Cref{assump:thm2-a2}.
\end{proposition}
\begin{proof}{Proof.}
    
    \textcolor{red}{Let $I_n(\theta):=\int_{\partial\Omega} \!\! \left|(\varphi(g(x,\theta)) \!-\! \varphi_n(g(x,\theta)))w(x,\theta)\right| \! d\sigma \!$. For any $x_1,x_2\in\Omega$, let $\dist(x_1,x_2)$ denote the Euclidean distance between them. Define $B_\delta(\theta):=\{x\in\partial\Omega \ : \ \dist(g(x,\theta),D_\varphi)\leq \delta\}$.
    First, we prove that $\lim_{\delta\to 0} \sup_{\theta\in\Theta}\int_{B_\delta(\theta)}|w(x,\theta)| d\sigma=0$. Suppose not. Then there exists $\epsilon>0$ and a sequence $\{\theta_k\}\subset \Theta$ such that $\int_{B_{1/k}(\theta_k)}|w(x,\theta_k)| d\sigma>\epsilon$ for all $k$. Since $\overline \Theta$ is compact, by passing to a subsequence, we may assume that $\theta_k\to\theta^*\in\overline \Theta$. Define $v(x,\theta)=\dist(g(x,\theta),D_\varphi)$. Since $D_\varphi$ is closed and $g$ is continuous, $\sup_{x\in\partial \Omega}|v(x,\theta_k)-v(x,\theta^*)|\to 0$ as $k\to\infty$. Let $r_k=1/k+\sup_{x\in\partial \Omega}|v(x,\theta_k)-v(x,\theta^*)|$. Then $r_k\to 0$ and $B_{1/k}(\theta_k)\subseteq B_{r_k}(\theta^*)$ by the triangle inequality. Therefore, $\int_{B_{r_k}(\theta^*)}|w(x,\theta_k)| d\sigma>\epsilon$. By the continuity of $w(x,\theta)$ in $\theta$ w.r.t. the $L^1$ norm, we have $\int_{B_{r_k}(\theta^*)}|w(x,\theta^*)| d\sigma>\epsilon/2$ for sufficiently large $k$. However, since $\1\{x\in B_{r_k}(\theta^*)\}\to \1\{x\in D_{\varphi\circ g}(\theta^*)\cap \partial\Omega\}$ as $k\to\infty$ and $\sigma(D_{\varphi\circ g}(\theta^*)\cap \partial\Omega)=0$, we have $\lim_{k\to\infty}\int_{B_{r_k}(\theta^*)}|w(x,\theta^*)| d\sigma=\int_{D_{\varphi\circ g}(\theta^*)\cap \partial\Omega}|w(x,\theta^*)| d\sigma=0$ by the DCT, contradicting the previous inequality. Thus, we conclude that $\lim_{\delta\to 0} \sup_{\theta\in\Theta}\int_{B_\delta(\theta)}|w(x,\theta)| d\sigma=0$.
    Now fix $\delta>0$ and define $K_\delta=\{g(x,\theta)\ : \ x\in\partial\Omega,\theta\in\overline\Theta,\dist(g(x,\theta),D_\varphi)\geq \delta\}$.Then $K_\delta$ is compact and $\varphi$ is continuous on $K_\delta$. By \Cref{prop:dense}, the smooth approximations $\{\varphi_n\}$ converge uniformly to $\varphi$ on $K_\delta$. Therefore, $a_n(\delta):=\sup_{y\in K_\delta}|\varphi(y)-\varphi_n(y)|\to 0$ as $n\to\infty$. Since $\varphi$ is bounded on $\tilde\Omega$, there exists $M>0$ such that $|\varphi(y)|+|\varphi_n(y)|\leq M,~\forall y\in\tilde\Omega$ for all sufficiently large $n$. Then, for any $\theta\in\Theta$, we have
    \begin{align*}
        I_n(\theta)&\leq\int_{B_\delta(\theta)} \!\! \left|(\varphi(g(x,\theta)) \!-\! \varphi_n(g(x,\theta)))w(x,\theta)\right| \! d\sigma
        +\int_{\partial\Omega\setminus B_\delta(\theta)} \!\! \left|(\varphi(g(x,\theta)) \!-\! \varphi_n(g(x,\theta)))w(x,\theta)\right| \! d\sigma\\
        &\leq M\int_{B_\delta(\theta)} |w(x,\theta)| d\sigma + a_n(\delta)\int_{\partial\Omega} |w(x,\theta)| d\sigma.
    \end{align*}
    Therefore, $\limsup_{n\to\infty} \sup_{\theta\in\Theta} I_n(\theta)\leq M \sup_{\theta\in\Theta}\int_{B_\delta(\theta)} |w(x,\theta)| d\sigma$. Taking $\delta\to 0$ and using the previous result, we conclude that $\lim_{n\to\infty} \sup_{\theta\in\Theta} I_n(\theta)=0$, which establishes \Cref{assump:thm2-a2}.
    \QEDB
    }
\end{proof}
\textcolor{red}{Replacing \Cref{assump:thm2-a5} with \Cref{assump:thm2-a6} below, a similar argument shows that \Cref{assump:thm2-a3,assump:thm2-a6} imply \Cref{assump:prop3-a1}; details are omitted.
\begin{assumption}\label{assump:thm2-a6}
    For each \textcolor{red}{$\theta\in\overline \Theta$}, the intersection $D_{\varphi\circ g}(\theta)\cap \Omega$ has measure zero, i.e., $\mu(D_{\varphi\circ g}(\theta)\cap \Omega)=0$. Moreover, the mapping $\theta\to (d(\cdot,\theta)+l(\cdot,\theta))f(\cdot,\theta)$ is continuous from $\overline \Theta$ to $L^1(\Omega, d\mu)$.
\end{assumption}
\vspace*{-0.5em}
\begin{proposition}\label{prop:relax-smooth-appr-2}
    If \Cref{assump:thm2-a1} holds, then \Cref{assump:thm2-a3,assump:thm2-a6} imply \Cref{assump:prop3-a1}.
\end{proposition}
}

\subsection{Unbounded Sample Space $\Omega$}\label{sec2-2:ubdd}
In this section, we assume $\Omega$ is a hyperrectangle as in \Cref{sec3-1:GLR}, but possibly unbounded, i.e., $\Omega=[a_1,b_1]\times\cdots\times[a_n,b_n]$, where $a_i=-\infty$ or $b_i=\infty$ is allowed. For any $L>0$, define the truncated domain $\Omega_L := \Omega \cap [-L,L]^n$ and boundary points $a_i^L := a_i \vee (-L)$, $b_i^L := b_i \wedge L$. We introduce the following notation:
\begin{align*}
    &A_{i,L}(\theta):=\E(\varphi(g(X,\theta))s(X,\theta)^Te_i\1\{X\in\Omega_L\}|X_i=a_i^L),~A_{i}(\theta):=\E(\varphi(g(X,\theta))s(X,\theta)^Te_i|X_i=a_i),\\
    &B_{i,L}(\theta):=\E(\varphi(g(X,\theta))s(X,\theta)^Te_i\1\{X\in\Omega_L\}|X_i=b_i^L),~B_{i}(\theta):=\E(\varphi(g(X,\theta))s(X,\theta)^Te_i|X_i=b_i),\\
    &C_{L}(\theta):=\E\left(\varphi(g(X,\theta)) (d(X,\theta)+l(X,\theta))\1\{X\in\Omega_L\}\right),~C(\theta):=\E\left(\varphi(g(X,\theta)) (d(X,\theta)+l(X,\theta))\right).
\end{align*}
Under the following assumption, we can extend \Cref{thm:glr1} to the unbounded hyperrectangle $\Omega$.
\vspace*{-0.5em}
\begin{assumption}\label{assump:unbdd}
    \begin{itemize}
        \item For $\mu$-a.e. $x \in \Omega$, the Jacobian $J_g(x,\theta)$ is invertible for almost every $\theta \in \Theta$. Additionally, $\int_{\Omega}\esssup_{\theta\in\Theta} \left|\varphi(g(x,\theta)) (d(x,\theta)+l(x,\theta)) f(x,\theta)\right|dx < \infty$.
        \item The sample performance $\varphi(g(X,\theta))$ is $L^1$-integrable.
        \item For each $i$, if $a_i = -\infty$, then $\lim_{L\rightarrow\infty}\sup_{\theta\in\Theta}f_{X_i}(a_i^L,\theta)= 0$, and $\sup_{\theta\in\Theta}\limsup_{L\rightarrow\infty} \left|A_{i,L}(\theta)\right|<\infty$; if $a_i<\infty$, then for $\sigma$-a.e. $x \in \prod_{j \ne i} [a_j, b_j]$, the Jacobian $J_g(x,\theta)$ is invertible in $x$ for almost every $\theta\in\Theta$, and $\E\left(\esssup_{\theta\in\Theta} \left|\varphi(g(X,\theta))s(X,\theta)^Te_i\right| \ \Big| \ X_i=a_i\right)<\infty$. Analogous conditions are assumed for $b_i$ and $B_{i,L}$.
    \end{itemize}
\end{assumption}
\vspace*{-0.5em}
\begin{proposition}
    Suppose \Cref{assump:thm2-a1,assump:thm2-a2,assump:thm2-a4} hold on $\Omega_L$ for any $L > 0$. If, in addition, \Cref{assump:unbdd} holds, then \Cref{thm:glr1} extends to the unbounded hyperrectangle $\Omega$.
\end{proposition}

\begin{proof}{Proof.}
    Since \Cref{assump:thm2-a1,assump:thm2-a2,assump:thm2-a4} hold on the truncated domain $\Omega_L$, applying \Cref{thm:glr1} and \Cref{eq:surface-orig} with $X$ restricted to $\Omega_L$ yields:
    \vspace*{-0.5em}
    \begin{align}\label{eq:trunc-deri}
        \frac{d}{d\theta}\E(\varphi(g(X,\theta))\1\{X\in\Omega_L\})&=C_L(\theta)
        +\sum_{i=1}^n B_{i,L}(\theta)f_{X_i}(b_i^L,\theta)
        -\sum_{i=1}^n A_{i,L}(\theta)f_{X_i}(a_i^L,\theta).
    \end{align}
    Each condition in \Cref{assump:unbdd} ensures the validity of taking the limit $L \to \infty$ in the corresponding terms in \Cref{eq:trunc-deri} via the DCT (details omitted for brevity):
    \begin{itemize}
        \item The first condition ensures that $C_L(\theta)$ converges to $C(\theta)$ uniformly over $\theta\in\Theta$. 
        \item The second condition ensures that $\E(\varphi(g(X,\theta))\1\{X\in\Omega_L\})$ converges to $\E(\varphi(g(X,\theta)))$ for each $\theta\in\Theta$.
        \item 
        For each $i$, if $a_i = -\infty$, the third condition implies that $A_{i,L}(\theta)f_{X_i}(a_i^L,\theta)$ converges to zero; if $a_i$ is finite, the third condition implies that $A_{i,L}(\theta)f_{X_i}(a_i^L,\theta)$ converges to $A_{i}(\theta)f_{X_i}(a_i,\theta)$. Both convergences are uniform over $\theta\in\Theta$. Similar results hold for $b_i$ and $B_{i,L}(\theta)f_{X_i}(b_i^L,\theta)$.
    \end{itemize}
    Taking the limit $L\rightarrow\infty$ in \Cref{eq:trunc-deri} and applying \citet[Theorem 4, Section 16.3.5]{zorich2004mathematical}, we recover \Cref{eq:glr}. \QEDB
\end{proof}


\section{Geometric Insights and Comparisons with GLR}\label{sec:compare}
In this section, we first compare the Leibniz integral and divergence estimators, focusing specifically on the origin of the surface integral term. Building on this geometric perspective, we highlight a major practical advantage of the Leibniz framework over GLR: its ability to identify conditions that eliminate the surface integral, thereby easing implementation and lowering computational costs. We show that the regularity conditions under the Leibniz framework are milder and more easily verifiable. Finally, we discuss the connections between the proposed Leibniz estimators and existing LR-based methods, and provide guidance on method selection.
\subsection{Comparison of the Leibniz Integral and Divergence Estimators}

In \Cref{exp:compare-leibs}, we use an indicator sample performance function to illustrate both Leibniz estimators. By clarifying the integration domains as the intersection of the underlying sample space and the region specified by the sample performance function, we highlight the connection between the two estimators, reveal how the surface integral arises, and motivate strategies for eliminating it within the Leibniz framework.
\begin{example}\label{exp:compare-leibs}
Consider a sample performance function $\psi(X,\theta)=\1\{g(X,\theta)\in U\}$, where $X$ is a two-dimensional random vector supported on a bounded sample space $\Omega \subseteq \R^2$ and with a density function $f:\Omega\times\Theta \to \R$ continuously differentiable in both arguments. The mapping $g:\Omega\times\Theta\mapsto \R^n$ is twice continuously differentiable in its first argument, continuously differentiable in its second, and admits an inverse $g^{-1}(\cdot,\theta)$ for each $\theta$. Assuming the region $\Omega\cap g^{-1}(U,\theta)$ can be smoothly parameterized by a differentiable map (i.e., \Cref{assump:thm1-a2} holds), we can apply the Leibniz divergence rule \Cref{eq:leibniz-div-sec1} to the expected performance $\int_{\Omega\cap g^{-1}(U,\theta)} f(x,\theta)dx$. The resulting estimator requires evaluating only a single volume integral over $\Omega\cap g^{-1}(U,\theta)$. To implement the Leibniz integral estimator, we can apply the change of variables $Y=g(X,\theta)$ and invoke \Cref{eq:leibniz-sec1} on the expected performance $\int_{g(\Omega,\theta)} \1\{y\in U\} \tilde f(y,\theta)dy$, where $\tilde{f}(y,\theta)$ is the density function of $Y$. Then, we must evaluate a volume integral over $g(\Omega,\theta)$ and a surface integral over its boundary $\partial g(\Omega,\theta)$ (alternatively, we can reverse the change of variables to evaluate the integrals over $\Omega$ and $\partial \Omega$).
To visualize this, consider the general case where $U \nsubseteq g(\Omega,\theta)$, as depicted in \Cref{figure:glr!=leibniz}. The Leibniz divergence estimator only requires evaluating the volume integral over the shaded region on the left-hand side (LHS), whereas the Leibniz integral estimator evaluates the volume integral over the shaded region on the RHS, alongside a line integral over the black boundary. However, because the integrand includes the indicator $\1\{y\in U\}$, this line integral only needs to be evaluated along the dashed portion. Since this segment does not form a closed curve, the line integral cannot be converted back into a volume integral using the Leibniz divergence rule. \QEDB
\end{example}

\newcommand{\LeftX}{0}
\newcommand{\ArrowStart}{4.45}
\newcommand{\ArrowEnd}{6.25}
\newcommand{\RShift}{0.55} 

\newcommand{\RcX}{8.85}
\newcommand{\RcY}{2.00}
\newcommand{\RR}{1.38}
\begin{figure}[bt]
\centering
\begin{subfigure}[bt]{0.45\textwidth}

\begin{tikzpicture}[line width=1pt]
  \path[use as bounding box] (0.2,0.2) rectangle (11.5,3.8);

  \begin{scope}[even odd rule]
    \clip (\LeftX-1,-1) rectangle (\LeftX+5,5)
          (\LeftX+3.00,2.0) ellipse [x radius=1.15, y radius=1.35];
    \draw (\LeftX+1.35,0.35) rectangle (\LeftX+3.65,3.65);
  \end{scope}

  \begin{scope}
    \clip (\LeftX+3.00,2.0) ellipse [x radius=1.15, y radius=1.35];
    \draw[dashed] (\LeftX+0.35,0.35) rectangle (\LeftX+3.65,3.65);
  \end{scope}

  \begin{scope}[even odd rule]
    \clip (\LeftX-1,-1) rectangle (\LeftX+5,5)
          (\LeftX+0.35,0.35) rectangle (\LeftX+3.65,3.65);
    \draw[red] (\LeftX+3.00,2.0) ellipse [x radius=1.15, y radius=1.35];
  \end{scope}

  \begin{scope}
    \clip (\LeftX+0.35,0.35) rectangle (\LeftX+3.65,3.65);
    \clip (\LeftX+3.00,2.0) ellipse [x radius=1.15, y radius=1.35];
    \fill[pattern=north east lines, pattern color=red]
      (\LeftX-0.2,-0.2) rectangle (\LeftX+4.2,4.2);
  \end{scope}

  \begin{scope}
    \clip (\LeftX+0.35,0.35) rectangle (\LeftX+3.65,3.65);
    \draw[red] (\LeftX+3.00,2.0) ellipse [x radius=1.15, y radius=1.35];
  \end{scope}

  \node at (\LeftX+1.00,3.20) {$\Omega$};
  \node[red] at (\LeftX+4.85,1.0) {$g^{-1}(U,\theta)$};

  \draw[-{Latex[length=3mm]}] (\ArrowStart,2) -- (\ArrowEnd,2);
  \node at ({(\ArrowStart+\ArrowEnd)/2},2.35) {$Y=g(X,\theta)$};

  \def\xeA{9.45}
  \def\yeA{2.48}
  \def\xradA{1.32}
  \def\yradA{1.12}

  \begin{scope}[even odd rule]
    \clip (6.9-\RShift,0.0) rectangle (11.8-\RShift,4.8)
          (\RcX-\RShift,\RcY) circle [radius=\RR];
    \draw[red] (\xeA-\RShift,\yeA) ellipse [x radius=\xradA, y radius=\yradA];
  \end{scope}

  \begin{scope}[even odd rule]
    \clip (6.9-\RShift,0.0) rectangle (11.8-\RShift,4.8)
          (\xeA-\RShift,\yeA) ellipse [x radius=\xradA, y radius=\yradA];
    \draw (\RcX-\RShift,\RcY) circle [radius=\RR];
  \end{scope}

  \begin{scope}
    \clip (\RcX-\RShift,\RcY) circle [radius=\RR];
    \clip (\xeA-\RShift,\yeA) ellipse [x radius=\xradA, y radius=\yradA];
    \fill[pattern=north east lines, pattern color=red]
      (7.0-\RShift,0.2) rectangle (11.7-\RShift,4.6);
  \end{scope}

  \begin{scope}
    \clip (\RcX-\RShift,\RcY) circle [radius=\RR];
    \draw[red] (\xeA-\RShift,\yeA) ellipse [x radius=\xradA, y radius=\yradA];
  \end{scope}
  \begin{scope}
    \clip (\xeA-\RShift,\yeA) ellipse [x radius=\xradA, y radius=\yradA];
    \draw[dashed] (\RcX-\RShift,\RcY) circle [radius=\RR];
  \end{scope}

  \node at (8.55-\RShift,1.12) {$g(\Omega,\theta)$};

  \node[red] at (10.35-\RShift,2.76) {$U$};

\end{tikzpicture}
\caption{$U\nsubseteq g(\Omega,\theta)$.}\label{figure:glr!=leibniz}
\end{subfigure}

\begin{subfigure}[bt]{0.45\textwidth}
\begin{tikzpicture}[line width=1pt]
  \path[use as bounding box] (0.2,0.2) rectangle (11.5,3.8);

  \draw (\LeftX+0.35,0.35) rectangle (\LeftX+3.65,3.65);
  \node at (\LeftX+3.10,3.20) {$\Omega$};

  \begin{scope}
    \clip (\LeftX+1.6,2.0) ellipse [x radius=1.1, y radius=1.4];
    \fill[pattern=north east lines, pattern color=red]
      (\LeftX+0.2,0.2) rectangle (\LeftX+3.8,3.8);
  \end{scope}

  \draw[red] (\LeftX+1.6,2.0) ellipse [x radius=1.1, y radius=1.4];

  \fill[white] (\LeftX+0.85,1.78) rectangle (\LeftX+2.35,2.22);
  \node[red] at (\LeftX+1.6,2.0) {$g^{-1}(U,\theta)$};

  \draw[-{Latex[length=3mm]}] (\ArrowStart,2) -- (\ArrowEnd,2);
  \node at ({(\ArrowStart+\ArrowEnd)/2},2.35) {$Y=g(X,\theta)$};

  \def\xeB{8.98}
  \def\yeB{2.15}
  \def\xradB{0.95}
  \def\yradB{0.82}

  \draw (\RcX-\RShift,\RcY) circle [radius=\RR];

  \begin{scope}
    \clip (\xeB-\RShift,\yeB) ellipse [x radius=\xradB, y radius=\yradB];
    \fill[pattern=north east lines, pattern color=red]
      (7.0-\RShift,0.4) rectangle (10.7-\RShift,4.0);
  \end{scope}

  \draw[red] (\xeB-\RShift,\yeB) ellipse [x radius=\xradB, y radius=\yradB];

  \node at (8.55-\RShift,1.12) {$g(\Omega,\theta)$};

  \fill[white] (8.82-\RShift,2.00) rectangle (9.14-\RShift,2.30);
  \node[red] at (8.98-\RShift,2.15) {$U$};

\end{tikzpicture}
\caption{$U\subseteq g(\Omega,\theta)$.}\label{figure:glr=leibniz}
\end{subfigure}

\caption{Illustration of the connection between the Leibniz divergence and Leibniz integral estimators. For the sample performance $\psi(X,\theta)=\1\{g(X,\theta)\in U\}$, the LHS of each subfigure is drawn in the original sample space: the black rectangle represents $\Omega$, the support of $X$, and the red ellipse represents $g^{-1}(U,\theta)$, the region specified by the indicator $\1\{g(X,\theta)\in U\}$. The Leibniz divergence method estimates an volume integral over $\Omega\cap g^{-1}(U,\theta)$, shown as the red shaded region on the LHS, whereas the Leibniz integral method applies the change of variables $Y=g(X,\theta)$, which transforms $\Omega$ and $g^{-1}(U,\theta)$ on the LHS into the black and red ellipses on the RHS, representing $g(\Omega,\theta)$ and $U$, respectively. The Leibniz integral estimator estimates a volume integral over $g(\Omega,\theta)$ plus a surface (line) integral over $\partial g(\Omega,\theta)$. However, since the transformed sample performance is $\tilde\psi(Y)=\1\{Y\in U\}$, the actual integration domain for the volume integral is only the red shaded region. Consequently, if $U\nsubseteq g(\Omega,\theta)$, then the surface (line) integral is taken only over the dashed black segment; if $U\subseteq g(\Omega,\theta)$, then the surface integral vanishes.
}
\end{figure}

\subsection{Elimination of Surface Integrals}\label{sec:elim-sur}
As \Cref{exp:compare-leibs} illustrates, the surface integral arises precisely from the $\theta$-dependence of the integration domain $g(\Omega,\theta)$. This geometric insight motivates a direct strategy for its elimination: selecting a change of variables $Y=g(X,\theta)$ that yields a $\theta$-independent domain inherently causes the surface integral to vanish \citep[see also][Section 2.2]{puchhammer2022likelihood}. While \citet{peng2018new} notes that GLR coincides with push-out LR when both (i) the surface integral vanishes and (ii) the domain $g(\Omega,\theta)$ is independent of $\theta$, the standard GLR framework does not establish that the latter condition implies the former. The following example illustrates this strategy. Additional examples of changes of variables that yield $\theta$-independent support can be found in \citet[Section 2.2]{wang2012new}.

\vspace*{-0.5em}
\begin{example}\label{exp6:glr-cov-smart}
    Consider the sample performance from \Cref{exp4:glr=leibniz}: $\psi(X,\theta)=\1\{0 \leq p(X) \leq w(\theta)\}$, where $X=(X_1,\cdots,X_n)$, $w: \Theta \to \R_+^n$, and $p: \R^n \to \R^n$ is invertible. Let $p_i$ and $w_i$ denote the $i^{\text{th}}$ component of $p$ and $w$, respectively. Assume $p_i(\Omega)=\R_+$ for each $i$. We will present a concrete instance of such sample performance in density estimation problems later. To apply the Leibniz integral estimator while avoiding the surface integral, define the change of variables $Y=g(X,\theta)$, where $Y=(Y_1,\cdots,Y_n)$ and each component is defined as $y_i=g_i(x,\theta) := p_i(x)/w_i(\theta)$. The sample performance then becomes $\varphi(g(X,\theta))$, where $\varphi(y) = \prod_{i=1}^{n}\1\{0 \leq y_i \leq 1 \}$. Since $g(\Omega,\theta)=\R_+^n$ is independent of $\theta$, the surface integral vanishes, and the Leibniz integral estimator is $\psi(X,\theta) \diver\left(-f(X,\theta) J_g^{-1}(X,\theta) \partial_\theta g(X,\theta)\right)/f(X,\theta)$. The same estimator can be obtained by applying the Leibniz divergence method \Cref{estimator:leibniz-div} with $U = \R_+^n$ and the same $g$. \QEDB
    \vspace*{-0.5em}
\end{example}
A special case of the general formulation in \Cref{exp6:glr-cov-smart}, applied to a density estimation problem for the completion time in a stochastic activity network (SAN), is considered in \citet{puchhammer2022likelihood}. The sample performance has the form $\prod_{i=1}^{n}\1\{0\leq \sum_{k=1}^n a_{i,k} X_k\leq \theta\}$, where $\{a_{i,k}\}$ are constants in $\{0,1\}$, $\{X_k\}$ are random variables supported on $\R_+$ representing edge lengths, and for each $i$, $\sum_{k=1}^n a_{i,k} X_k$ corresponds to the length of a path connecting the source to the sink. This fits into the framework of \Cref{exp6:glr-cov-smart} by taking $p_i(x)=\sum_{k=1}^n a_{i,k} x_k$ and $w_i(\theta)=\theta$. Two Leibniz integral estimators are proposed, corresponding to two different changes of variables. The first, based on $g_i(x,\theta) := \sum_{k=1}^n a_{i,k} x_k / \theta$, aligns with the construction in \Cref{exp6:glr-cov-smart} and yields a single-run estimator. The second, based on $g_i(x,\theta): = \sum_{k=1}^n a_{i,k} x_k - \theta$, results in a $\theta$-dependent domain $g_i(\Omega,\theta) = [-\theta, \infty)$ and requires estimating a surface integral. \citet{peng2022variance} studies the same problem and proposes a GLR estimator that coincides with the one based on the second change of variables. 

In their numerical experiments, \citet{puchhammer2022likelihood} demonstrate that the first estimator, which avoids surface integrals, consistently achieves lower variance than the second. Because their specific setup assumes the inputs $\{X_k\}$ are i.i.d., evaluating the surface integral does not require simulating extra sample paths. However, as we illustrate in \Cref{sec:simulation}, the presence of a surface integral can complicate implementation or even render the estimator computationally prohibitive, particularly when the input variables are correlated. In such scenarios, the number of extra densities (paths) required for simulation grows linearly with the number of inputs. Furthermore, these densities often lack closed-form expressions, making them difficult to derive and sample from.

Selecting a change of variables that yields a $\theta$-independent support is not the only way to eliminate the surface integral. Even when $g(\Omega,\theta)$ depends on $\theta$, analyzing the actual integration domain can reveal conditions to either eliminate the surface integral entirely or convert it into a volume integral. The Leibniz divergence estimator (\Cref{thm:glr}) exemplifies the latter, providing a systematic approach for this conversion by smoothly parameterizing the actual integration domain. In other scenarios, the surface integral may simply vanish on its own, as illustrated by the following example.
\begin{example}\label{exp:act-domain}
    Revisiting \Cref{exp:compare-leibs}, suppose that $U\subseteq g(\Omega,\theta)$, as depicted in \Cref{figure:glr=leibniz}.  Applying the change of variables $Y=g(X,\theta)$ reduces the problem to differentiating an integral over the fixed set $U$: $\frac{d}{d\theta}\int_{U} \tilde f(y,\theta)dy$. Consequently, the surface integral vanishes completely under the Leibniz integral rule, despite the $\theta$-dependence of $g(\Omega,\theta)$. Notably, applying the Leibniz divergence estimator (as in \Cref{exp4:glr=leibniz} with $h:=g^{-1}$) yields the exact same estimator. \QEDB
\end{example}
In summary, the surface integral may vanish under three distinct conditions: (1) for general discontinuous sample performance functions, if the transformed domain $g(\Omega,\theta)$ is independent of $\theta$; (2) for indicator-type discontinuities, if the actual integration domain (e.g., \Cref{exp:act-domain}) is independent of $\theta$ or admits a smooth parameterization in the sense of \Cref{assump:thm1-a2}; or (3) if the density of the input vector $X$ vanishes on the boundary of its support, as noted in \Cref{sec3-1:GLR}. A suitable change of variables can often achieve these conditions: the first two are demonstrated in this section, while the third can occur when transforming the sample space to $\R^n$ \citep{peng2018new}.

\subsection{Milder Regularity Conditions}
The Leibniz-based derivation made in this paper establishes regularity conditions that are milder and easier to verify than those assumed in \citet{peng2018new,peng2020generalized}. To illustrate this, we examine two types of conditions required by both the Leibniz integral and GLR methods. First, in the initial stage of the derivation, both methods require verifying convergence in expectation for smooth approximations of discontinuous sample performances. This verification can be challenging \citep[see][Remark 5]{peng2018new}; for specific condition statements, see \Cref{assump:thm2-a2,assump:prop3-a1}, \citet[(A.4)(iii)]{peng2018new}, and \citet[(A.1)]{peng2020generalized}. Although \citet[Proposition 1]{peng2020generalized} offers more tractable alternatives, these conditions may still be violated in applications including density estimation, such as \Cref{exp1:max}. In contrast, the corresponding alternative conditions under the Leibniz framework (\textcolor{red}{\Cref{assump:thm2-a3,assump:thm2-a5,assump:thm2-a6}}) are readily satisfied, as discussed in \Cref{sec:easy-verifiable} and the example below.
\begin{example}
    Reconsider the sample performance from \Cref{exp1:max}: $\psi(X,\theta) = \1\{\max\{X_1,X_2\} \leq \theta\} = \1\{X_1 \leq \theta\} \1\{X_2 \leq \theta\}$, where $X=(X_1,X_2)$ consists of two independent uniform random variables on $(0,1)$, and $\Theta := [a,b] \subset (0,1)$. To apply the Leibniz integral estimator, we write $\psi(X,\theta) = \varphi(g(X,\theta))$, where $\varphi:\R^2\mapsto \R$ is defined as $\varphi(y) = \1\{y_1 < 1\}\1\{y_2 < 1\}$ for $y=(y_1,y_2)$, and $g:\R^2 \mapsto \R^2$ is defined as $g(x,\theta) = (g_1(x,\theta), g_2(x,\theta))$ with $g_i(x,\theta) = x_i/\theta$ for $i=1,2$. 
    \citet{puchhammer2022likelihood} observe that the regularity conditions required in GLR papers are violated in density estimation problems of this type. First, because $X$ is supported on the bounded domain $(0,1)^2$, the conditions required by \citet{peng2018new} do not apply, as their framework requires $X$ to be supported on the entire $\R^2$. Thus, this example falls under the scope of \citet{peng2020generalized}, which requires at least one of the following to hold for some fixed $\epsilon \in (0,1)$:
    \begin{align*}
        &\inf_{\theta\in[a,b],x_i\in[0,1]\setminus[\epsilon,1-\epsilon]} |g_i(x,\theta)|>0,~i=1,2,~\text{(Condition A.1/ Proposition 1), or}\\
        &\lim_{x_i\rightarrow1^-} \sup_{\theta\in[a,b]} |g_i(x,\theta)|=\lim_{x_i\rightarrow0^+} \sup_{\theta\in[a,b]} |g_i(x,\theta)|=0,~i=1,2,~\text{(Condition A.4)}.
    \end{align*}
    Clearly, neither condition holds.
    In contrast, the conditions proposed in this paper can be easily verified. Based on the discussion in \Cref{prop:relax-smooth-appr,prop:relax-smooth-appr-2}, we only need to check \textcolor{red}{\Cref{assump:thm2-a1,assump:thm2-a4,assump:thm2-a3,assump:thm2-a5,assump:thm2-a6}}. It is straightforward to see that \Cref{assump:thm2-a1} holds. For \Cref{assump:thm2-a4}, notice that the transformed density is $\tilde f(y,\theta)=\theta^2 \1\{y\in (0,1/\theta)^2\}$; thus, the integrability conditions are clearly satisfied. \Cref{assump:thm2-a3,assump:thm2-a5} are also readily verified: $\varphi$ is an indicator function and is therefore bounded; the transformed domain $\tilde{\Omega}$ is a bounded rectangle, and $D_{\varphi}=\{y|y_1=1,\text{ or }y_2=1\}$ is the union of two straight lines \textcolor{red}{and is therefore closed}. The discontinuity set of $\varphi\circ g$ is $D_{\varphi\circ g}(\theta)=\{x\in\Omega : x_1=\theta \text{ or } x_2=\theta  \}$, which intersects the boundary $\partial\Omega$ at exactly four points. Finally, since $s(x,\theta) = (-x_1/\theta, -x_2/\theta)$ and $d(x,\theta)=2/\theta$, it is straightforward to \textcolor{red}{verify the $L^1$-continuity conditions in both \Cref{assump:thm2-a5,assump:thm2-a6}} . \QEDB
\end{example}
The second type of condition concerns local invertibility. As established in \Cref{prop:local-inv}, the Leibniz integral estimator is valid provided the Jacobian $J_g(\cdot,\theta)$ is invertible a.e. (\Cref{assump:prop3-a2}), i.e., $g(x,\theta)$ is locally invertible by the inverse function theorem. In contrast, the papers on GLR methods impose stricter invertibility requirements. For instance, \citet[Condition (A.3)]{peng2018new} requires partitioning the sample space $\Omega$ into fixed, $\theta$-independent subsets over which $g(\cdot,\theta)$ is invertible for any $\theta$, while \citet[Condition (A.4)]{peng2020generalized} demands strict global invertibility. A simple parameter-dependent quadratic mapping, such as $g(x,\theta) = (x - \theta)^2$, violates both conditions: Because the vertex (i.e., the boundary of monotonicity) shifts continuously with $\theta$, it is impossible to construct a $\theta$-independent partition of $\R$ where the function remains strictly invertible. However, since its derivative vanishes at only a single point for any given $\theta$, the Jacobian is invertible a.e., thereby satisfying the local invertibility required by the Leibniz framework.


\subsection{Connections with Existing LR-Based Methods}\label{sec:ipalr-leibniz}
In this section, we summarize the connections between the Leibniz derivative estimators and existing LR-based methods. For a more detailed summary, including comparisons with PA-based methods, see \ref{appd:prac-implem}. First, both the Leibniz integral and Leibniz divergence estimators can be viewed as generalizations of push-out LR. In particular, in the derivation of the Leibniz integral estimator in \Cref{sec2:inv}, Step 1 applies a change of variables (push-out), and Step 3 differentiates the resulting density. Under the reverse change of variables in Step 4, the resulting LR term is $d(X,\theta)+l(X,\theta)$ in \Cref{eq:glr1}. Likewise, \Cref{estimator:leibniz-div} also contains terms of a similar form, indicating its LR nature. The main advantages of the proposed Leibniz methods over push-out LR and SLRIPA are twofold: they do not require an explicit change of variables, and they can handle parameter-dependent sample spaces after the change of variables.

In practice, method selection among LR-based approaches may proceed as follows. For discontinuous sample performance functions, to which neither IPA nor the standard LR method applies, one may first ask whether a change of variables is available that smooths the sample performance. If so, and if the transformed random vector has parameter-independent support, then push-out LR or SLRIPA is applicable. Otherwise, one should turn to the Leibniz estimators. One way to avoid the surface integral in the Leibniz integral estimator is to choose a change of variables such that the transformed sample space is parameter-independent or the input density vanishes on its boundary. If the Leibniz integral estimator still involves a surface integral that is difficult to evaluate, for example in the correlated-input case in \Cref{sec:simulation}, then the Leibniz divergence estimator is often the more suitable choice.

The Leibniz integral estimator reduces to the classical IPA-LR estimator only when the discontinuities of the sample performance w.r.t. the input vector are negligible in the sense required by the divergence theorem. This requires more than Lebesgue measure zero; in particular, the discontinuity set must have suitable capacity zero. In the examples considered in this paper, that condition typically fails, so the surface integral term cannot in general be converted into a volume integral. Thus, the surface integral should be viewed as the term that accounts for the contribution of discontinuities. Further details are given in \ref{appd:ipalr-leibniz}.

\section{Simulation Example: An Inventory-Price Control Problem}\label{sec:simulation}
In this section, we conduct simulation experiments to evaluate and compare the proposed Leibniz estimators by analyzing the sensitivity of the long-run average inventory level in a periodically reviewed inventory-price system over an infinite planning horizon, as discussed in \citet{huh2008s}.

Let $\{X_t\}$ be a sequence of (possibly correlated) random variables. In each period $t$, the demand is given by $z(X_t, \theta)$, where $z: \R \times \Theta \mapsto \R_+$. The parameter of interest, $\theta \in \Theta$, represents factors such as price or advertising cost. The setting of \citet{huh2008s} assumes $z(x, \theta)$ is concave in $\theta$ for any fixed $x$, so in our experiments, we take $z(x, \theta) = \log(x + \theta)$ for $x \in \R_+$, and define $\Theta := (1, \infty)$. They also assume zero lead time for order arrivals and allow backlogging. Within each period, orders are fulfilled before the demand arrives.

Under suitable conditions (on costs), \citet{huh2008s} established the optimality of $(s, S)$ policies for this setting. An $(s, S)$ policy operates as follows: whenever the inventory level falls below a reorder point $s$, an order is placed to restore it to a maximum level $S$. Let $q := S - s$, and let $W_t$ denote the inventory level at the beginning of period $t$. Under an $(s, S)$ policy, the dynamics of $W_t$ are governed by:
\begin{align*}
    W_t=\begin{cases}
        W_{t-1}-z(X_t,\theta) &\text{if } W_{t-1}-z(X_t,\theta)>s,\\
        S &\text{if } W_{t-1}-z(X_t,\theta)\leq s.
    \end{cases}
\end{align*}
Our objective is to estimate the derivative of the long-run average inventory level w.r.t. the parameter $\theta$:
\begin{align*}
    \frac{d}{d\theta} \E \left(\lim_{T\rightarrow\infty} \frac{1}{T} \sum_{t=1}^T W_t \right).
\end{align*}
\subsection{I.I.D. Demand}
In this section, we assume the sequence $\{X_t\}$ is i.i.d. We can utilize the regenerative property of the system to express the long-run average inventory level. Specifically, renewals occur whenever a replenishment order restores the inventory level to $S$. Let $Y$ denote the total inventory in a regenerative cycle and $Z$ denote the cycle length. The detailed expressions for $Y$ and $Z$ are given in \Cref{eq:sample-org-W-T}. By the renewal reward theorem, the average inventory level is given by:
\begin{align}\label{eq:sample-org-ratio}
    \E \left( \lim_{T\rightarrow\infty} \frac{1}{T} \sum_{t=1}^T W_t \right)= \frac{\E(Y)}{\E(Z)}.
\end{align}
To compute the derivative of the average inventory level w.r.t. $\theta$, it suffices to estimate the expectations in the numerator and denominator, along with their respective derivatives. \textcolor{red}{For further details on estimating derivatives of ratios, see \citet{glynn1991gradient}.} As this example aims to illustrate the proposed derivative estimators, we focus specifically on estimating these derivatives. Detailed derivations of these Leibniz gradient estimators are provided in \ref{appd:details-simu}. Specifically, for constructing the Leibniz divergence estimator, both $Y$ and $Z$ fall within the class of sample performance functions discussed in \Cref{exp5:inventory}, and the techniques developed there are directly applicable.

In our simulation experiments, we implement both the Leibniz divergence and Leibniz integral estimators, comparing their performance against FD estimators with common random numbers across various perturbation sizes. The simulation results are presented as point estimates, with corresponding standard errors enclosed in parentheses. The number of replications $N$, the system parameters $s$, $S$, and $\theta$, the distribution parameters for $X_t$, and the estimated expected regenerative cycle length $\E(Z)$ are provided in the respective tables. For the results in \Cref{tb:indp-gamma-combined}, we assume $X_t$ follows a Gamma distribution and set $N=10^6$. Additional simulations using $N=10^4$ replications, as well as cases where $X_t$ follows a log-normal distribution, yield similar conclusions; see \ref{appd:details-simu} for full details.
\begin{table}[tb]
\centering
\caption{Comparison of derivative estimators for $N=10^6$, $s=1$, $S=3$, and $\theta=1.1$ under two different Gamma distributions. Values represent point estimates (standard errors). The expected cycle lengths are $\E(Z)=4.142$ for the first case and $\E(Z)=2.869$ for the second.}
\label{tb:indp-gamma-combined}
\begin{tabularx}{\textwidth}{l *{4}{Y}}
\toprule
 & \multicolumn{2}{c}{$X_t\sim\text{Gamma}(1.5,0.5)$} & \multicolumn{2}{c}{$X_t\sim\text{Gamma}(0.5,5)$} \\
\cmidrule(lr){2-3} \cmidrule(lr){4-5}
Estimator & $d\E(Y)/d\theta$ & $d\E(Z)/d\theta$ & $d\E(Y)/d\theta$ & $d\E(Z)/d\theta$ \\
\midrule
FD(0.1)               & $-7.403(0.006)$ & $-3.553(0.005)$ & $-3.303(0.005)$ & $-1.379(0.004)$\\
FD(0.05)              & $-7.926(0.009)$ & $-3.801(0.008)$ & $-3.469(0.006)$ & $-1.439(0.005)$ \\
FD(0.01)              & $-8.421(0.020)$ & $-4.045(0.020)$ & $-3.629(0.012)$ & $-1.507(0.012)$ \\
FD(0.001)             & $-8.570(0.064)$ & $-4.133(0.064)$ & $-3.642(0.039)$ & $-1.498(0.039)$ \\
Leibniz divergence    & $-8.479(0.018)$ & $-4.045(0.010)$ & $-3.864(0.008)$ & $-1.718(0.004)$ \\
Leibniz integral (GLR) & $-8.056(0.192)$ & $-3.849(0.106)$ & $10^6(10^6)$ & $10^6(10^6)$  \\
\bottomrule
\end{tabularx}
\end{table}
We have the following observations:
\begin{itemize}
    \item For Gamma marginal distributions with a shape parameter less than $1$, the Leibniz integral estimator becomes numerically unstable because the estimator involves the term $\partial_{x_i}f_t/f_t$, which is not integrable, where $f_t$ denotes the joint density of $X_1, \dots, X_t$. The Leibniz divergence estimator remains numerically stable and yields lower-variance estimates than the Leibniz integral estimator across all applicable scenarios.
    \item For $X_t\sim\text{Gamma}(1.5,0.5)$, the system has a larger expected cycle length and lower input variance than in the case $X_t \sim \text{Gamma}(0.5, 5)$. In this setting, an FD estimator with perturbation size $0.01$ achieves performance comparable to that of the Leibniz divergence estimator, although the comparison for FD reflects its bias-variance trade-off, whereas the Leibniz divergence estimator is unbiased. Note, however, that FD requires simulating two separate sample paths for each derivative estimate, which makes it more difficult and costly to implement for high-dimensional parameters. In \ref{appd:details-simu}, we provide additional simulation experiments that separately isolate the effects of a longer expected cycle length and lower input variance. The results show that each factor individually degrades the performance of the Leibniz divergence estimator. 
    \item For $X_t \sim \text{Gamma}(0.5, 5)$, the FD estimator exhibits a poor bias-variance trade-off across the tested perturbation sizes. For example, when estimating $d\E(Z)/d\theta$, a perturbation size of $0.05$ yields an FD estimator with variance comparable to the Leibniz divergence estimator, but with substantial bias. Reducing the perturbation size to $0.01$ slightly decreases the bias but triples the standard deviation relative to the Leibniz divergence estimator. A further reduction to $0.001$ inflates the variance, significantly deteriorating the overall estimation quality. \Cref{tb:fd-results} reports additional results for perturbation sizes from $10^{-4}$ to $10^{-8}$ for $N=10^6$, $10^7$, and $10^8$ replications, showing that, as the perturbation size decreases further, the FD estimators become extremely noisy and finally fail to provide reliable estimates. Consequently, the Leibniz divergence estimator delivers the most reliable performance in this setting.
\end{itemize}


\subsection{Two-State Regime-Switching Model}
In this section, we consider a correlated sequence $\{X_t\}$. For each period $t$, we assume the distribution of $X_t$ is determined by a hidden state $S_t$, where $\{S_t\}$ is a discrete-time Markov chain representing the underlying state of the demand process. This formulation is known as a Markov regime-switching model \citep{zucchini2009hidden}, which can effectively capture real-world demand arrivals that shift between different operational modes, rather than fluctuating randomly around a single mean. In this example, we assume a state space of $\mathcal{S}=\{1, 2\}$, where state $1$ represents a low-demand regime and state $2$ represents a high-demand regime. The switching between regimes is governed by the transition probability matrix $P$:
\begin{align*}
        P=\begin{bmatrix}
            p_{11} &1-p_{11}\\
            1-p_{22} &p_{22}
        \end{bmatrix}.
\end{align*}
Given the current state $S_t = s$, the random variable $X_t$ is drawn from a regime-dependent positive distribution:
        \begin{align*}
            X_t|S_t=s \sim f(\cdot,\alpha_s),
        \end{align*}
where $f(\cdot, \alpha_s)$ is a density on $(0, \infty)$ parameterized by a vector $\alpha_s \in \R^k$ for $k \geq 1$ (e.g., the shape and scale parameters of a Gamma distribution).

To estimate the derivative of the long-run average inventory level w.r.t. $\theta$, we can still apply the regenerative property as expressed in \Cref{eq:sample-org-ratio}. However, unlike the standard $(s, S)$ system with i.i.d. demand discussed previously, the condition $W_t = S$ alone is insufficient to trigger a renewal, and a single regenerative cycle may contain multiple replenishments. Nonetheless, the \textcolor{red}{derivative can still be estimated} by evaluating the expectations of the sample performances $Y$ and $Z$ from \Cref{eq:sample-org-W-T} \textcolor{red}{together with their derivatives, under different initial conditions ($S_0=1$ or $S_0=2$). In addition, we need to estimate the proportions of replenishment corresponding to these initial conditions and their derivatives. In this section, we focus on the sample performances in \Cref{eq:sample-org-W-T} under the two initial conditions; further details are provided in \ref{appd:reg-switch}.} Both Leibniz estimators retain their previous forms. The key difference is that for each $t$, the demand sequence $(X_1, \dots, X_t)$ is mutually correlated with the following joint density \citep{zucchini2009hidden}:
\begin{align*}
f_t(x_1,\cdots,x_t)= e_0 P \begin{bmatrix}
    f(x_1,\alpha_1) &0\\ 0 &f(x_1,\alpha_2)
\end{bmatrix}
P \begin{bmatrix}
    f(x_2,\alpha_1) &0\\ 0 &f(x_2,\alpha_2)
\end{bmatrix} \cdots P \begin{bmatrix}
    f(x_t,\alpha_1) &0\\ 0 &f(x_t,\alpha_2)
\end{bmatrix} (1~1)^T,
\end{align*}
where $e_0$ is a two-dimensional vector representing the initial state $S_0$ (i.e., $e_0 = (1~ 0)$ if $S_0 = 1$, and $e_0 = (0~1)$ if $S_0 = 2$).

\textcolor{red}{Let $\tau$ denote the first replenishment epoch.} As shown in \ref{sec:simu-int-est}, implementing the Leibniz integral estimator requires sampling from the conditional densities $f_t(x_1, \dots, x_t \mid x_i)$ for every \textcolor{red}{$t \leq \tau $} and $i \leq t$ in order to estimate the surface integral term. While the unconditional sequence ${X_t}$ is straightforward to simulate by first generating the Markov chain ${S_t}$ and then drawing each $X_t$ conditional on $S_t$, generating ${X_t}$ conditional on a specific $X_i=0$ is significantly more complex because the conditional density $f_t(x_1,\dots,x_t\mid x_i)$ is not readily available. \textcolor{red}{In particular, there are $\tau(\tau+1)/2$ conditional densities to sample from. When the sample of $\tau$ is large, implementing the Leibniz integral estimator becomes computationally cumbersome} without introducing approximations or truncation. Given these computational hurdles, we omit the Leibniz integral estimator from our simulation experiments unless $f(0, \alpha_1) = f(0, \alpha_2) = 0$, which causes the surface integral to vanish.
In the simulation experiments, we assume $\{X_t\}$ follow a conditional Gamma distribution given the hidden state $S_t$, and $S_0=1$. We implement the Leibniz divergence estimator and, where computationally feasible, the Leibniz integral estimator, comparing their performance against FD estimators across various perturbation sizes. The state transition probabilities are set to $p_{11} = 0.8$ and $p_{22} = 0.3$. The simulation results for $N=10^6$ are presented in \Cref{tb:cor-gamma-combined} in the same format as in the previous section, with additional results for $N=10^4$ provided in \ref{appd:details-simu}. The trends in \Cref{tb:cor-gamma-combined} are similar to those in \Cref{tb:indp-gamma-combined}. In particular, the Leibniz divergence estimator continues to perform best among all estimators considered: it remains effective and robust even when the Leibniz integral estimator becomes computationally prohibitive due to the presence of nonvanishing surface integrals. 
\begin{table}[tb]
\centering
\caption{Comparison of derivative estimators for state-dependent Gamma distributions with $N=10^6$, $s=1$, and $\theta=1.1$. Values represent point estimates (standard errors). Case 1 uses $S=5$ with $X_t | S_t=1 \sim \text{Gamma}(1.5,1.5)$ and $X_t | S_t=2 \sim \text{Gamma}(3,1.5)$, yielding $\E(Z)=3.394$. Case 2 uses $S=3$ with $X_t | S_t=1 \sim \text{Gamma}(0.5,5)$ and $X_t | S_t=2 \sim \text{Gamma}(3,5)$, yielding $\E(Z)=2.635$.}
\label{tb:cor-gamma-combined}
\begin{tabularx}{\textwidth}{l *{4}{Y}}
\toprule
 & \multicolumn{2}{c}{Case \textcolor{red}{1} ($S=5$)} & \multicolumn{2}{c}{Case \textcolor{red}{2} ($S=3$)} \\
\cmidrule(lr){2-3} \cmidrule(lr){4-5}
Estimator & $d\E(Y)/d\theta$ & $d\E(Z)/d\theta$ & $d\E(Y)/d\theta$ & $d\E(Z)/d\theta$ \\
\midrule
FD(0.1)                & $-2.037(0.003)$ & $-0.622(0.002)$ & $-2.798(0.005)$ & $-1.048(0.003)$ \\
FD(0.05)               & $-2.078(0.004)$ & $-0.630(0.003)$ & $-2.931(0.006)$ & $-1.092(0.005)$ \\
FD(0.01)               & $-2.119(0.008)$ & $-0.642(0.008)$ & $-3.031(0.011)$ & $-1.114(0.010)$ \\
FD(0.001)              & $-2.179(0.026)$ & $-0.696(0.033)$ & $-2.995(0.033)$ & $-1.060(0.033)$ \\
Leibniz divergence     & $-2.144(0.009)$ & $-0.655(0.003)$ & $-3.159(0.008)$ & $-1.224(0.003)$ \\
Leibniz integral (GLR) & $-2.370(0.264)$ & $-0.780(0.141)$ & N/A\textsuperscript{*} & N/A\textsuperscript{*} \\
\bottomrule
\multicolumn{5}{@{}l@{}}{\footnotesize \textsuperscript{*} Inapplicable due to the presence of a non-vanishing surface integral.} \\
\end{tabularx}
\end{table}
These results not only confirm our previous findings from the independent input case but also demonstrate the validity of the Leibniz divergence estimator for correlated input scenarios.

\section{Conclusions}\label{sec:conc}
We propose two types of stochastic derivative estimators for discontinuous sample performance functions based on the multidimensional Leibniz rules: the Leibniz integral estimator, for general discontinuous sample performance functions, and the Leibniz divergence estimator, for discontinuities induced by indicator functions. The proposed estimators extend existing LR-based methods to a broader range of settings. Compared with push-out LR and SLRIPA, both Leibniz estimators can handle parameter-dependent sample spaces and apply even when no explicit change of variables is available. Compared with GLR, the main contributions are twofold. First, the Leibniz integral estimator includes GLR as a special case, while its analysis relies on milder and easier-to-verify conditions. Second, both Leibniz estimators can provide efficient single-run estimation in settings where GLR may need to evaluate a surface integral, which can make its implementation costly or even infeasible. In particular, the Leibniz divergence estimator is always single-run, and although the Leibniz integral estimator has the same general form as GLR, the derivation in this paper identifies more cases in which the surface integral can be avoided. We illustrate these advantages through simulation experiments with an inventory-price model.

As discussed in \Cref{sec2:indicator}, discontinuities arising from indicator functions appear in a wide range of applications. In such cases, if the Leibniz integral estimator requires the evaluation of multiple sample paths for the surface integral, then the Leibniz divergence estimator is often preferable, since it is always single-run. However, parameterizing the integration domain after embedding the indicator function is problem-specific. \Cref{exp4:glr=leibniz,exp5:inventory} illustrate two representative scenarios, but they do not cover all possibilities. Constructing the Leibniz integral estimator is often more straightforward, provided that a suitable change of variables exists. To avoid the surface integral, as suggested in \Cref{sec:elim-sur}, one may look for a change of variables such that the transformed integration domain is parameter-independent or the input density vanishes on its boundary.



Future research directions include extending the Leibniz framework to a wider range of applications, and developing systematic techniques for parameterizing the integration domain and identifying effective changes of variables. A more comprehensive comparison between the Leibniz divergence and integral estimators would also help clarify when each method is more effective. Given that the push-out LR method has been successfully applied to train artificial neural networks with discontinuous loss and activation functions \citep{peng2022new}, exploring the Leibniz framework within the broader context of machine learning is a highly promising avenue for future work. Finally, integrating the Leibniz estimators with conditioning techniques \citep{l2022monte,peng2022variance} is a natural extension that could significantly improve computational efficiency.

\newpage
\bibliographystyle{informs2014} 
\bibliography{leibniz-ref} 



\newpage
\ECSwitch
\vspace*{-1em}

\noindent\textbf{\large E-Companion}
\section{A Geometric Interpretation of the Leibniz Integral Rule}\label{appd:geo-leibniz}
We provide a geometric interpretation of the surface integral term in the Leibniz integral rule for $\R^2$. Let $D_\theta \subset \R^2$ be a compact domain with smooth boundary, and let $G: \Omega \to \R$ be a smooth function independent of $\theta$, where $\Omega$ is an open set containing $D_\theta$. Our goal is to compute $\frac{d}{d\theta} \int_{D_\theta} G(x,y),dxdy$. For small $\Delta\theta$, suppose the domain $D_\theta$ moves to $D_{\theta+\Delta\theta}$, as shown in \Cref{fig:domain}. As in \Cref{thm:leibniz}, we assume that $D_\theta$ is characterized by a smooth function $\phi:\R^2\times\Theta\mapsto \R^2$ and a fixed domain $U\subset\R^2$, i.e., $D_{\theta}=\phi(U,\theta)$. Consider the difference $\int_{D_{\theta+\Delta\theta}} G(x,y)dxdy - \int_{D_\theta} G(x,y)dxdy$. The integral over the intersection $D_{\theta+\Delta\theta}\cap D_\theta$ cancels out, leaving only two strips surrounding the boundary $\partial D_\theta$ contributing to the difference. We zoom in on a small segment of this strip around a point $x\in\partial D_\theta$, illustrated by the blue region in \Cref{fig:domain}. Here, $d\sigma$ is the arc length element, $\vec n$ is the normal vector of the boundary, 
and $\vec v$ is the velocity vector. 
For sufficiently small $\Delta\theta$, this region is approximately a rectangle of length $d\sigma$ and width $(\vec v\cdot \vec n) \Delta \theta$, the displacement of the domain along the normal vector. Therefore, the area of the blue region is $(\vec v\cdot \vec n) \Delta \theta d\sigma$, and
\begin{align*}
    \frac{d}{d\theta}\int_{D_\theta} G(x,y)dxdy= \lim_{\Delta\theta\rightarrow 0}\frac{1}{\Delta\theta} \left(\int_{U_{\theta+\Delta\theta}} G(x,y)dxdy - \int_{D_\theta} G(x,y)dxdy\right)= \int_{\partial D_\theta} G(x,y) (\vec v\cdot \vec n) d\sigma.
\end{align*}

\begin{figure}[h]
    \centering
    \begin{subfigure}[b]{0.4\textwidth}
        \centering
        \begin{tikzpicture}[scale=0.75, transform shape]
            \draw[thick] (0,2) circle (1.5);
            \draw[red,thick,dashed] (0.3,1.8) circle (1.5);
            \filldraw[fill=cyan, draw=blue] (1.8,1.61) -- (1.45,1.61) arc (-15:15:1.5) -- (1.68,2.39) -- (1.68,2.39) arc (24:-6.5:1.5);
            \draw [->] (-1.5,3.5) -- (-1.3,3) node at (-1.5,3.8){$D_\theta$}; 
            \draw [->,red] (1.6,3.4) -- (1.4,2.9) node at (1.6,3.7){\textcolor{red}{$D_{\theta+\Delta\theta}$}}; 
        \end{tikzpicture}
    \end{subfigure}
    \begin{subfigure}[b]{0.4\textwidth}
        \centering
        \begin{tikzpicture}[scale=0.75, transform shape]
            \filldraw[fill=cyan, draw=blue , thick] (3.2,-0.78) -- (2.6,-0.78) arc (-15:15:3) -- (2.96,0.78) -- (2.96,0.78) arc (24:-6.5:3);
            \draw [->] (2.7,0) -- (3.8,0) node[right=0.1]{$\vec n$}; 
            \draw [->] (2.7,0) -- (4,-0.6) node[right=0.1]{$\vec v$}; 
            \draw [<->] (2.5,-1) -- (3.3,-1) node at (2.9,-1.3){$(\vec v\cdot \vec n )\Delta \theta$}; 
            \draw [<->] (2.4,-0.83) -- (2.4,0.83) node at (2.15,0){$d\sigma$};
        \end{tikzpicture}       
    \end{subfigure}
    \vspace*{1mm}
    \caption{The original domain $D_\theta$ and the perturbed domain $D_{\theta+\Delta\theta}$.}\label{fig:domain}
\end{figure}


\section{Derivations of Estimators in Examples}\label{appd:estimator-derive}
\subsection{Leibniz Estimators in \Cref{exp1:max}}
Consider the sample performance $\psi(X,\theta)=\1\{\max\{X_1,X_2\}\leq \theta\}=\1\{X_1\leq \theta\}\1\{X_2\leq \theta\},~\theta\in(0,1).$ We can rewrite this as $\psi(X,\theta)=\1\{g(X,\theta)\in U\}$, where $g(X,\theta) = X/\theta$ and $U = [0,1]^2$. Note that this corresponds to the case $U \subseteq g(\Omega, \theta)$ discussed in \Cref{exp4:glr=leibniz}. To derive the Leibniz divergence estimator, we set $h := g^{-1}$ in \Cref{estimator:leibniz-div}, which yields
\begin{align*}
    \psi(X,\theta)\diver (f(X,\theta) X /\theta)/f(X,\theta).
\end{align*}
To derive the GLR or Leibniz integral estimator, we write $\psi(X,\theta):=\varphi(g(X,\theta))$, with $\varphi(y) = \prod_{i=1}^2 \1\{y_i \leq 0\}$ and $g_i(x,\theta) = x_i - \theta,~i=1,2$, and substitute these into \Cref{thm:glr1}.

\subsection{Leibniz Estimators in \Cref{exp2:log,exp4:leibniz}}
In \Cref{exp2:log,exp4:leibniz}, we consider the sample performance
\begin{align*}
    \psi(X,\theta)=\1\bigg\{\sum_{i=1}^{2}\log(X_i+\theta)<q\bigg\},~\theta\in[1,\infty),
\end{align*}
which, as shown in \Cref{exp4:leibniz}, is equivalent to
\begin{align*}
    \psi(X,\theta)=\1\bigg\{\log(X_1+\theta)<q\bigg\}\1\bigg\{\log(X_1+\theta)+\log(X_2+\theta)<q\bigg\}.
\end{align*}
We develop both Leibniz divergence and Leibniz integral derivative estimators. In \Cref{sec:simulation}, their performance is evaluated under four different joint distributions for $(X_1,X_2)$: (1) Independent exponential; (2) FGM copula with exponential marginals; (3) Joint log-normal; (4) Clayton copula with Gamma marginals. In the following, we provide details on the derivation of the estimators

\subsubsection*{Leibniz Divergence Estimator in \Cref{exp4:leibniz}}
Let $g_1(x,\theta)=\log(x_1+\theta)$, $g_2(x,\theta)=\log(x_1+\theta)+\log(x_2+\theta)$, and $U=(-\infty,q)^2$. Now we apply the method from \Cref{exp4:leibniz} to construct a function $h(v,\theta)$ and domain $V$, such that $g(\Omega,\theta)\cap U=h(V,\theta)$. The region defined by the indicator function $\1\{g(x,\theta)\in U\}$ corresponds to:
\begin{align*}
    x_1\in \left(0,\frac{1}{\theta}e^q-\theta\right),~x_2\in\left(0,\frac{1}{x_1+\theta}e^q-\theta\right).
\end{align*}
To match the ranges of each dimension, we let $V_1=V_2=(0,1)$, and define
\begin{align*}
    h_1(v,\theta)=\left(\frac{1}{\theta}e^q-\theta\right)v_1,~h_2(v,\theta)=\left(\frac{1}{h_1(v,\theta)+\theta}e^q-\theta\right)v_2.
\end{align*}
Note that $h(v,\theta)$ is invertible in $v$. In particular, solving $x=h(v,\theta)$ yields
\begin{align*}
    v_1=\frac{\theta x_1}{e^q-\theta^2},~v_2=\frac{x_2(x_1+\theta)}{e^q-\theta^2-\theta x_1}.
\end{align*}
We now compute the partial derivatives needed for the Leibniz estimator:
\begin{align*}
    &\partial_\theta h_1(v) |_{v=h^{-1}(x,\theta)}=-\frac{e^q+\theta^2}{\theta(e^q-\theta^2)}x_1,\\
    &\partial_\theta h_2(v) |_{v=h^{-1}(x,\theta)}=\left(-\frac{\partial_\theta h_1(v) |_{v=h^{-1}(x,\theta)}+1}{(x_1+\theta)^2}e^q-1\right)\frac{(x_1+\theta)x_2}{e^q-\theta^2-\theta x_1}.
\end{align*}
By \Cref{thm:glr}, the Leibniz estimator is given by:
\begin{align*}
    \1\{g(X,\theta)\in U\}\diver (f(X)\partial_\theta h(v) |_{v=h^{-1}(X,\theta)})/f(X),
\end{align*}
Carrying out the differentiation yields the explicit form of the Leibniz divergence estimator used in \Cref{sec:simulation}.

\subsubsection*{Leibniz Integral Estimator in \Cref{exp2:log}} Let $g_1(x,\theta)=\log(x_1+\theta)$, $g_2(x,\theta)=\log(x_2+\theta)$, and $\varphi(y)=\1\{y_1+y_2<q\}$. Then $J_g(x,\theta)$, $J_g^{-1}(x,\theta)$, and $\partial_\theta g(x,\theta)$ are given by:
\begin{align*}
    J_g(x,\theta)=\begin{bmatrix}
        \frac{1}{x_1+\theta} &0 \\ 0 &\frac{1}{x_2+\theta}
    \end{bmatrix},~
    J^{-1}_g(x,\theta)=\begin{bmatrix}
        x_1+\theta & 0 \\ 0 & x_2+\theta
    \end{bmatrix},
    \partial_\theta g(x,\theta)=\begin{bmatrix}
        \frac{1}{x_1+\theta} \\ \frac{1}{x_2+\theta}
    \end{bmatrix}.
\end{align*}
Then, we can compute $s(x,\theta)$ and $d(x,\theta)$:
\begin{align*}
    s(x,\theta)=J^{-1}_g(x,\theta)\partial_\theta g(x,\theta)=(1,1)^T,~d(x,\theta)=\diver(-f(x)s(x,\theta))=-(\partial_{x_1}f(x)+\partial_{x_2}f(x))/f(x).
\end{align*}
Substituting these expressions into \Cref{thm:glr1} yields the Leibniz integral estimator.


\section{The Relationship Between IPA-LR and Leibniz Integral Estimators} \label{appd:ipalr-leibniz}
\citet[Theorem 1]{shapiro1958divergence} provides necessary and sufficient conditions for the divergence theorem to hold for an a.e. differentiable vector field in $\R^n$. We restate this result in \ref{appd:div-discont}. Specifically, the discontinuity set has logarithmic capacity zero for $n = 2$ and Newtonian capacity zero for $n \geq 3$. Note that ``capacity zero'' is a stronger condition than ``measure zero''. For instance, in $\R^3$, both a 2-dimensional disk and a line segment have Lebesgue measure zero, but only the line segment has zero Newtonian capacity; the disk’s capacity is positive \citep{landkof1972foundations}.

It is tempting to assume that we can apply \citet[Theorem 1]{shapiro1958divergence} to convert the surface integral in \Cref{thm:glr1} into a volume integral, thereby obtaining a single-run estimator. Unfortunately, the conditions for this theorem are rarely satisfied by the sample performance functions discussed in this paper. Assuming an $n$-dimensional input, the discontinuity sets in our examples typically have an $n$-dimensional Lebesgue measure of zero but a strictly positive capacity. For example, on the RHS of \Cref{figure:glr!=leibniz}, where $n=2$, the discontinuity set forms a dashed line segment, which has a positive logarithmic capacity. Therefore, the divergence theorem cannot be directly applied. Nonetheless, applying \citet[Theorem 1]{shapiro1958divergence} hypothetically provides valuable insight into the origin of the surface integral and its connection to discontinuities.

For each $\theta$, suppose the vector-valued function $x\mapsto\varphi(g(x,\theta))s(x,\theta)f(x,\theta)$ satisfies the conditions of \citet[Theorem 1]{shapiro1958divergence}.
Then, by the divergence theorem, the following equation holds:
\begin{align*}
    \int_{\partial\Omega} \varphi(g(x,\theta))s(x,\theta)^T \vec n(x) f(x,\theta)d\sigma=\int_\Omega \diver(\varphi(g(x,\theta))s(x,\theta)f(x,\theta))dx.
\end{align*}
Substituting this equation into \Cref{eq:glr}, we obtain:
\begin{align*}
    \frac{d}{d\theta}\E(\varphi(g(X,\theta)))&=  \E\left(\varphi(g(X,\theta))l(X,\theta)\right) \\
    &+\int_\Omega\left(\varphi(g(x,\theta)) \diver(-s(x,\theta)f(x,\theta))  + \diver(\varphi(g(x,\theta))s(x,\theta)f(x,\theta))\right)dx.
\end{align*}
Let $\nabla_x$ denote the gradient operator w.r.t. $x$. For any differentiable scalar-valued function $h(x)$ and vector-valued function $\vec v(x)$, the identity $\diver(h(x)\vec v(x))=\nabla_x h(x)^T \vec v(x) + h(x)\diver( \vec v(x))$ holds \citep{frankel2011geometry}, allowing us to compute
\begin{align}\label{eq:glr=ipalr}
    \begin{split}
        &\varphi(g(x,\theta)) \diver(-s(x,\theta)f(x,\theta))  + \diver(\varphi(g(x,\theta))s(x,\theta)f(x,\theta))\\
        &=f(x,\theta)\nabla_x \varphi(g(x,\theta))^T J_g^{-1}(x,\theta)\partial_\theta g(x,\theta)
        =f(x,\theta)\left(\nabla_y \varphi(y)|_{y=g(x,\theta)}\right)^T \partial_\theta g(x,\theta)
        = f(x,\theta)\partial_\theta \varphi(g(x,\theta)).
    \end{split}
\end{align}
Therefore, $\frac{d}{d\theta}\E(\varphi(g(X,\theta)))=\E\left(\varphi(g(X,\theta))l(X,\theta)+\partial_\theta\varphi(g(X,\theta))\right)$, leading to the IPA-LR estimator \citep{l1990unified}. This implies that if the discontinuity set of the sample performance w.r.t. the \textit{input vector} has capacity zero, then its impact is negligible, and the Leibniz integral estimator is equivalent to the IPA-LR estimator. Otherwise, the surface integral term arises to account for the effect of discontinuities.

\section{The Divergence Theorem for Discontinuous Vector Fields}\label{appd:div-discont}
In this section, we restate Theorem 1 from \citet{shapiro1958divergence}.
\begin{theorem}
    Suppose $\Gamma\subset\R^n$ is a bounded set and its boundary $\partial \Gamma$ is a simple closed curve. If the following conditions hold:
    \begin{itemize}
        \item $F$ is continuous on $\overline \Gamma\setminus D_F$ and is $L^2$-integrable on $\Gamma$.
        \item $\diver F$ exists a.e. and is integrable on $\Gamma$.
        \item $\diver_* F$ and $\diver^* F$ are finite on $\Gamma\setminus D_F$, with $\diver _* F(y):=\liminf_{t\rightarrow 0}\frac{1}{\text{vol}(B(y,t))}\int_{\partial B(y,t)} F(y)^T \vec n(y)dy,$ where $B(y,t)=\{y'\in\R^n \ | \ \|y'-y\|_\infty<t \}$ is an open ball centered at $y$ with radius $t$, and $\text{vol}(B(y,t))$ is its $n$-dimensional volume. $\diver^* F$ is defined similarly by replacing $\liminf$ with $\limsup$.
        \item The set $D_F$ has logarithmic capacity zero if $n=2$, or Newtonian capacity zero if $n\geq3$. For a compact set $K$, the logarithmic capacity is given by $\exp\left(-\min_\mu \int_K\int_K \log(|x-y|^{-1})d\mu(x)d\mu(y)\right)$, and the Newtonian capacity is given by $\left(\min_\mu \int_K\int_K |x-y|^{-(n-2)}d\mu(x)d\mu(y)\right)^{-1}$, where the minimum is taken over all Borel probability measures on $K$. 
    \end{itemize}
    Then, the divergence theorem holds on $\Gamma$: $\int_\Gamma \diver(F(y))dy =\int_{\partial\Gamma}F(y)^T \vec n(y)dy.$
\end{theorem}

\section{Additional Simulation Example: Correlated Inputs Modeled by Copulas}\label{appd:exp-copula}
In this section, we conduct simulation experiments to evaluate and compare the proposed Leibniz estimators using the sample performance $\psi(X,\theta)=\1\{\sum_{j=1}^{2}\log(X_j+\theta)<q\}$, where $X=(X_1,X_2)$, $\theta\in[1,\infty)$, and $q\in\R$ is a constant. The corresponding Leibniz integral and Leibniz divergence estimators have been derived in \Cref{exp2:log,exp4:leibniz}, respectively. 
We consider the following settings for the joint distribution of $(X_1,X_2)$ (for background on copulas and detailed derivations of the estimators, see \ref{appd:copula}): 
\begin{enumerate}
    \item \textbf{Independent exponential:} $X_1$ and $X_2$ are independent $\exp(1)$ random variables. In this case, the conditional density $f_{X \mid X_i = 0}$ reduces to the unconditional marginal, allowing the Leibniz integral estimator to be sampled using a single sample path.
    \item \textbf{Farlie–Gumbel–Morgenstern (FGM) copula with exponential marginals:} The pair $(X_1,X_2)$ has dependence modeled by an FGM copula with parameter $1$ and $\exp(1)$ marginals. In this case, the Leibniz integral estimator requires additional samples from the conditional densities $f_{X|X_i=0},~i=1,2$.
    \item \textbf{Joint log-normal:} $\log X_1$ and $\log X_2$ are jointly normal with correlation $\rho\in\{0.1,0.9\}$ and marginally standard normal. Since $f_i(0,\theta)=0,~i=1,2$, no additional sampling is required.
    \item \textbf{Clayton copula with Gamma marginals:} The pair $(X_1,X_2)$ has dependence modeled by a Clayton copula with parameter $1$ and Gamma$(a,1)$ marginals, for $a\in\{0.5,1,2\}$. Due to strong left-tail dependence of the Clayton copula, $X_2 = 0$ almost surely given $X_1 = 0$, and vice versa; thus, sampling from conditional densities is unnecessary.
\end{enumerate}

For each setting, we set $q=0.5$ and evaluate the derivative estimators at $\theta=1$, using $10000$ independent replications. For the FGM copula, each conditional distribution $f_{X|X_i=0}$ is also simulated with $10000$ samples. We compare the Leibniz estimators with an FD estimator using a perturbation size of $0.02$, which gives the smallest mean square error (MSE) in our experiments. Results are presented in \Cref{table:simulation}, with the best for each of the seven cases indicated by the entry with the smallest standard error. 

\begin{table}[tb]
    \centering
    \begin{tabularx}{\textwidth}{c *{3}{Y}}
    \toprule
    Estimator &FD &Leibniz integral &Leibniz divergence \\
    \midrule
    Independent \& $\exp(1)$   &$-0.710(0.059)$ & $-0.723$\hl{$(0.006)$} & $-0.705(0.020)$\\
    FGM \& $\exp(1)$   &$-0.775(0.062)$ & $-0.848$\hl{$(0.015)$} &  $-0.853(0.020)$\\
    Log-normal ($\rho=0.1$)    &$-0.335(0.041)$ &  $-0.318(0.031)$ & $-0.323$\hl{$(0.019)$}\\
    Log-normal ($\rho=0.9$)    &$-0.565(0.053)$ &  $-0.580(0.041)$ & $-0.587$\hl{$(0.020)$}\\
    Clayton \& Gamma$(0.5,1)$    &$-0.975(0.069)$ &  7.022E+3(1.991E+3) & $-0.977$\hl{$(0.011)$}\\
    Clayton \& Gamma$(1,1)$    &$-0.665(0.058)$ &  $0.105(1.502)$ & $-0.676$\hl{$(0.014)$}\\
    Clayton \& Gamma$(2,1)$    &$-0.170(0.028)$ &  $-0.162(0.029)$ & $-0.165$\hl{$(0.010)$}\\
    \bottomrule
    \end{tabularx}

    \caption{Simulation results: point estimates (standard errors). The FD estimator uses a perturbation size of $0.02$.}
    \label{table:simulation}
\end{table}

The Leibniz integral estimator performs well under independent or weakly dependent inputs (e.g., FGM), but its variance increases under strong dependencies, such as the Clayton copula. Moreover, for Gamma marginals with shape parameter $a\leq 1$, the Leibniz integral estimator becomes unstable, because the term $\sum_{i=1}^2\partial_{x_i}f(X)/f(X)$ is not integrable (for details, see \ref{appd:copula}). In contrast, the Leibniz divergence estimator remains numerically stable and consistently yields low-variance estimates across all scenarios. The numerical results highlight the robustness of the Leibniz divergence estimator, particularly when the Leibniz integral estimator fails and the FD estimator suffers from high variance.



\section{Copulas}\label{appd:copula}
Copulas provide a way for constructing and analyzing scale-free measures of dependence and are particularly useful in simulation. In this section, we briefly review the basic concepts and key properties of two-dimensional copulas \citep{nelsen2006introduction}.

A copula is a multivariate distribution function whose one-dimensional margins are uniform on the interval $[0,1]$. Given any copula $C(u,v)$ and marginal distribution functions $F_1$ and $F_2$, one can construct a joint distribution with those marginals via: $F(x_1,x_2)=C(F_1(x_1),F_2(x_2))$. While this construction is straightforward, the converse result---known as Sklar’s Theorem---is more remarkable: for any joint distribution $F(x_1,x_2)$ with marginals $F_1$ and $F_2$, there exists a copula $C$ such that
\begin{align*}
    F(x_1,x_2)=C(F_1(x_1),F_2(x_2)).
\end{align*}
If the copula $C$ has a density $c(u,v)=\frac{\partial^2C(u,v)}{\partial u\partial v}$, and $F_1,F_2$ have densities $f_1,f_2$, then the joint density $f(x_1,x_2)$ is given by
\begin{align*}
    f(x_1,x_2)=c(F_1(x_1),F_2(x_2))f_1(x_1)f_2(x_2).
\end{align*}
The conditional distribution function can also be derived from the copula. In particular:
\begin{align*}
    F_2(x_2|X_1=x_1):=\P(X_2\leq x_2 |X_1=x_1)= \partial_u C(F_1(x_1),F_2(x_2)).
\end{align*}
In \ref{appd:exp-copula}, we consider two specific copulas:
\begin{itemize}
    \item Clayton Copula - exhibits strong lower tail dependence and no upper tail dependence:
    \begin{align*}
        &C(u,v)=(u^{-\alpha}+v^{-\alpha}-1)^{-\frac{1}{\alpha}},\\
        &c(u,v)=(1+\alpha)(uv)^{-1-\alpha}(-1+u^{-\alpha}+v^{-\alpha})^{-2-\frac{1}{\alpha}}.
    \end{align*}
    \item Farlie-Gumbel-Morgenstern (FGM) copula - allows only weak dependence:
    \begin{align*}
        &C(u,v)=uv+\alpha uv(1-u)(1-v),\\
        &c(u,v)=1+\alpha(1-2u)(1-2v).
    \end{align*}
\end{itemize}
In \ref{appd:exp-copula}, we fix $\alpha=1$ for both copulas.

\subsection{Computational Details for Copula-Based Densities}
We discuss two important implementation details related to copula densities. First, both Leibniz estimators contain the common term $\partial_{x_i} f(X)/f(X)$ (i.e., $\partial_{x_i} \log f(X)$). When testing different joint distributions, this is the only term that changes, while the rest of the estimator remains fixed. For general copulas, a direct computation gives:
\begin{align*}
    \frac{\partial_{x_1} f(x)}{f(x)} 
    = \frac{\partial_u c(F_1(x_1), F_2(x_2))}{c(F_1(x_1), F_2(x_2))} f_1(x_1) + \frac{\partial_{x_1} f_1(x_1)}{f_1(x_1)}.
\end{align*}

Second, implementing the Leibniz integral estimator also requires simulating conditional distributions such as $f_{X|X_1=0}$ and $f_{X|X_2=0}$ to estimate the surface integral. For general copulas, we have:
\begin{align*}
    F_2(x_2|X_1=0)=\partial_u C(0,F_2(x_2)).
\end{align*}

We now present computational details for both $\partial_{x_1} f(X)/f(X)$ and $F_2(x_2|X_1=0)$ in the cases of the Clayton and FGM copulas. The independent exponential and log-normal cases are more straightforward and omitted here.

\subsection{Clayton Copula with Gamma Marginals.}
For the Clayton copula, we have:
\begin{align*}
    \frac{\partial_u c(u,v)}{c(u,v)}=-\frac{2}{u}+\frac{3}{(-1+u^{-1}+v^{-1})u^2}.
\end{align*}
In practice, we substitute $u=F_1(x_1),v=F_2(x_2)$. Let $f_1\sim$Gamma$(\alpha,1)$, so:
\begin{align*}
    &f_1(x_1)=\frac{1}{\Gamma(\alpha)}x_1^{\alpha-1}e^{-x_1},~
    \frac{\partial_{x_1}f_1(x_1)}{f_1(x_1)}=(\alpha-1)x_1^{-1}-1.
\end{align*}

We now provide a rough argument for why the Leibniz integral estimator becomes unstable when $\alpha<1$. Since the integration region defined by $\1\{g(\Omega,\theta)\cap U\}$ is bounded, we consider the expectation of $\frac{\partial_{x_1}f_1(X_1)}{f_1(X_1)}$ over a finite range. For any fixed $z>0$,
\begin{align*}
    \E\left(\1\{X_1<z\}\frac{\partial_{x_1}f_1(X_1)}{f_1(X_1)}\right)&=\int_0^z ((\alpha-1)x_1^{-1}-1) \frac{1}{\Gamma(\alpha)}x_1^{\alpha-1}e^{-x_1} dx_1\\ 
    &\geq\frac{\alpha-1}{\Gamma(\alpha)e^z}\int_0^z x_1^{\alpha-2} dx_1 -1
\end{align*}
The integral diverges when $\alpha<1$, implying that the expectation is unbounded and leading to instability in the Leibniz integral estimator.

In contrast, the Leibniz divergence estimator includes this same term, but it is multiplied by $\partial_\theta h_1(v) |_{v=h^{-1}(x,\theta)}$, yielding:
\begin{align*}
    \frac{\partial_{x_1}f_1(X_1)}{f_1(X_1)}\partial_\theta h_1(v) |_{v=h^{-1}(X,\theta)}=-\frac{e^q+\theta^2}{\theta(e^q-\theta^2)}(-X_1+\alpha -1),
\end{align*}
which has finite expectation for all $\alpha>0$, ensuring the stability of the Leibniz divergence estimator.

To simulate from the conditional distribution $f_{X|X_1=0}$ and $f_{X|X_2=0}$, note that for the Clayton copula:
\begin{align*}
    F_2(x_2|X_1=0)=\partial_u C(0,F_2(x_2))=1
\end{align*}
for all $x_2>0$, which implies that $ X_2 = 0 $ almost surely when $ X_1 = 0 $. This illustrates the strong lower tail dependence of the Clayton copula.

\subsection{FGM Copula with Exponential Marginals}

For the FGM copula, we have:
\begin{align*}
    \frac{\partial_u c(u,v)}{c(u,v)}=\frac{-2(1-2v)}{1+(1-2u)(1-2v)}.
\end{align*}
Let $f_1\sim\exp(1)$, so $f_1(x_1)=\exp(-x_1)$, and $\frac{\partial_{x_1}f_1(x_1)}{f_1(x_1)}=-1$.

Again, to implement the Leibniz integral estimator, we need to simulate from $f_{X|X_1=0}$. The conditional CDF can be obtained via:
\begin{align*}
    \partial_u C(u,v)=(2u-1)v^2+(2-2u)v.
\end{align*}
Thus, when $X_1=0$, we have $F_2(x_2|X_1=0)=\partial_u C(0,F_2(x_2))=2F_2(x_2)-F_2^2(x_2)$.

\section{Derivations of Estimators and Additional Simulation Results for \Cref{sec:simulation}}\label{appd:details-simu}
\subsection{Derivations of Estimators for the I.I.D. Demand Case}\label{appd:deri-simu}
\subsubsection{Leibniz Divergence Estimator}
The total inventory in a regenerative cycle, $Y$, and the cycle length, $Z$, can be written explicitly as follows:
\begin{align}\label{eq:sample-org-W-T}
    \begin{split}
        &Z=1+\sum_{t=1}^\infty \prod_{i=1}^t \1\left\{\sum_{j=1}^i \log(X_j+\theta)\leq q\right\},\\
        &Y=S+\sum_{t=1}^\infty \left(S-\sum_{i=1}^t\log(X_i+\theta)\right)\prod_{i=1}^t \1\left\{\sum_{j=1}^i \log(X_j+\theta)\leq q\right\}.
    \end{split}
\end{align}
We note in \Cref{eq:sample-org-W-T} that each term in the summations for both $Y$ and $Z$ falls into the class of sample performances discussed in \Cref{exp5:inventory}. Specifically, we can write $Y:=S + \sum_{t=1}^\infty \psi_t(X,\theta)$, where
\begin{align}
    \begin{split}\label{eq:sample-perform-W}
        &\psi_t(X,\theta)=\varphi_t(X_1,\cdots,X_t,\theta)\prod_{j=1}^{t}\1\Bigg\{\sum_{i=1}^{j}z_i(X_i,\theta)< q\Bigg\},\\
        &\varphi_t(X_1,\cdots,X_t,\theta)= S-\sum_{i=1}^t z_t(X_t,\theta).
    \end{split}
\end{align}
We focus on the derivative of $\E(\psi_t(X,\theta))$. As developed in \Cref{exp5:inventory}, the $i$-th component of a feasible mapping $h: \R^n \times \Theta \to \R^n$ satisfying $h(V, \theta) = g(\Omega, \theta) \cap U$, where $V = (0,1)^n$, can be recursively defined as follows:
\begin{align*}
    h_i(v,\theta)=\left(\exp(q)\prod_{j=1}^{i-1}\left(h_j(v,\theta)+\theta\right)^{-1}\theta^{-(t-i)}-\theta\right)v_i.
\end{align*}
To evaluate $\partial_\theta h_i (v,\theta)|_{v=h^{-1}(x,\theta)}$, we first solve the equation $x=h(v,\theta)$ for $v$. This yields:
\begin{align*}
    x_i=h_i(v,\theta)=\left(\exp(q)\prod_{j=1}^{i-1}\left(x_j+\theta\right)^{-1}\theta^{-(t-i)}-\theta\right)v_i,
\end{align*}
which we can then rearrange to obtain:
\begin{align*}
    v_i=x_i\left(\exp(q)\prod_{j=1}^{i-1}\left(x_j+\theta\right)^{-1}\theta^{-(t-i)}-\theta\right)^{-1}.
\end{align*}
Next, we compute $\partial_\theta h_i (v,\theta)$ and $\partial_\theta h_i (v,\theta)|_{v=h^{-1}(x,\theta)}$ as follows:
\begin{align*}
    &\partial_\theta h_i (v,\theta)= \left( \exp(q)\prod_{j=1}^{i-1}\left(h_j(v,\theta)+\theta\right)^{-1}\theta^{-(t-i)} \left( \sum_{j=1}^{i-1}\frac{-\partial_\theta h_j (v,\theta)-1}{h_j(v,\theta)+\theta} -\frac{t-i}{\theta} \right) -1 \right)v_i,\\
    &\partial_\theta h_i (v,\theta)|_{v=h^{-1}(x,\theta)}= \left( \exp(q)\prod_{j=1}^{i-1}\left(x_j+\theta\right)^{-1}\theta^{-(t-i)} \left( \sum_{j=1}^{i-1}\frac{-\partial_\theta h_j (v,\theta)|_{v=h^{-1}(x,\theta)}-1}{x_j+\theta} -\frac{t-i}{\theta} \right) -1 \right)\\
    &\times x_i\left(\exp(q)\prod_{j=1}^{i-1}\left(x_j+\theta\right)^{-1}\theta^{-(t-i)}-\theta\right)^{-1}.
\end{align*}
Let $f_t:\R_+^t\mapsto\R_+$ denote the joint density of $(X_1,\cdots,X_t)$. Finally, applying \Cref{thm:glr}, the Leibniz divergence estimator for $\frac{d}{d\theta}\E(\psi_t(X,\theta))$ is given by
\begin{align*}
    &\1\{\varphi_t(X,\theta)>s\} \Bigg( -\sum_{i=1}^t(X_i+\theta)^{-1} - \sum_{i=1}^t(X_i+\theta)^{-1}\partial_\theta h_i (v,\theta)|_{v=h^{-1}(x,\theta)} \\
    &+ \varphi_t(X,\theta)\sum_{i=1}^t \left( \partial_\theta h_i (v,\theta)|_{v=h^{-1}(x,\theta)}\partial_{x_i}f_t(x_1,\cdots,x_t)/f_t(x_1,\cdots,x_t) + \partial_{x_i}\left(\partial_\theta h_i (v,\theta)|_{v=h^{-1}(x,\theta)}\right) \right) \Bigg).
\end{align*}
The Leibniz divergence estimator for $\frac{d}{d\theta}\E(Y)$ is obtained by summing the above terms over $t$, which can be expressed as:
\begin{align*}
    \sum_{t=1}^\tau &\1\{\varphi_t(X,\theta)>s\} \Bigg( -\sum_{i=1}^t(X_i+\theta)^{-1} - \sum_{i=1}^t(X_i+\theta)^{-1}\partial_\theta h_i (v,\theta)|_{v=h^{-1}(x,\theta)} \\
    &+ \varphi_t(X,\theta)\sum_{i=1}^t \left( \partial_\theta h_i (v,\theta)|_{v=h^{-1}(x,\theta)}\partial_{x_i}f_t(x_1,\cdots,x_t)/f_t(x_1,\cdots,x_t) + \partial_{x_i}\left(\partial_\theta h_i (v,\theta)|_{v=h^{-1}(x,\theta)}\right) \right) \Bigg),
\end{align*}
where $\tau=\min\{t\geq 1 | \varphi_t(X,\theta)\leq s\}$. The Leibniz divergence estimator for $\frac{d}{d\theta}\E(Z)$ can be derived similarly (using the same mapping $h$):
\begin{align*}
    \sum_{t=1}^\tau \1\{\varphi_t(X,\theta)>s\} \sum_{i=1}^t \left( \partial_\theta h_i (v,\theta)|_{v=h^{-1}(x,\theta)}\partial_{x_i}f_t(x_1,\cdots,x_t)/f_t(x_1,\cdots,x_t) + \partial_{x_i}\left(\partial_\theta h_i (v,\theta)|_{v=h^{-1}(x,\theta)}\right) \right).
\end{align*}
\subsubsection{Leibniz Integral Estimator}\label{sec:simu-int-est}
To apply the Leibniz integral estimator for $\frac{d}{d\theta}\E(Y)$, we return to the expression for the sample performance in \Cref{eq:sample-perform-W} and apply \Cref{thm:glr1} to estimate $\frac{d}{d\theta}\E(\psi_t(X,\theta))$ for each $t$. We write $\psi_t(X,\theta)=\phi_t(g(X,\theta))$, where
\begin{align*}
    &g:\R^t\times\Theta\mapsto \R^t,~ g_i(x,\theta)=z_i(x,\theta)=\log(x_i+\theta),~i=1,\cdots,t,~x=(x_1,\cdots,x_t),\\
    &\phi_t:\R^t\mapsto\R,~\phi_t(y)=\left(S-\sum_{i=1}^t y_i\right) \prod_{i=1}^t \1\left\{\sum_{k=1}^i y_k <q\right\},~y=(y_1,\cdots,y_t).
\end{align*}
Next, we determine the following quantities:
\begin{align}\label{eq:num-glr-comp}
    \begin{split}
        &J_g(x,\theta)=\diag\left((x_1+\theta)^{-1},\cdots,(x_t+\theta)^{-1}\right),\\ 
        &s(x,\theta)=(1,\cdots,1),~ d(x,\theta)= \sum_{i=1}^t\partial_{x_i}f_t(x_1,\cdots,x_t)/f_t(x_1,\cdots,x_t).
    \end{split}
\end{align}
Let $f_{X|X_i=x}$ denote the conditional density of $X$ given $X_i=x$, and let $f$ denote the marginal density of each $X_i$. Substituting \Cref{eq:num-glr-comp} into \Cref{thm:glr1}, the Leibniz integral estimator for $\frac{d}{d\theta}\E(\psi_t(X,\theta))$ is then given by:
\begin{align*}
    \psi_t(X,\theta)d(x,\theta) + \sum_{i=1}^t \psi_t(X,\theta)f(0)|_{X\sim f_{X|X_i=0}}.
\end{align*}
The Leibniz divergence estimator for $\frac{d}{d\theta}\E(Y)$ is obtained by summing the above terms over $t$, which can be expressed as:
\begin{align*}
    \sum_{t=1}^{\tau}\left(\psi_t(X,\theta)d(x,\theta) + \sum_{i=1}^t \psi_t(X,\theta)f(0)|_{X\sim f_{X|X_i=0}}\right).
\end{align*}
The Leibniz integral estimator for $\frac{d}{d\theta}\E(Z)$ can be derived similarly (using the same mapping $g$):
\begin{align*}
    \sum_{t=1}^{\tau}\left(\1\Bigg\{\sum_{i=1}^{j}z_i(X_i,\theta)< q\Bigg\}d(x,\theta) + \sum_{i=1}^t \1\Bigg\{\sum_{i=1}^{j}z_i(X_i,\theta)< q\Bigg\}f(0)|_{X\sim f_{X|X_i=0}}\right).
\end{align*}

\subsection{Derivative Estimation for the Regime-Switching Model}\label{appd:reg-switch}
For the regime-switching model, the joint process $(W_t, S_t)$ is regenerative only when two conditions are satisfied simultaneously:
\begin{itemize}
    \item Inventory reset: The inventory level is restored to $S$.
    \item Regime reset: The underlying demand state returns to a fixed, designated state (e.g., $S_t = 1$).
\end{itemize}
Consequently, a single regenerative cycle may contain multiple replenishments. This cycle consists of two different types of path components:
\begin{itemize}
    \item Type $1$: Starts at $\{W_0=S,S_0=1\}$, ends at first replenishment.
    \item Type $2$: Starts at $\{W_0=S,S_0=2\}$, ends at first replenishment.
\end{itemize}
For a type $i$ path, we define $Y_i$ as the cumulative inventory level and $Z_i$ as its duration. \textcolor{red}{Let $\pi_i$ denote the proportion of Type $i$ paths. Then, $\E(Y)=\sum_{i=1}^2\pi_i \E(Y_i)$, $\E(Z)=\sum_{i=1}^2\pi_i \E(Z_i)$. The derivative of the ratio can be expressed as:
\begin{align*}
    \frac{d}{d\theta}\frac{\E(Y)}{\E(Z)}
    =\frac{\sum_{i=1}^2\pi_i \frac{d}{d\theta}\E(Y_i)+\sum_{i=1}^2\frac{d}{d\theta}\pi_i \E(Y_i)}{\sum_{i=1}^2\pi_i \E(Z_i)}-\frac{\sum_{i=1}^2\pi_i \E(Y_i)(\sum_{i=1}^2\frac{d}{d\theta}\pi_i \E(Z_i)+\sum_{i=1}^2\pi_i \frac{d}{d\theta}\E(Z_i))}{(\sum_{i=1}^2\pi_i \E(Z_i))^2}.
\end{align*}} 
To estimate this, it suffices to estimate $\E(Y_i)$, $\E(Z_i)$, \textcolor{red}{$\pi_i$}, and their respective derivatives w.r.t. $\theta$. To illustrate the application of the Leibniz derivative estimators, \Cref{sec:simulation} focuses on estimating $\frac{d}{d\theta}\E(Y_1)$ and $\frac{d}{d\theta}\E(Z_1)$; the derivatives associated with Type 2 paths can be estimated similarly. Notice that the sample performances take exactly the same form as in \Cref{eq:sample-org-W-T}:
\begin{align*}
    &\overline{W_i}=S+\sum_{t=1}^\infty (S-\sum_{i=1}^t\log(X_i+\theta))\prod_{i=1}^t \1\{\sum_{j=1}^i \log(X_j+\theta)\leq q\},\\
    &\overline{Z_i}=1+\sum_{t=1}^\infty \prod_{i=1}^t \1\{\sum_{j=1}^i \log(X_j+\theta)\leq q\}.
\end{align*}
So the derivative estimators developed in \ref{appd:deri-simu} still apply.

\textcolor{red}{
We now provide a method for estimating the derivative of $\pi_i$. This method is similar to the conditional density estimation approach in \citet{l2022monte} and the conditional Leibniz method to be discussed in \ref{appd:cmccov-leibniz}. Define the replenishment epochs by $\{\tau_i\}$. Because every path begins with the same inventory level $S$, $\{S_{\tau_i}\}$ is a two-state Markov chain. Let $Q$ denote its transition matrix, where $Q_{ij}=\P(S_{\tau_{k+1}}=j|S_{\tau_k}=i)$. Assuming this Markov chain is irreducible, its stationary distribution is exactly $\pi:=(\pi_1,\pi_2)$, where $\pi_1=\frac{Q_{21}}{Q_{12}+Q_{21}}$ and $\pi_2=\frac{Q_{12}}{Q_{12}+Q_{21}}$. To estimate the derivative of $\pi_i$, it suffices to estimate the derivatives of $Q_{ij}$ w.r.t. $\theta$. We consider $Q_{12}$ as an example. Conditioning on $S_0=1$, we have $Q_{12}=\P(S_{\tau_1}=2|S_0=1)=\E(\1\{S_{\tau_1}=2\}|S_0=1)$. Let $C_t(\theta)= \sum_{i=1}^t z_i(X_i,\theta)$. We can express $\1\{S_{\tau_1}=2\}$ by the following telescoping sum:
\begin{align*}
    \1\{S_{\tau_1}=2\}&=\1\{S_1=2\}+\sum_{t=1}^\infty (\1\{S_{t+1}=2\}- \1\{S_{t}=2\})\1\{\tau_1>t\}\\ 
    &=\1\{S_1=2\}+\sum_{t=1}^\infty (\1\{S_{t+1}=2\}- \1\{S_{t}=2\})\1\{C_t(\theta)<q\}.
\end{align*}
Recall that given $S_t=s$, we have $X_t|\{S_t=s\} \sim f(\cdot,\alpha_s)$. Let $F(\cdot,\alpha_s)$ denote the corresponding CDF. Let $\mathcal H_t$ denote the history up to time $t-1$ together with $S_t$ and $S_{t+1}$. Conditioning on $\mathcal H_t$, we can write
\begin{align*}
    \E((\1\{S_{t+1}=2\}- \1\{S_{t}=2\})\1\{C_t(\theta)<q\}|\mathcal H_t)=(\1\{S_{t+1}=2\}- \1\{S_{t}=2\}) F(q-C_{t-1}(\theta),\alpha_{S_t}).
\end{align*}
Differentiating gives
\begin{align*}
    \frac{d}{d\theta}F(q-C_{t-1}(\theta),\alpha_{S_t}) = -f(q-C_{t-1}(\theta),\alpha_{S_t})\frac{d C_{t-1}(\theta)}{d\theta},
\end{align*}
where
\begin{align*}
    \frac{d C_{t-1}(\theta)}{d\theta}=\sum_{i=1}^{t-1}\frac{d z_i(X_i,\theta)}{d\theta}=\sum_{i=1}^{t-1} \partial_\theta z_i(X_i,\theta).
\end{align*}
Let $A_{t}:=-(\1\{S_{t+1}=2\}-\1\{S_t=2\})f(q-C_{t-1}(\theta),\alpha_{S_t})\frac{d C_{t-1}(\theta)}{d\theta}.$ Notice that $A_t$ depends on the history up to time $t-1$ and on $S_t,S_{t+1}$. Notice that the first term in the telescoping sum has zero derivative. Given $S_0=1$, we can therefore express $Q'_{12}(\theta)=\sum_{t=1}^\infty \E(A_{t})$. We can convert this infinite sum into an expectation evaluated at the stopping time $\tau_1$ as follows. On the event $\{\tau_1\geq t\}$, given the history $\mathcal H_t$, we have $\tau_1=t\Leftrightarrow C_t(\theta)\geq q$. Therefore, $\P(\tau_1=t|\mathcal H_t)=\1\{\tau_1\geq t\}(1-F(q-C_{t-1}(\theta),\alpha_{S_t}))=\1\{\tau_1\geq t\}\overline{F}(q-C_{t-1}(\theta),\alpha_{S_t}),$ 
where $\overline F=1-F$. Consequently,
\begin{align*}
    \E\left(\1\{\tau_1=t\}\frac{A_t}{\overline{F}(q-C_{t-1}(\theta),\alpha_{S_t})}\bigg|\mathcal H_t\right)=A_t \frac{\P(\tau_1=t|\mathcal H_t)}{\overline{F}(q-C_{t-1}(\theta),\alpha_{S_t})}=A_t.
\end{align*}
where the last equality follows because $A_t=0$ on $\{\tau_1<t\}$. Thus,
\begin{align*}
    \E(A_t|S_0=1)=\E\left(\1\{\tau_1=t\}\frac{A_t}{\overline{F}(q-C_{t-1}(\theta),\alpha_{S_t})}\bigg|S_0=1\right).
\end{align*}
Now substitute this into the infinite sum and assume that we can interchange the order of summation and expectation:
\begin{align*}
    Q'_{12}(\theta)= \E\left(\sum_{t=1}^\infty \1\{\tau_1=t\}\frac{A_t}{\overline{F}(q-C_{t-1}(\theta),\alpha_{S_t})}\bigg|S_0=1\right).
\end{align*}
On every sample path, exactly one indicator is nonzero, namely the one with $t=\tau_1$. Therefore, we can write
\begin{align*}
    Q'_{12}(\theta)&= \E\left(\frac{A_{\tau_1}}{\overline{F}(q-C_{\tau_1-1}(\theta),\alpha_{S_{\tau_1}})}\bigg|S_0=1\right)\\
    &=\E\left((\1\{S_{\tau_1}=2\}- \1\{S_{\tau_1+1}=2\})\frac{f(q-C_{\tau_1-1}(\theta),\alpha_{S_{\tau_1}})}{\overline{F}(q-C_{\tau_1-1}(\theta),\alpha_{S_{\tau_1}})}\frac{d C_{\tau_1-1}(\theta)}{d\theta}\bigg|S_0=1\right).
\end{align*}
An unbiased estimator for $Q'_{12}(\theta)$ is thus given by
\begin{align*}
    (\1\{S_{\tau_1}=2\}- \1\{S_{\tau_1+1}=2\})\frac{f(q-C_{\tau_1-1}(\theta),\alpha_{S_{\tau_1}})}{\overline{F}(q-C_{\tau_1-1}(\theta),\alpha_{S_{\tau_1}})}\sum_{i=1}^{\tau_1-1} \partial_\theta z_i(X_i,\theta).
\end{align*}
}

\subsection{Additional Simulation Results}
Additional results for the i.i.d. demand case are provided in \Cref{tb:indp-gamma-1,tb:fd-results,tb:indp-gamma-2,tb:indp-log-1,tb:indp-log-2,tb:indp-log-3}, and for the two-state regime-switching model in \Cref{tb:cor-gamma-1,tb:cor-gamma-2}.

\begin{table}[H]
    \centering
    \begin{tabularx}{\textwidth}{c *{4}{Y}}
    \toprule
    & \multicolumn{2}{c}{$N=10^4$} & \multicolumn{2}{c}{$N=10^6$} \\
    \cmidrule(lr){2-3} \cmidrule(lr){4-5}
    Estimator & $d\E(Y)/d\theta$ & $d\E(Z)/d\theta$ & $d\E(Y)/d\theta$ & $d\E(Z)/d\theta$ \\
    \midrule
    FD(0.1) & $-3.374(0.051)$ & $-1.433(0.037)$ & $-3.303(0.005)$ & $-1.379(0.004)$ \\
    FD(0.05) & $-3.546(0.066)$ & $-1.498(0.053)$ & $-3.469(0.006)$ & $-1.439(0.005)$ \\
    FD(0.01) & $-3.675(0.129)$ & $-1.530(0.123)$ & $-3.629(0.012)$ & $-1.507(0.012)$ \\
    FD(0.001) & $-4.268(0.460)$ & $-2.100(0.458)$ & $-3.642(0.039)$ & $-1.498(0.039)$ \\
    Leibniz divergence & $-3.903(0.083)$ & $-1.734(0.041)$ & $-3.864(0.008)$ & $-1.718(0.004)$ \\
    Leibniz integral (GLR) & $10^4(10^3)$ & $10^3(10^3)$ & $10^6(10^6)$ & $10^6(10^6)$ \\
    \bottomrule
    \end{tabularx}
    \caption{Parameters: $s=1$, $S=3$, $\theta=1.1$, $X_t\sim$Gamma$(0.5,5)$. $\E(Z)=2.869$.} 
    \label{tb:indp-gamma-1}
    \vspace*{-3em}
\end{table}

\begin{table}[H]
    \centering
    \begin{tabularx}{\textwidth}{c *{6}{Y}}
    \toprule
    & \multicolumn{2}{c}{$N=10^6$} & \multicolumn{2}{c}{$N=10^7$} & \multicolumn{2}{c}{$N=10^8$} \\
    \cmidrule(lr){2-3} \cmidrule(lr){4-5} \cmidrule(lr){6-7}
    Estimator & $d\E(Y)/d\theta$ & $d\E(Z)/d\theta$ & $d\E(Y)/d\theta$ & $d\E(Z)/d\theta$ & $d\E(Y)/d\theta$ & $d\E(Z)/d\theta$ \\
    \midrule
    FD($10^{-4}$) & $-3.727(0.126)$ & $-1.580(0.126)$ & $-3.705(0.039)$ & $-1.557(0.039)$ & $-3.668(0.039)$ & $-1.522(0.039)$ \\
    FD($10^{-5}$) & $-4.247(0.458)$ & $-2.100(0.458)$ & $-3.518(0.117)$ & $-1.370(0.117)$ & $-3.546(0.374)$ & $-1.400(0.374)$ \\
    FD($10^{-6}$) & $-3.154(1.000)$ & $-1.000(1.000)$ & $-4.048(0.436)$ & $-1.900(0.436)$ & $-3.626(0.122)$ & $-1.480(0.122)$ \\
    FD($10^{-7}$) & $-2.155(0.003)$ & $0(0)$ & $-5.148(1.732)$ & $-3.000(1.732)$ & $-3.446(0.361)$ & $-1.300(0.361)$ \\
    FD($10^{-8}$) & $-2.155(0.003)$ & $0(0)$ & $-2.148(0.001)$ & $0(0)$ & $-4.146(1.414)$ & $-2.000(1.414)$ \\
    \bottomrule
    \end{tabularx}
    \caption{Parameters: $s=1$, $S=3$, $\theta=1.1$, $X_t\sim$Gamma$(0.5,5)$. $\E(Z)=2.869$.}
    \label{tb:fd-results}
    \vspace*{-3em}
\end{table}

\begin{table}[H]
    \centering
    \begin{tabularx}{\textwidth}{c *{4}{Y}}
    \toprule
    & \multicolumn{2}{c}{$N=10^4$} & \multicolumn{2}{c}{$N=10^6$} \\
    \cmidrule(lr){2-3} \cmidrule(lr){4-5}
    Estimator & $d\E(Y)/d\theta$ & $d\E(Z)/d\theta$ & $d\E(Y)/d\theta$ & $d\E(Z)/d\theta$ \\
    \midrule
    FD(0.1) & $-7.369(0.062)$ & $-3.529(0.050)$ & $-7.403(0.006)$ & $-3.553(0.005)$ \\
    FD(0.05) & $-7.920(0.089)$ & $-3.806(0.079)$ & $-7.926(0.009)$ & $-3.801(0.008)$ \\
    FD(0.01) & $-8.500(0.203)$ & $-4.130(0.199)$ & $-8.421(0.020)$ & $-4.045(0.020)$ \\
    FD(0.001) & $-8.029(0.601)$ & $-3.600(0.599)$ & $-8.570(0.064)$ & $-4.133(0.064)$ \\
    Leibniz divergence & $-8.586(0.184)$ & $-4.113(0.105)$ & $-8.479(0.018)$ & $-4.045(0.010)$ \\
    Leibniz integral (GLR) & $-9.113(1.270)$ & $-4.284(0.661)$ & $-8.056(0.192)$ & $-3.849(0.106)$ \\
    \bottomrule
    \end{tabularx}
    \caption{Parameters: $s=1$, $S=3$, $\theta=1.1$, $X_t\sim$Gamma$(1.5,0.5)$. $\E(Z)=4.142$.} 
    \label{tb:indp-gamma-2}
    \vspace*{-3em}
\end{table}

\begin{table}[H]
    \centering
    \begin{tabularx}{\textwidth}{c *{4}{Y}}
    \toprule
    & \multicolumn{2}{c}{$N=10^4$} & \multicolumn{2}{c}{$N=10^6$} \\
    \cmidrule(lr){2-3} \cmidrule(lr){4-5}
    Estimator & $d\E(Y)/d\theta$ & $d\E(Z)/d\theta$ & $d\E(Y)/d\theta$ & $d\E(Z)/d\theta$ \\
    \midrule
    FD(0.1) & $-2.585(0.049)$ & $-0.453(0.024)$ & $-2.661(0.005)$ & $-0.487(0.002)$ \\
    FD(0.05) & $-2.659(0.054)$ & $-0.460(0.032)$ & $-2.751(0.006)$ & $-0.504(0.003)$ \\
    FD(0.01) & $-2.700(0.081)$ & $-0.440(0.066)$ & $-2.829(0.009)$ & $-0.518(0.007)$ \\
    FD(0.001) & $-2.673(0.205)$ & $-0.400(0.200)$ & $-2.832(0.023)$ & $-0.507(0.022)$ \\
    Leibniz divergence & $-2.761(0.077)$ & $-0.536(0.020)$ & $-2.897(0.009)$ & $-0.576(0.002)$ \\
    Leibniz integral (GLR) & $10^5(10^4)$ & $10^4(10^3)$ & $10^5(10^4)$ & $10^4(10^4)$ \\
    \bottomrule
    \end{tabularx}
    \caption{Parameters: $s=1$, $S=5$, $\theta=1.1$, $X_t\sim$Lognormal$(1,5)$. $\E(Z)=2.708$.}
    \label{tb:indp-log-1}
    \vspace*{-3em}
\end{table}

\begin{table}[H]
    \centering
    \begin{tabularx}{\textwidth}{c *{4}{Y}}
    \toprule
    & \multicolumn{2}{c}{$N=10^4$} & \multicolumn{2}{c}{$N=10^6$} \\
    \cmidrule(lr){2-3} \cmidrule(lr){4-5}
    Estimator & $d\E(Y)/d\theta$ & $d\E(Z)/d\theta$ & $d\E(Y)/d\theta$ & $d\E(Z)/d\theta$ \\
    \midrule
    FD(0.1) & $-2.060(0.029)$ & $-0.623(0.024)$ & $-2.084(0.003)$ & $-0.636(0.002)$ \\
    FD(0.05) & $-2.127(0.039)$ & $-0.656(0.036)$ & $-2.135(0.004)$ & $-0.651(0.004)$ \\
    FD(0.01) & $-2.149(0.082)$ & $-0.650(0.080)$ & $-2.171(0.008)$ & $-0.659(0.008)$ \\
    FD(0.001) & $-2.206(0.265)$ & $-0.700(0.265)$ & $-2.185(0.026)$ & $-0.666(0.026)$ \\
    Leibniz divergence & $-2.096(0.090)$ & $-0.632(0.032)$ & $-2.176(0.009)$ & $-0.663(0.003)$ \\
    Leibniz integral (GLR) & $-2.188(0.183)$ & $-0.666(0.061)$ & $-2.181(0.019)$ & $-0.662(0.006)$ \\
    \bottomrule
    \end{tabularx}
    \caption{Parameters: $s=1$, $S=5$, $\theta=1.1$, $X_t\sim$Lognormal$(1,1)$. $\E(Z)=3.398$.}
    \label{tb:indp-log-2}
    \vspace*{-3em}
\end{table}

\begin{table}[H]
    \centering
    \begin{tabularx}{\textwidth}{c *{4}{Y}}
    \toprule
    & \multicolumn{2}{c}{$N=10^4$} & \multicolumn{2}{c}{$N=10^6$} \\
    \cmidrule(lr){2-3} \cmidrule(lr){4-5}
    Estimator & $d\E(Y)/d\theta$ & $d\E(Z)/d\theta$ & $d\E(Y)/d\theta$ & $d\E(Z)/d\theta$ \\
    \midrule
    FD(0.1) & $-5.959(0.089)$ & $-0.754(0.031)$ & $-6.078(0.009)$ & $-0.766(0.003)$ \\
    FD(0.05) & $-6.200(0.098)$ & $-0.834(0.044)$ & $-6.265(0.010)$ & $-0.789(0.004)$ \\
    FD(0.01) & $-6.308(0.127)$ & $-0.800(0.089)$ & $-6.429(0.013)$ & $-0.814(0.009)$ \\
    FD(0.001) & $-6.239(0.282)$ & $-0.700(0.265)$ & $-6.467(0.030)$ & $-0.819(0.028)$ \\
    Leibniz divergence & $-6.095(0.22)$ & $-0.778(0.041)$ & $-6.460(0.025)$ & $-0.827(0.004)$ \\
    Leibniz integral (GLR) & $10^5(10^5)$ & $10^5(10^4)$ & $10^5(10^5)$ & $10^5(10^4)$ \\
    \bottomrule
    \end{tabularx}
    \caption{Parameters: $s=1$, $S=10$, $\theta=1.1$, $X_t\sim$Lognormal$(1,5)$. $\E(Z)=4.508$.}
    \label{tb:indp-log-3}
    \vspace*{-3em}
\end{table}

\begin{table}[H]
    \centering
    \begin{tabularx}{\textwidth}{c *{4}{Y}}
    \toprule
    & \multicolumn{2}{c}{$N=10^4$} & \multicolumn{2}{c}{$N=10^6$} \\
    \cmidrule(lr){2-3} \cmidrule(lr){4-5}
    Estimator & $d\E(Y)/d\theta$ & $d\E(Z)/d\theta$ & $d\E(Y)/d\theta$ & $d\E(Z)/d\theta$ \\
    \midrule
    FD(0.1) & $-2.904(0.050)$ & $-1.094(0.034)$ & $-2.798(0.005)$ & $-1.048(0.003)$ \\
    FD(0.05) & $-3.016(0.061)$ & $-1.112(0.047)$ & $-2.931(0.006)$ & $-1.092(0.005)$ \\
    FD(0.01) & $-3.031(0.108)$ & $-1.040(0.101)$ & $-3.031(0.011)$ & $-1.114(0.010)$ \\
    FD(0.001) & $-2.804(0.285)$ & $-0.800(0.283)$ & $-2.995(0.033)$ & $-1.060(0.033)$ \\
    Leibniz divergence & $-3.345(0.082)$ & $-1.323(0.040)$ & $-3.159(0.008)$ & $-1.224(0.003)$ \\
    Leibniz integral (GLR) & N/A & N/A & N/A & N/A \\
    \bottomrule
    \end{tabularx}
    \caption{Parameters: $s=1$, $S=3$, $\theta=1.1$; $X_t \mid S_t=1 \sim \text{Gamma}(0.5,5)$ and $X_t \mid S_t=2 \sim \text{Gamma}(3,5)$. $\E(Z)=2.635$. Note that the Leibniz integral estimator is inapplicable here due to the presence of a non-vanishing surface integral.} 
    \label{tb:cor-gamma-1}
    \vspace*{-3em}
\end{table}

\begin{table}[H]
    \centering
    \begin{tabularx}{\textwidth}{c *{4}{Y}}
    \toprule
    & \multicolumn{2}{c}{$N=10^4$} & \multicolumn{2}{c}{$N=10^6$} \\
    \cmidrule(lr){2-3} \cmidrule(lr){4-5}
    Estimator & $d\E(Y)/d\theta$ & $d\E(Z)/d\theta$ & $d\E(Y)/d\theta$ & $d\E(Z)/d\theta$ \\
    \midrule
    FD(0.1) & $-2.065(0.030)$ & $-0.637(0.025)$ & $-2.037(0.003)$ & $-0.622(0.002)$ \\
    FD(0.05) & $-2.109(0.039)$ & $-0.646(0.035)$ & $-2.078(0.004)$ & $-0.630(0.003)$ \\
    FD(0.01) & $-2.082(0.078)$ & $-0.590(0.077)$ & $-2.119(0.008)$ & $-0.642(0.008)$ \\
    FD(0.001) & $-2.698(0.347)$ & $-1.200(0.346)$ & $-2.179(0.026)$ & $-0.696(0.033)$ \\
    Leibniz divergence & $-2.257(0.087)$ & $-0.693(0.033)$ & $-2.144(0.009)$ & $-0.655(0.003)$ \\
    Leibniz integral (GLR) & $-2.061(0.496)$ & $-0.582(0.150)$ & $-2.370(0.264)$ & $-0.780(0.141)$ \\
    \bottomrule
    \end{tabularx}
    \caption{Parameters: $s=1$, $S=5$, $\theta=1.1$; $X_t \mid S_t=1 \sim \text{Gamma}(1.5,1.5)$ and $X_t \mid S_t=2 \sim \text{Gamma}(3,1.5)$. $\E(Z)=3.394$.} 
    \label{tb:cor-gamma-2}
\end{table}

\section{Comparison of PA, LR and Their Variants}\label{appd:prac-implem}
In this section, we compare the required assumptions, sampling costs, and applicability of various derivative estimators. By clarifying their respective advantages and limitations, we aim to provide concrete guidance for practical implementation. A summary of these comparisons is provided in \Cref{tab:estimator_comparison}.
\subsection{LR Variants and the Leibniz Framework} 
The standard LR estimator does not require sample performance continuity, but it is limited to distributional parameters. To handle structural parameters within a discontinuous sample performance, several LR variants have been developed, such as push-out LR, SLRIPA, Leibniz integral (GLR), and Leibniz divergence estimators. The general principle behind these methods is to use a change of variables to shift the parameter dependence into the input density function and/or its support. As a result, the transformed sample performance either becomes continuous w.r.t. the parameter or independent of it entirely. Notice that Augmented PA (APA) \citep{gaivoronski1992augmented} and Discontinuous PA (DPA) \citep{shi1996discontinuous,shi1996variance}, despite their names, can be viewed as special one-dimensional cases of the Leibniz integral estimator. Next, we present a detailed comparison of these LR variants. Since, as discussed in \Cref{sec2:inv,sec:compare}, GLR is a special case of the Leibniz integral estimator, we do not discuss it separately.
\begin{itemize}
    \item Push-out LR vs. SLRIPA: Both push-out LR and SLRIPA require a global change of variables, and a transformed density with parameter-independent support. SLRIPA is more general than push-out LR: the latter requires the transformed sample performance to be fully parameter-independent, whereas SLRIPA allows parameter dependence as long as the transformed sample performance is continuous.
    \item The Leibniz framework vs. SLRIPA: The Leibniz framework is more general than SLRIPA, since it requires only local changes of variables and can accommodate parameter-dependent supports.
    \item Leibniz integral vs. Leibniz divergence: The main challenge for the Leibniz integral estimator is the evaluation of the surface integral, which can be complicated and computationally demanding, especially when the input random variables are correlated. The Leibniz divergence estimator avoids the surface integral entirely, but requires the identification of a smooth parameterization of the integration domain (\Cref{assump:thm1-a2}). While the examples in the paper suggest that piecewise smooth boundaries are often sufficient, such a parameterization is still problem-specific in general. 
\end{itemize}

\begin{table}[bt]
\centering
\caption{Comparison of derivative estimators that can handle discontinuous sample performance.}
\label{tab:estimator_comparison}
\renewcommand{\arraystretch}{1.3} 
\begin{tabular}{@{} l l l p{7cm} @{}}
\toprule
Estimator & Parameter & Sample Cost & Key Assumptions \& Limitations \\
\midrule
PA Variants (e.g., SPA) & General & Single/Multi-run & Highly problem-specific; requires identifying suitable random variables to condition on. \\
Standard LR & Distributional & Single-run & Cannot handle structural parameters. \\
Push-out LR & General & Single-run & Requires global changes of variables; transformed density must have parameter-independent support; transformed sample performance must be parameter-independent. \\
SLRIPA & General & Single-run & Requires global changes of variables; transformed density must have parameter-independent support; allows continuous parameter dependence in sample performance. \\
Leibniz Integral (GLR) & General & Single/Multi-run & Requires only local changes of variables; accommodates parameter-dependent supports; surface integral can be computationally expensive to estimate. \textsuperscript{\dag} \\
Leibniz Divergence & General & Single-run & Accommodates parameter-dependent supports; avoids surface integral; requires smooth parameterization of the integration domain. \\
WD & Distributional & Multi-run & May require multiple sample paths, with the required number potentially increasing with the input dimension. \\
\bottomrule
\multicolumn{4}{@{}l@{}}{\footnotesize \textsuperscript{\dag} Among the compared methods, only this estimator involves a surface integral.} \\
\end{tabular}
\end{table}

\subsection{PA vs. LR} 
In this section, we review existing PA estimators and compare them with LR. IPA, the foundational PA method, is typically low-variance but applies only to continuous sample performance functions. To address discontinuities, various PA extensions have been developed \citep{fu2012conditional}, including Smoothed PA (SPA) \citep{gong1987smoothed,fu1992extensions,fu1993second,fu1994sample,miyoshi1997smoothed,heidergott1999optimisation,fu2009conditional}, Structural IPA (SIPA) \citep{dai1995structural}, Rare PA (RPA) \citep{bremaud1992pathwise}, Generalized IPA \citep{chen2014american}, and CMC-CoV \citep{feng2016conditional}. Although these variants often retain the low-variance advantages of IPA, their implementation is highly problem-specific. For instance, SPA and CMC-CoV require identifying suitable random variables to condition on to smooth the sample performance, restricting their applicability. Moreover, some PA variants, such as SPA and RPA, often require simulating multiple sample paths.

As discussed in \citet{l1990unified}, the distinction between PA- and LR-based methods is not always clearly defined, since either class can often be reformulated as the other through an appropriate representation of the sample performance or choice of probability space. Under suitable conditions, PA estimators often have lower variance than LR estimators and their variants \citep{cui2020variance}, whereas the variance of LR estimators may grow linearly with the number of input random variables whose distributions depend on the parameter. More generally, however, variance behavior, applicability, and sampling cost remain highly problem-dependent.

\section{Comparison of CMC-CoV and the Leibniz Framework}\label{appd:cmccov-leibniz}
The CMC-CoV method of \citet{feng2016conditional} considers sample performance functions of the form $l(X(\theta))\1\{h(X(\theta))\leq\xi\}$, where $l,h:\R^n\to\R,~n\geq1$, $\xi\in\R$ is a given constant, and $X=(X_1,\dots,X_n)$ is a random vector that may depend on a parameter of interest $\theta\in\Theta\subseteq\R$ and has a density function $f(x)$ supported on $\Omega\subset\R^n$. \citet{feng2016conditional} consider the problem of estimating the derivative
\begin{align*}\label{eq:cmc-1}
    \frac{d}{d\theta}\E(l(X(\theta))\1\{h(X(\theta))\leq\xi\}),
\end{align*}
and show that it fundamentally reduces to the problem of estimating the following derivative instead:
\begin{align*}
    \frac{d}{d\xi}\E(l(X(\theta))\1\{h(X(\theta))\leq\xi\}).
\end{align*}
Through a change of variables followed by conditioning, their CMC-CoV method expresses the original expectation as the expectation of a one-dimensional integral with an upper limit of $\xi$, a constant lower limit, and an integrand independent of $\xi$. This specific formulation allows for the direct application of the one-dimensional Leibniz integral rule. The primary connections and distinctions between the CMC-CoV method and the Leibniz integral estimator are as follows:
\begin{itemize}
    \item Origin of the surface integral: Both methods involve a surface integral term, but they arise from fundamentally different mathematical mechanisms. In the Leibniz integral estimator, the surface integral arises from differentiating a parameter-dependent domain after a change of variables; the CMC-CoV method applies a change of variables from a lower-dimensional space to a higher-dimensional space, transforming the original expectation (a volume integral in the lower-dimensional space) into a surface integral in the higher-dimensional space.
    \item Applicability: While both methods address indicator-type discontinuities, the formulation of the Leibniz framework in \Cref{thm:glr,thm:glr1} is more general in several respects. First, while the Leibniz framework can accommodate parameter-dependent sample spaces, CMC-CoV assumes that the dependence of $X(\theta)$ on $\theta$ is continuous, which may preclude its use for sample spaces with $\theta$-dependent boundaries. Second, CMC-CoV assumes that $h$ in sample performance (1) of \citet{feng2016conditional} is a scalar-valued function, whereas both Leibniz methods allow vector-valued counterparts. Consequently, CMC-CoV cannot be directly applied to \Cref{exp2:log} or to the inventory-pricing model in \Cref{sec:simulation} of the current paper.
    \item Efficiency \& Variance: Conceptually, CMC-CoV is close to SPA and may be viewed as a PA variant, whereas, as clarified in \Cref{sec:ipalr-leibniz}, the Leibniz framework is LR-based. Roughly speaking, suitable conditioning can reduce variance, and PA-based methods are often expected to have lower variance and greater computational efficiency, whereas LR-based methods are often more straightforward to implement \citep{l1990unified,cui2020variance}. At the same time, the relative performance of derivative estimators remains highly problem-specific, depending, for example, on the problem formulation, the choice of probability space, and the quantity chosen for conditioning \citep{l2022monte}. In the simulation experiment later in this section, CMC-CoV does achieve better variance reduction than the original Leibniz estimators. However, we also find that the Leibniz integral estimator can be combined with conditioning techniques, and the resulting conditional Leibniz estimator attains variance reduction comparable to that of CMC-CoV. Thus, neither method uniformly dominates the other in general.
\end{itemize}

Next, we conduct simulation experiments on an example of estimating the derivative of a chance-constrained program from \citet[Section 5.2]{feng2016conditional} to compare both methods in detail. Consider the following sample performance:
\begin{align*}
    \1\left\{\theta X_1 + \sum_{i=2}^n X_i \leq b \right\},
\end{align*}
where $X=(X_1,\dots,X_n)$ follows a multivariate normal distribution with zero mean and covariance matrix $\Sigma$, and $b$ is a constant. The CMC-CoV estimator is given by:
\begin{align*}
    -\frac{b|b|^{n-1}}{\Gamma(n/2)2^{n/2}} Z_1(Z^T\Sigma^{-1}Z)^{n/2} \exp\left(-\frac{1}{2}Z^T\Sigma^{-1}Zb^2\right) ,
\end{align*} 
where $Z=X/\left(\theta X_1 + \sum_{i=2}^n X_i\right)$ and $\Gamma$ is the gamma function.

To apply the Leibniz divergence estimator, we can rewrite the indicator function as $\1\left\{\theta X_1 + \sum_{i=2}^n X_i \leq b \right\}=\1\{g(X,\theta)\in U\}$, where $g_1(x,\theta)=\theta x_1 + \sum_{i=2}^n x_i$, $g_k(x,\theta)=x_k$ for $k\geq 2$, $U=U_1\times\dots\times U_n$, $U_1=(-\infty,b]$, and $U_k=\R$ for $k\geq 2$.  This aligns with the case discussed in \Cref{exp4:glr=leibniz}, which yields the following Leibniz divergence estimator:
\begin{align*}
    \1\left\{\theta X_1 + \sum_{i=2}^n X_i \leq b \right\}\left(-\frac{X_1\partial_{x_1}f(X)}{\theta f(X)}-\frac{1}{\theta}\right).
\end{align*}

To apply the Leibniz integral estimator, \Cref{thm:glr1} requires the dimension of the differentiable function inside the discontinuity to match the dimension of the input. As discussed at the end of \Cref{sec2:inv}, if the input has a higher dimension, we can condition on a subset of the variables so the dimensions match. In this example, the input is $n$-dimensional, but the function inside the indicator is scalar-valued. Therefore, we must condition on $n-1$ input random variables. A natural approach is to condition on $X_2,\dots,X_n$ and express the expected performance as:
\begin{align*}
    \E\left(\1\left\{\theta X_1 + \sum_{i=2}^n X_i \leq b \right\} \Bigg| X_2,\dots,X_n\right) 
    &=\E\left(\1\left\{ X_1 \leq \frac{1}{\theta}\left(b-\sum_{i=2}^n X_i\right) \right\} \Bigg| X_2,\dots,X_n\right)\\
    &=\int^{\frac{1}{\theta}\left(b-\sum_{i=2}^n X_i\right)}_{-\infty} f_{1|2\dots n}(x_1 | X_2,\dots,X_n) dx_1,
\end{align*}
where $f_{1|2\dots n}$ is the conditional density of $X_1$ given $X_2,\dots,X_n$, which can be expressed explicitly using the components of $\Sigma$ (details omitted here). As a result, the conditional Leibniz integral estimator can be obtained by applying the univariate Leibniz integral rule:
\begin{align*}
    -\frac{1}{\theta^2}\left(b-\sum_{i=2}^n X_i\right) f_{1|2\dots n}\left( \frac{1}{\theta}\left(b-\sum_{i=2}^n X_i\right) \Bigg| X_2,\dots,X_n \right).
\end{align*}
Notice that this sample performance is essentially identical to those in density estimation problems. Consequently, the conditional Leibniz integral estimator derived above can also be obtained using the methods in \citet{l2022monte}.

In the simulation experiment, we set $n=5$, $b=1$, and $\theta=2$. We define $\Sigma$ as an autoregressive covariance matrix with a parameter of $0.1$ and perform $N=10^4$ independent replications. The performance results for the Leibniz divergence, Leibniz integral, and CMC-CoV estimators are displayed in \Cref{tb:leibniz-cmc}. We observe that the CMC-CoV method outperforms the Leibniz divergence estimator, with only about one third of its variance. This aligns with our earlier analysis that appropriate conditioning can reduce variance, and that CMC-CoV, as a PA-based method, often exhibits lower variance than LR-based methods. \textcolor{red}{The conditional Leibniz integral estimator achieves a further variance reduction relative to CMC-CoV in this example. However, this improvement is partly attributable to the choice of conditioning and should not be interpreted as a generic advantage of the Leibniz framework.} These results suggest that a more systematic integration of Leibniz estimators with conditioning techniques is a promising direction for future research.
\vspace*{-1em}
\begin{table}[htbp]
    \centering
    \caption{Simulation results comparing the Leibniz divergence, Leibniz integral, and CMC-CoV estimators ($n=5, b=1, \theta=2, N=10^4$). Values represent point estimates (standard errors).}
    \label{tb:leibniz-cmc}
    \begin{tabular}{lc}
        \toprule
        Estimator & Estimate (Standard error) \\
        \midrule
        Leibniz divergence & -0.0299 (0.0053) \\
        Leibniz integral (conditional) & -0.0279 (0.0007) \\
        CMC-CoV & -0.0283 (0.0017) \\
        \bottomrule
    \end{tabular}
\end{table}


\end{document}